\newif\ifpre
\pretrue

\ifpre
\documentclass[twocolumn,pre,tightenlines,nofootinbib,amsmath,amssymb,amsfonts,superscriptaddress,eqsecnum]{revtex4}
\usepackage{subeqnarray}
\usepackage{xcolor}
\usepackage{hyperref}
\usepackage{lipsum}
\usepackage{accents}
\usepackage{yhmath} 
\usepackage{array}
\usepackage{graphicx}
\usepackage{bm}
\else

\documentclass{jfm}
\usepackage{mathtools}
\usepackage{bm}
\usepackage{accents}
\usepackage{yhmath} 
\usepackage{array}
\usepackage{graphicx}
\usepackage{bm}
\usepackage{amssymb, amsmath} 
\fi

\newcommand{\prebool}[2]{
\ifpre
#1
\else
#2
\fi
}
\newcommand{\myquad}{\prebool{\quad}{\qquad}}
\ifpre
\newcommand{\mytfrac}{\tfrac}
\else
\newcommand{\mytfrac}{\tfrac}
\fi

\newcommand{\be}{\begin{equation}}
\newcommand{\ee}{\end{equation}}
\newcommand{\bea}{\begin{eqnarray}}
\newcommand{\eea}{\end{eqnarray}}

\newcommand{\eps}{\varepsilon}

\newcommand{\beas}{\begin{subeqnarray}}
\newcommand{\eeas}{\end{subeqnarray}}

\newcommand{\dd}{{\rm d}}
\newcommand{\rb}{{\cal R}}
\newcommand{\hrb}{\widehat{\cal R}}
\newcommand{\gr}[1]{{\bm #1}}

\newcommand{\dimless}[1]{{\underline{#1}}}

\newcommand{\aver}[1]{{\overline{#1}}}

\newcommand{\com}[1]{}

\newcommand{\galifr}[1]{{{}^{\rm G}{#1}}}
\newcommand{\rotfr}[1]{{{}^{\rm R}{#1}}}

\newcommand{\ii}{{\rm i}}
\newcommand{\tetr}[1]{{\underline{#1}}}

\newcommand{\totV}{{\cal V}}

\newcommand{\calO}[1]{{{\cal O}(#1) }}

\newcommand{\Nv}{N}
\newcommand{\ra}{{R}}

\newcommand{\myes}{\varnothing}

\newcommand{\visc}{{\mu}} 
\newcommand{\ST}{{\nu}}
\newcommand{\myalpha}{{\alpha}}

\newcommand{\Sv}{v}
\newcommand{\Sw}{{\dot{\phi}}}

\newcommand{\EqC}{ {{}^C {\cal E}} }

\newcommand{\EqD}{ {{\cal D}} }

\newcommand{\EqCP}{ {{}^{p} {\cal C}} }
\newcommand{\EqCr}{ {{}^{r} {\cal C}} }
\newcommand{\EqCtheta}{ {{}^{\theta} {\cal C}} }

\newcommand{\EqCD}{ {{}^{D} {\cal C}} }

\newcommand{\EqBC}{ {{\cal C}} }
\newcommand{\EqBD}{ {{}^\rb {\cal E}} }

\newcommand{\myvort}{{\varpi}}

\newcommand{\Ls}{{L}}

\newcommand{\VF}{{\cal F}}

\newcommand{\yu}{{n}}
\newcommand{\tyu}{{\widetilde{\yu}}}

\newcommand{\para}[1]{{\overline{#1}}} 
\newcommand{\stf}[1]{{\widehat{#1}}} 
\newcommand{\mysh}[1]{{\wideparen{#1}}} 

\newcommand{\shV}{u}

\newcommand{\interior}[1]{\accentset{\smash{\raisebox{-0.12ex}{$\scriptstyle\circ$}}}{#1}\rule{0pt}{2.3ex}}
\newcommand{\omeg}[1]{{\interior{#1}}}

\newcommand{\tkappa}{{\widetilde{\kappa}}}

\makeatletter
\def\l@subsubsection#1#2{}
\makeatother

\ifpre

\usepackage{natbib}
\setcitestyle{authoryear,round}
\makeatletter
\def\references{%
\addcontentsline{toc}{section}{\numberline {}References}%
\vskip 2.5pc
\begin{center}
\noindent{\fontseries{bx}\fontshape{n}\selectfont REFERENCES}
\end{center}
\vskip 1.5pc
\list{}%
{
\leftmargin9pt  \itemindent-9pt
\ifdim\baselinestretch pt>1 pt %
\parsep  4pt\relax %
\else %
\parsep  0pt\relax %
\fi
\itemsep\parsep %
}%
\let\newblock\relax %
\sloppy\clubpenalty4000\widowpenalty4000
\sfcode`\.=1000\relax
\small
}

\def\endreferences{%
\def\@noitemerr{\@warning{Empty `thebibliography' environment}}%
\endlist
}

\makeatother
\def\thebibliography#1{\references}

\begin{document}
\title{One-dimensional reduction of viscous jets. I. Theory}

\author{Cyril Pitrou}
\email{pitrou@iap.fr}
\affiliation{Institut d'Astrophysique de Paris, CNRS-UMR 7095, Universit\'e Pierre~\&~Marie Curie - Paris VI, Sorbonne
  Universit\'es, 98 bis Bd Arago, 75014 Paris, France}
\affiliation{Saint-Gobain Recherche, 39 Quai Lucien Lefranc, 93300
  Aubervilliers, France}

\date{31 March 2017}

\else

\begin{document}

\shorttitle{1D reduction of viscous jets} 
\shortauthor{C. Pitrou} 

\title{One-dimensional reduction of viscous jets. I. Theory}

\author
 {
 C. Pitrou\aff{1,2}
  \corresp{\email{pitrou@iap.fr}}
  }

\affiliation
{
\aff{1}
Institut d'Astrophysique de Paris, CNRS-UMR 7095, Universit\'e Pierre~\&~Marie Curie - Paris VI, Sorbonne
  Universit\'es, 98 bis Bd Arago, 75014 Paris, France
\aff{2}
Saint-Gobain Recherche, 39 Quai Lucien Lefranc, 93300
  Aubervilliers, France
}
\fi

\ifpre
\else
\maketitle
\fi

\begin{abstract}
We build a general formalism to describe thin viscous
jets as one-dimensional objects with an internal structure. We present
in full generality the steps needed to describe the viscous jets
around their central line, and we argue that the Taylor expansion of all
fields around that line is conveniently expressed in terms of
symmetric trace-free tensors living in the two dimensions of the fiber
sections. We recover the standard results of axisymmetric jets and we report the first and second corrections to the lowest
order description, also allowing for a rotational component around the axis
of symmetry. When applied to generally curved fibers, the lowest order
description corresponds to a viscous string model whose sections are circular. However, when including the
first corrections we find that curved jets generically develop
elliptic sections. Several subtle effects imply that the first corrections cannot be
described by a rod model, since it amounts to selectively discard
some corrections. However, in a fast rotating frame we find that the dominant
effects induced by inertial and Coriolis forces should be correctly
described by rod models. For completeness, we also recover the constitutive relations for
forces and torques in rod models and exhibit a missing term in the
lowest order expression of viscous torque. Given that our method is
based on tensors, the complexity of all computations has been beaten down by using an
appropriate tensor algebra package such as {\it xAct}, allowing us to
obtain a one-dimensional description of curved
viscous jets with all the first order corrections consistently
included. Finally, we find a description for straight fibers with
elliptic sections as a special case of these results, and recover that
ellipticity is dynamically damped by surface tension. An application
to toroidal viscous fibers is presented in the companion paper
[Pitrou, Phys. Rev. E 97, 043116 (2018)].
\end{abstract}

\ifpre
\maketitle
\else
\fi

\tableofcontents

\section{Introduction}

Solving exactly the non-linear fluid equations for long viscous jets
is extremely complicated and one needs to resort to an
approximation scheme to study the dynamics of these systems. Due to
the elongated shape, there is an obvious simplification which consists
in considering a one-dimensional description.  A body is considered as being slender if its radius $\ra$ is typically much
smaller than the inverse size of velocity gradients $L$, that is if
the velocity field changes on length scales which are larger than the fiber radius. 
Hence the one-dimensional reduction induces naturally an expansion in
the slenderness parameter $\epsilon_\ra\equiv \ra/L$. Given that at lowest order, a solid is approximated by a
point particle, then we expect that a slender jet is approximated
at lowest order by some type of string. Furthermore, as extended objects are described as point particles with an
internal structure encoded in various moments (e.g. in the moment of inertia), the internal
structure of the one-dimensional object which approximates a viscous
jet is encoded in some moments which vary continuously along the
one-dimensional fiber. From the perturbative expansion in the small parameter $\epsilon_\ra$ we show how this
series of moments must be truncated at a given order of corrections around the
string description. The various moments describing the viscous jet
happen to separate naturally into moments which evolve dynamically and
moments which are related by constraints to the former ones.

For simplicity, we restrict our analysis to incompressible Newtonian fluids whose
internal forces are captured entirely by a constant viscosity
parameter, and we allow for surface tension
effects. These ingredients are sufficient to describe the
dynamics of drop formation from the Rayleigh-Plateau instability~\citep{Plateau1873,Rayleigh1878,EggersRMP}.
However, concerning the global shape of the viscous jet, our aim is to
remain as general as possible, allowing for curved fibers (that is
curved central lines) with possibly non-circular cross sections.  Indeed there are a series of geometrical
simplifications which are usually performed given the symmetries of
specific problems. From the most restrictive to the most general, we
find the axisymmetric case, the straight fiber case with non-circular
sections, the curved fiber case with circular sections, and the curved
fiber case with non-circular sections. 

\begin{itemize}
\item {\it Axisymmetric fibers:} the fiber central line is a straight line and the cross sections around that central line are  disks only. The
  one-dimensional reduction of viscous jets for this geometry has
  been extensively studied in previous literature with several
  non-equivalent methods. A first method consists in using the
  Cosserat theory \citep{Bogy}, and it has been shown that this method
  is in fact equivalent to expanding the velocity fields along a
  suitable basis of functions \citep{EggersRMP,Eggers2008}. The second
  method is based on a radial expansion (mathematically a Taylor
  expansion) of velocity fields and it has been developed in,
  e.g., \mbox{\citet{GarciaCastellanos}}, \mbox{\citet{EggersDupont}}, or \citet{BBCF} when
  allowing for a possible angular rotation around the axis of
  symmetry. The validity of these methods has been studied in details
  in the subsequent literature, e.g. in
  \citet{PeralesVega}, \citet{Ferrera2011}, \citet{Montanero2011} or \citet{Vincent2014}. In~\S~\ref{SecApplications},  we recover the standard lowest order results plus first corrections
using the radial expansion method. We also report a general method to
obtain recursively its corrections up to any order and report the
second set of corrections. In this geometry, once the constraints from
the stress tensor on the fiber side have been used, the fundamental dynamical variables
appear to be the velocity along the axis $\Sv$, the local rotation rate around
that axis $\Sw$, and the radius $\ra$. 
\item {\it Straight fibers:} the fiber central line is still a
  straight line, but the cross sections can have more general
  shapes. We find that the section shape is most conveniently expanded into shape
  multipoles which are symmetric trace free tensors. The lowest multipole describes for instance the
  elliptical modulation of the cross sections~\citep{BFHL,BLF}. Under this description,
  the shape multipoles are additional fundamental variables.
\item {\it Curved fibers with circular sections:} the fiber central  line can have any general shape as long as the curvature radius remains larger than the typical extension of the cross sections. A formalism was initially developed in \citet{EntovYarin} and further
summarized in \citet{Yarin1993,Yarin2011}. Curved fibers were
considered with surface tension effects in~\citet{Wilmott1992,Cummings} and also in~\citet{Fraunhofer1,Fraunhofer2} to study rotational
spinning processes such as those used in the production of
glass wool or candy floss. A similar viscous rod model, based on curvilinear
coordinates adapted to the problem, has been developed
by~\citet{Ribe2004,Ribe2006,Ribe2012} to study the coiling of
viscous jets, and numerical methods were developed by
e.g. \citet{Audoly2013,BergouAudolyVega} to obtain general solutions. In this article, we develop a formalism based on a $2+1$ splitting~\citep{Miyamoto:2010ga}
of equations, that is, a separation between the two-dimensional fiber
sections and its one-dimensional central line, to reduce curved viscous jets as one-dimensional object. We first describe in full generality the central line
along which the jet is described by following essentially the method developed by
Ribe. The tangential direction of this central line naturally determines a fiber
direction and a fiber section which is orthogonal to it, along which our $2+1$ splitting of
equations is performed. Then, using the irreducible representation of
${\rm SO}(2)$, we build an expansion of the velocity field. It is based on symmetric trace-free tensors which are lying in
the fiber sections and we show that these tensors are the moments
which naturally take into account the internal structure of the
fiber. Eventually, the fundamental variables are the same as for
straight fibers with non-circular sections (velocity along the axis $\Sv$, local rotation rate around
that axis $\Sw$, fiber radius $\ra$, and shape multipoles), since all
other velocity moments can be obtained as constraints from these
variables. These fundamental variables must also be supplemented by
the fiber central line position and velocity. The difference
with the straight case lies mainly in the fact that circular sections are only compatible with the
lowest order description, that is with the viscous string model. Indeed, as soon as
corrections are included, the shape multipoles are necessarily sourced. For instance, terms which are quadratic in the central
line curvature generically source the ellipticity.
\end{itemize}

Our consistent description of elongated but possibly curved viscous
fibers allows to find a number of qualitative results on the structure
of the one-dimensional models which contrast with past literature.
Most importantly, we find that the first corrections for curved fiber
geometries cannot be encompassed by the rod model
of~\citet{Fraunhofer1,Ribe2004,Ribe2006}. These methods are based on the observation that, when considering
extended solid objects instead of point particles, we must supplement the momentum balance
equation by an angular momentum equation. It is thus expected that to
go beyond the string approximation which can also be obtained from a
momentum balance equation, we should use some form of angular momentum
balance equation. However, this method inspired from solids fails for viscous
fluids for a number of reasons which are absent in non-deformable solids.

At lowest order in $\epsilon_\ra$, the rotation of the fiber section follows the rotation
of the fluid  on the fiber central line. Since the dynamics of the central line is
determined from the momentum balance equation, its local rotation rate
is also derived from it, implying that the angular momentum method
cannot bring any new dynamical information about fiber section
rotation. In fact, it is precisely because the rotation of sections is
determined from the rotation of the central line at lowest order that
the angular momentum method is instead a constraint on the sectional
component of the viscous forces which appear as corrections to the
lowest order description. Eventually, the coupling between the momentum balance
and angular momentum balance equation amounts to selecting only some
corrections and discarding the other ones as several order $\epsilon_\ra^2$ effects are
  missing. For instance, the sectional component of the velocity {\it on the central
  line} differs from the sectional component of the velocity {\it of the central line} by corrections of
order $\epsilon_\ra^2$. Additionally, the longitudinal velocity develops
a Hagen-Poiseuille profile (that is a parabolic profile in terms of the radial
distance), which blurs the notion of solid displacement of fiber
sections. 

For these various reasons we find that we should not build a one-dimensional reduction of viscous fibers from the usual methods
which have been developed to describe the continuous deformation of
solids, but we should instead start from a Taylor expansion of the
velocity field and find a consistent truncation at any given
order. When deriving corrections to a viscous string model, this
requires to abandon the hypothesis of circular sections and to derive
the dynamical equations for the shape moments as well.

\subsubsection{Outline}

In \S~\ref{SecGeometry} we review the general formalism to describe
the fiber central line and fiber sections. We introduce a
coordinates system and a vector basis which are adapted to the
description of curved fibers. In \S~\ref{SecTaylors}, we review in
details how scalar fields  and vector field such as velocity can be
expanded in multipoles using an adapted $2+1$ decomposition. This
leads us to introduce irreducible representations of ${\rm SO}(2)$ according
to which these multipoles are classified. With this formalism clearly
established, it is then possible to explore in full generality all the kinematical relations of fluid fibers
in~\S~\ref{SecKinematics} and then the dynamical laws
in~\S~\ref{SecDynamics}. The formalism is then applied for the two
main geometries of interest. First, in~\S\ref{SecApplications} we rederive the axisymmetric results up to
first corrections, including the interplay between the velocity
along the axis of symmetry, and rotation around that same axis. Second
corrections of the axisymmetric case are also reported in
Appendix~\ref{AppOrder2}. In~\S\ref{SecCurvedJets} we then tackle the
problem of curved fibers, first deriving the lowest order string model
and then discussing the general method to obtain corrections. We
report the first corrections in the curved case and show that elliptical shapes are necessarily sourced at that order. We also discuss the special case of straight but non-circular
sections in ~\S~\ref{SecElliptic}. Finally, we compare our method with the rod models where dynamical equations
are usually obtained from momentum/angular momentum balance and we
discuss the range of validity of these methods. Several technical developments are gathered in the Appendices,
among which the symmetric trace-free tensors in Appendix~\ref{STF}
which we use throughout the article. The formalism is applied to study
toroidal viscous fibers in \citet{PitrouPRE2}.

\subsubsection{Notation} 

Assuming incompressibility, the mass density $\rho$ is
constant. Hence, in the remainder of this article we will
use the notation
\be\label{mutomurho}
\visc/\rho \to \visc\,,\myquad \ST/\rho \to \ST\,,\myquad P/\rho \to P,
\ee
where $\visc$ is the viscosity of the fluid, $\ST$ the surface tension
parameter, and $P$ the pressure. \\
We use Einstein summation convention whenever indices are placed in
pairs with one index up and one index down as, e.g., in $x^\mu x_\mu$ or
$X^i Y_i$.

\section{Geometry}\label{SecGeometry}

For axisymmetric fibers, it is natural to use cylindrical
coordinates. The third coordinate ($z$) is naturally associated with
the axis of symmetry, and the other coordinates ($r,\theta$) parameterze the
two-dimensional space which is orthogonal to this axis.
However, for generally curved fibers, we must use fiber
adapted coordinates. They are closely related to cylindrical
coordinates in the sense that we choose a central line inside the
fiber and we use the natural coordinate of this line as the third
coordinate. The two other coordinates are then used to describe the
planes which are orthogonal to this central line. In this section,
we construct fiber adapted coordinates and the orthonormal
basis which is naturally associated with it.

\subsection{Description of the fiber central line}\label{SecCentralLine}

Throughout this article, we use Cartesian coordinates $x^\mu$, with
the corresponding canonical basis of vector and co-vectors (forms) 
$\gr{e}_\mu$ and $\gr{e}^\mu = \dd x^\mu$. Greek indices refer to
components in this canonical basis. The scalar and wedge products of
two vectors $\gr{X}=X^\mu \gr{e}_\mu$ and $\gr{Y}=Y^\mu \gr{e}_\mu$
are simply
\be
\gr{X}\cdot \gr{Y} \equiv X^\mu Y_\mu\,,\qquad [\gr{X}\times
\gr{Y}]^\alpha \equiv \eps^{\alpha\mu\nu}X_\mu Y_\nu\,,
\ee
where $\eps^{\alpha\mu\nu}$ is the totally antisymmetric tensor with $\eps^{123}=1$.

We assume that it is possible to define a fiber central line (FCL) as an
approximation to the fiber shape. We postpone to~\S~\ref{GaugeFix} the
discussion about the geometrical construction of this line. The position of this FCL and its tangent vector $\gr{T} = T^\mu \gr{e}_\mu$ are given by 
\be\label{TTun}
R^\mu(s,t)\,,\qquad T^\mu \equiv \partial_s R^\mu\,, \qquad T^\mu T_\mu =1\,.
\ee
The last condition ensures that the coordinate $s$ can also be used
to measure lengths along the FCL, and $t$ is the absolute time. A parallel projector, also named
longitudinal projector, and an orthogonal projector, also named sectional projector, can be defined as
\be
{P_\parallel}^\mu_\nu=T^\mu T_\nu\,,\myquad
{P_\perp}^\mu_\nu=\delta^\mu_\nu-T^\mu T_\nu=\delta^\mu_\nu - {P_\parallel}^\mu_\nu\,,
\ee
and can be used to project any vectorial quantity along and
orthogonally to the fiber tangential direction $T^\mu$. 

The velocity of the FCL is given by 
\be\label{DefU}
\gr{U} \equiv \partial_t \gr{R}
\ee
from which we deduce that the evolution of the tangent vector obeys
\be
\partial_s\partial_t \gr{R} = \partial_t \partial_s \gr{R}\myquad
\Rightarrow \myquad \partial_t\gr{T} = \partial_s \gr{U} \,.
\ee
From the normalization condition~\eqref{TTun} of $T^\mu$, we
deduce that
\be\label{dsUortho}
\gr{T}\cdot \partial_t\gr{T} = 0 \myquad \Rightarrow\myquad
\gr{T}\cdot\partial_s \gr{U} =0\,.
\ee
This last relation means that the FCL velocity can only rotate the
FCL direction $\gr{T}$, but without stretching it. Indeed, since it is
only a geometrical line there would be no physical information
contained in such stretching and we thus chose the
normalization~\eqref{TTun}. The FCL can be understood as a
non-extensible but flexible wire that would be inside the bulk of the
viscous jet, and the coordinate $s$ can be thought of as the distance
along that wire from a reference point $s=0$.

The curvature of the FCL is defined as the rate of change of the
tangential direction along the line for a fixed time, that is 
\be\label{DefCurvature}
\gr{\kappa} \equiv \gr{T} \times \partial_s \gr{T}\,,\qquad\partial_s \gr{T} = \gr{\kappa} \times \gr{T} \,,
\ee
and it depends on $(s,t)$.
Note that we have chosen deliberately $\gr{\kappa} \cdot \gr{T} = 0$ and our convention thus differs from
\citet{Ribe2004,Ribe2006,Fraunhofer1} where the longitudinal component
of curvature does not necessarily vanish.

Finally, to alleviate the notation, for any vector $\gr{X}$ we will
use the notation
\be\label{DefTilde}
\widetilde{\gr{X}} \equiv  \gr{T}\times \gr{X}\,.
\ee
Note that the vector $\widetilde{\gr{\kappa}}$ points toward the exterior
of the FCL curvature.

\subsection{Local orthonormal basis}\label{SecNormalBasis}

On each fiber section we can consider an orthonormal basis
$(\gr{d}_\tetr{1},\gr{d}_\tetr{2},\gr{d}_\tetr{3} \equiv \gr{T})$
where the vectors of the basis depend only on $(s,t)$. We use
coordinates $\tetr{i},\tetr{j},\tetr{k},\dots$ to refer to components
along this basis. It is oriented such that
\be
\gr{d}_\tetr{i} \times \gr{d}_\tetr{j} = {\eps_{\tetr{i} \tetr{j}}}^\tetr{k} \gr{d}_\tetr{k}\,,
\ee
where $\eps_{\tetr{i} \tetr{j} \tetr{k}}=\eps^{\tetr{i} \tetr{j} \tetr{k}}={\eps_{\tetr{i} \tetr{j}}}^{\tetr{k}}$ with $\eps_{\tetr{1} \tetr{2} \tetr{3}} = 1 $ is the alternating
symbol which is fully antisymmetric. This orthonormal basis is related
to the canonical Cartesian basis by a change of basis
\be
\gr{d}_\tetr{i} = {d_\tetr{i}}^\mu \gr{e}_\mu\,.
\ee
Similarly, the co-basis which depends also only on $(s,t)$ is related to the canonical co-basis by a
change of co-basis
\be
\gr{d}^\tetr{i} = {d^\tetr{i}}_\mu \gr{e}^\mu \,.
\ee
The change of basis and co-basis satisfy the basic properties
\prebool{\beas\label{BasicPropertyBasis}
 {{d}^\tetr{i}}_\mu&=& \delta^{\tetr{i}\tetr{j}} \delta_{\mu\nu}
 {{d}_\tetr{j}}^\nu={{d}_\tetr{i}}^\mu\\
 {d_\tetr{3}}^\mu &=& T^\mu\myquad {d^\tetr{3}}_\mu = T_\mu=\delta_{\mu\nu}T^\nu\,.
\eeas}
{\be\label{BasicPropertyBasis}
 {{d}^\tetr{i}}_\mu = \delta^{\tetr{i}\tetr{j}} \delta_{\mu\nu} {{d}_\tetr{j}}^\nu={{d}_\tetr{i}}^\mu\,,\qquad {d_\tetr{3}}^\mu = T^\mu\myquad {d^\tetr{3}}_\mu = T_\mu=\delta_{\mu\nu}T^\nu\,.
\ee}
Any vector on the central line, is decomposed as
\be\label{XtoXi}
\gr{X}(s,t) = X^\tetr{i}(s,t) \gr{d}_\tetr{i}(s,t)\,.
\ee
The fiber central line and its associated
orthonormal basis are illustrated in Fig.~\ref{FigFAC}. 
Note that for components in the orthonormal basis, the position of
indices does not matter. Indeed from~\eqref{BasicPropertyBasis}, we find that for any vector
$X^{\tetr{i}} = X_{\tetr{i}}$. In the orthonormal basis, the scalar products and wedge products
of two vectors $\gr{X}$ and $\gr{Y}$ are
\be\label{scalwedge}
\gr{X}\cdot\gr{Y}=X^\tetr{i} Y_\tetr{i}\,,\myquad [\gr{X}\times
\gr{Y}]^\tetr{i} = \eps^{\tetr{i}\tetr{j}\tetr{k}}X_\tetr{j} Y_\tetr{k}\,.
\ee

\ifpre
\begin{figure}[!htb]
\else
\begin{figure}
\begin{center}
\fi
\ifpre
\includegraphics[width=\columnwidth]{FAC.eps}
\else
\includegraphics[width=0.6\columnwidth]{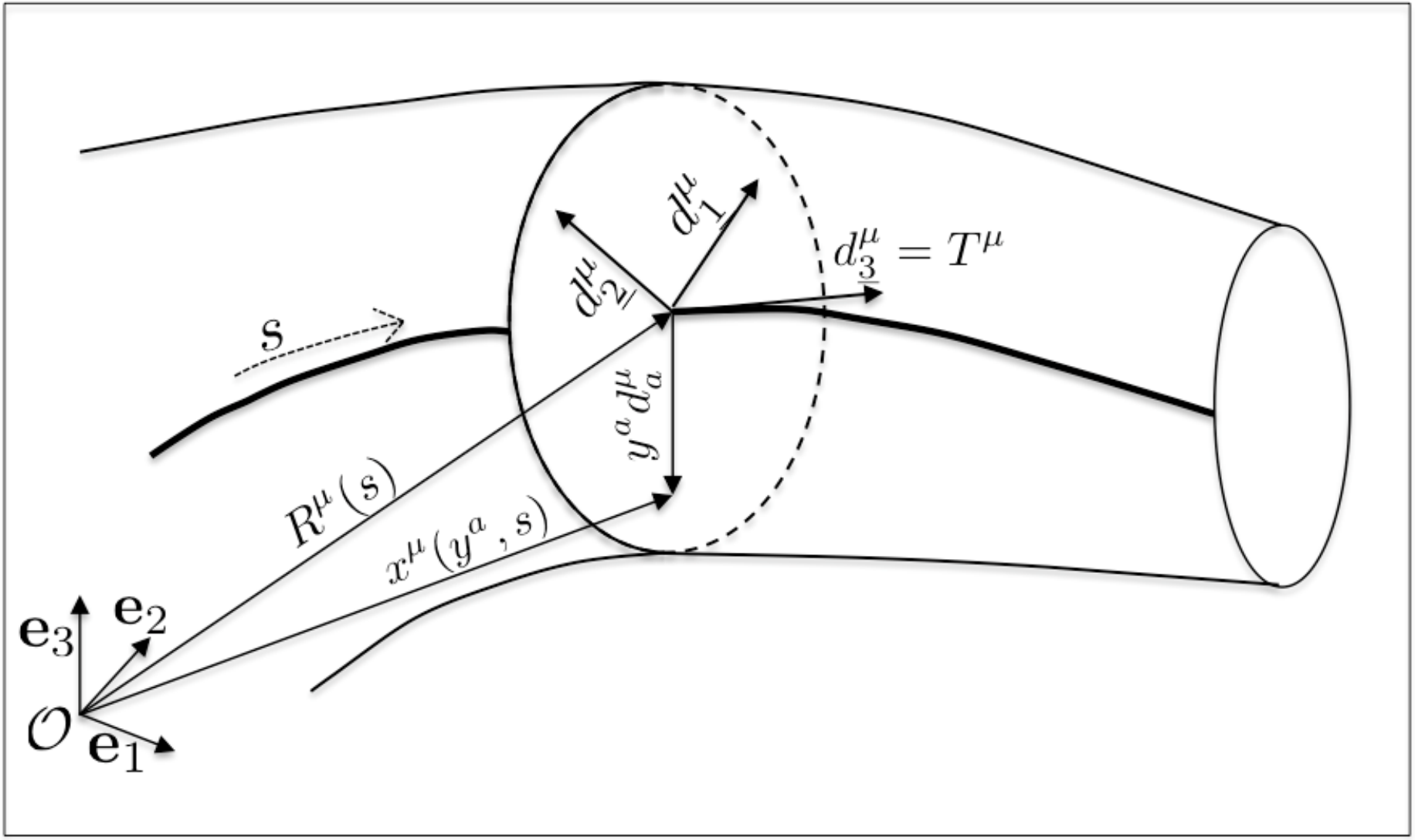}
\fi
\caption{Fiber central line position  $R^\mu$, and the associated
  orthonormal basis $d_\tetr{i}^\mu$. A point lying in a fiber section
is then labeled by $x^\mu$ defined in \eqref{xRy}.}
\label{FigFAC}
\ifpre
\else
\end{center}
\fi
\end{figure}

Following the evolution of the tangent vector along the
fiber~\eqref{DefCurvature}, we choose to transport this basis along the FCL according to
\be\label{Propkappa}
\partial_s \gr{d}_\tetr{i} = \gr{\kappa} \times \gr{d}_\tetr{i} \myquad
\Leftrightarrow\myquad \kappa^\tetr{i} = \mytfrac{1}{2}\epsilon^{\tetr{i}\tetr{j}\tetr{k}}(\partial_s \gr{d}_\tetr{j}) \cdot \gr{d}_\tetr{k} \,,
\ee
which is consistent with~\eqref{DefCurvature} since $\gr{T}=\gr{d}_\tetr{3}$.

\subsection{Rotation of the orthonormal basis}

The rotation rate of the orthonormal basis is defined by
\be\label{DefRotation}
\partial_t \gr{d}_\tetr{i} = \gr{\omega} \times \gr{d}_\tetr{i}\myquad
\Leftrightarrow\myquad \omega^\tetr{i} = \mytfrac{1}{2}\eps^{\tetr{i}\tetr{j}\tetr{k}}(\partial_t \gr{d}_\tetr{j}) \cdot \gr{d}_\tetr{k}\,.
\ee
In particular $\partial_t \gr{T} = \gr{\omega} \times \gr{T}$ and the sectional projection of rotation satisfies
\be\label{omegaperp}
P_\perp(\gr{\omega} )= \gr{T} \times \partial_t \gr{T}= \gr{T} \times \partial_s \gr{U}\,.
\ee

From the definitions~\eqref{DefCurvature} and~\eqref{DefRotation} of curvature and
rotation, we get the structure relation
\be\label{RiemannCurvature}
\partial_t \gr{\kappa}-\partial_s \gr{\omega} = \gr{\omega}\times\gr{\kappa}\,.
\ee
This relation is a consequence of the flatness of the classical
structure of spacetime as explained in Appendix~\ref{AppCartan}. When
projected on $\gr{T}$, it also implies
\be\label{BonusEquation}
(\partial_s \gr{\omega})\cdot \gr{T}=0\,.
\ee

We use the indices $a,b,\dots=1,2$ for the sectional components, that is which are orthogonal to the fiber tangential direction. For instance the
curvature $\gr{\kappa}$ is projected since $P_\perp(\gr{\kappa}) = \gr{\kappa}$,
but the rotation is not a projected vector, so their decompositions in
components read as
\be
\gr{\kappa} = \kappa^\tetr{a} \gr{d}_\tetr{a}\,,\myquad \gr{\omega} = \omega^\tetr{i} \gr{d}_\tetr{i}\,,\myquad P_\perp(\gr{\omega}) = \omega^\tetr{a} \gr{d}_\tetr{a}\,.
\ee
The notation~\eqref{DefTilde} in components is simply
\be
\widetilde{{X}}^\tetr{a} = (\gr{T}\times \gr{X})^\tetr{a} = {X}_\tetr{b}\eps^{\tetr{b}\tetr{a}} \,,
\ee
where $\eps_{\tetr{a} \tetr{b} }=\eps^{\tetr{a} \tetr{b} }\equiv
\eps_{\tetr{a}\tetr{b}\tetr{3}}$ is the two-dimensional alternating symbol ($\eps_{\tetr{1}\tetr{2}}=1$). Since
$\partial_s \eps^{\tetr{a}\tetr{b}}=\partial_t \eps^{\tetr{a}\tetr{b}}=0$ the tilde operation and the
$\partial_s$ or $\partial_t$ derivatives commute.

From \eqref{RiemannCurvature} the relations between the
derivatives of components of curvature and rotation read as
\be\label{kawa}
\partial_t \kappa^\tetr{i}-\partial_s \omega^\tetr{i}=\eps^{\tetr{i}\tetr{j}\tetr{k}}\kappa_\tetr{j}\omega_{\tetr{k}}\,.
\ee
Separating the sectional and longitudinal components these relations
are just
\beas\label{Reldtdsko}
\partial_t \kappa^\tetr{a}-\partial_s
\omega^\tetr{a}&=&\eps^{\tetr{a}\tetr{b}}\kappa_\tetr{b}
\omega^\tetr{3}=-\tkappa^a \omega^\tetr{3}\,,\slabel{Reldtdsko1}\\
\partial_s \omega^\tetr{3} &=&
\eps^{\tetr{a}\tetr{b}}\omega_\tetr{a} \kappa_\tetr{b}=-\omega_a
\tkappa^a=\widetilde{\omega}_a \kappa^a\slabel{Reldtdsko2}\,.
\eeas
This last relation \eqref{Reldtdsko2}, which is the component version
of \eqref{BonusEquation}, states that once the curvature
and the sectional projection of the rotation are fixed at a given time
along the FCL, then the longitudinal component of rotation $\omega_\tetr{3}$ is also determined at that
time along the FCL, as a consequence of the structure
relation~\eqref{RiemannCurvature}.
We call this type of relation a {\it constraint equation}.

\subsection{Essential relations for components}\label{SecEssential}

For a vector $\gr{X}(s,t)$, the component of the derivatives
[e.g., $(\partial_t X)^\tetr{i}\equiv\gr{d}^\tetr{i} \cdot \partial_t  \gr{X}$] are not the derivatives of the
components [e.g., $\partial_t X^\tetr{i} \equiv \partial_t (\gr{d}^\tetr{i} \cdot \gr{X})$]. Indeed, they are related by
\beas\label{dtXidtXi}
(\partial_t X)^\tetr{i} &=&\partial_t X^\tetr{i} +(\omega \times
X)^\tetr{i}\nonumber\\
&=&\partial_t X^\tetr{i} +\eps^{\tetr{i}\tetr{j}\tetr{k}}
\omega_\tetr{j} X_\tetr{k}\,, \slabel{dtXidtXi1}\\
(\partial_s X)^\tetr{i} &=&\partial_s X^\tetr{i} +(\kappa \times
X)^\tetr{i} \nonumber\\
&=&\partial_s X^\tetr{i} +\eps^{\tetr{i}\tetr{j}\tetr{k}}
\kappa_\tetr{j} X_\tetr{k}\,.\slabel{dtXidtXi2}
\eeas
In particular, for the projected (or sectional) indices, this reads as
\beas\label{dtXdtX}
(\partial_t X)^\tetr{a} &=&\partial_t X^\tetr{a} -\widetilde{\omega}^\tetr{a} X_\tetr{3}+\widetilde{X}^\tetr{a}\omega_\tetr{3} \,,\slabel{dtXdtX1}\\
(\partial_s X)^\tetr{a} &=&\partial_s X^\tetr{a}-\widetilde{\kappa}^\tetr{a} X_\tetr{3}\,. \slabel{dtXdtX2}
\eeas
If a vector $\gr{X}$ is projected, that is orthogonal to the tangential direction ($X_\tetr{3}=0$), then in that special case, $(\partial_s X)^\tetr{a}
=\partial_s X^\tetr{a}$ but note that we still have $(\partial_t
X)^\tetr{a} \neq \partial_t X^\tetr{a}$.
From the derivatives of the components, that is, from $\partial_t
X^\tetr{i}$, $\partial_s X^\tetr{i}$, we can conversely obtain the derivatives
of the vector using \eqref{dtXidtXi} with the simple relations
\beas\label{Covariantization}
\partial_s \gr{X}&=&(\partial_s X)^\tetr{i}
\gr{d}_\tetr{i}=(\partial_s X^\tetr{i})
\gr{d}_\tetr{i}+\gr{\kappa}\times \gr{X}\,,\\
\partial_t \gr{X}&=&(\partial_t X)^\tetr{i} \gr{d}_\tetr{i}=(\partial_t X^\tetr{i})
\gr{d}_\tetr{i}+\gr{\omega}\times \gr{X}\,.
\eeas

Let us report in coordinates some relations previously
obtained in a covariant form. Eq.~\eqref{dsUortho} reads as in components
\be\label{dsU3}
\partial_s U^{\tetr{3}} +\eps^{\tetr{a}\tetr{b}} \kappa_\tetr{a} U_\tetr{b}= \partial_s
U^{\tetr{3}} +\widetilde{\kappa}^\tetr{b} U_\tetr{b} = 0\,.
\ee
As for the property~\eqref{omegaperp} for the projection of the
rotation, it reads as simply
\be\label{omegacomponent}
\omega^\tetr{a} =\partial_s \widetilde{U}^\tetr{a}+ \kappa^\tetr{a} U^{\tetr{3}}\,. 
\ee
This relation which is also a constraint equation states that once the
sectional projection of the central line
velocity ($U^\tetr{a}$) is known at a given time along the central line, then the projection
of rotation ($\omega^\tetr{a}$) is determined. Since the longitudinal
component of rotation $\omega^\tetr{3}$ is determined from
\eqref{Reldtdsko2}, we can then determine the time evolution of the
components of curvature from~\eqref{Reldtdsko1}.

\subsection{Fiber adapted coordinates}

We use Cartesian coordinates $y^a=y^1,y^2$  inside the
fiber section labeled by $(s,t)$ so as to parametrize points which
do not lie exactly on the FCL. With $y^3\equiv s$, the fiber adapted (FA)
coordinates are the set of $(y^i)=(y^a,s)$. The FA coordinates of a
point in the fiber are related to the Cartesian coordinates $x^\mu$ by
\be\label{xRy}
x^\mu(y^i) = R^\mu(s,t) + y^a {d_\tetr{a}}^\mu(s,t)\,.
\ee
These coordinates are illustrated in Fig.~\ref{FigFAC}. The canonical basis and co-basis associated with the FA coordinates are
\be
\gr{e}_i = {e_i}^\mu \gr{e}_\mu,\myquad \gr{e}^i = {e^i}_\mu
\gr{e}^\mu,\myquad {e_i}^\mu= \frac{\partial x^\mu}{\partial y^i},\myquad {e^i}_\mu= \frac{\partial y^i}{\partial x^\mu}
\ee
and they are related to the orthonormal basis by 
\beas\label{etod}
\gr{e}_3 &=& h \gr{T} = h \gr{d}_\tetr{3}\,,\qquad \gr{e}^3 =\frac{1}{h} \gr{d}^\tetr{3},\\
\gr{e}_a &=& \gr{d}_\tetr{a}\,,\qquad\qquad\,\quad \gr{e}^a =\gr{d}^\tetr{a}\,,\slabel{avsb}
\eeas
where we defined
\be\label{Defh}
h \equiv 1+\widetilde{\kappa}_a y^a\,.
\ee
Given the relations~\eqref{avsb}, we will ignore in the rest of this
article the difference between indices referring to $\gr{e}_a$ (resp. $\gr{e}^a$) and $\gr{d}_{\tetr{a}}$ (resp. $\gr{d}^{\tetr{a}}$) since they are equal
and refer to unit vectors in both cases. The distinction is only meaningful for the third components, that is, the components along the
tangential direction, since $\gr{d}_\tetr{3}$ is normalized whereas $\gr{e}_{3}$ is not.

The relation between the FA coordinates canonical basis and the
orthonormal basis can also be written in the form
\be
\gr{d}_{\tetr{i}} = {d_\tetr{i}}^j \gr{e}_j\,,\qquad \gr{d}^{\tetr{i}} =
{d^{\tetr{i}}}_j \gr{e}^j\,
\ee
with
\beas
&&{d_\tetr{3}}^3=h^{-1},\myquad {d_\tetr{3}}^a=0,\myquad {d_a}^3=0,\myquad
{d_a}^b = \delta_{a}^b\\
&&{d^\tetr{3}}_3=h,\qquad {d^\tetr{3}}_a=0,\myquad {d^a}_3=0,\myquad
{d^a}_b = \delta^{a}_b\,.
\eeas

The metric and its inverse in the canonical basis of the FA
coordinates are simply
\prebool{\beas\label{gij}
g_{ij}&\equiv& \gr{e}_i \cdot \gr{e}_j=\left( \begin{array}{ccc}
1 & 0 & 0 \\
0 & 1 &0 \\
 0 & 0 & h^2 \end{array} \right),\slabel{gijdown}\\
g^{ij}&\equiv&\gr{e}^i \cdot \gr{e}^j=\left( \begin{array}{ccc}
1 & 0 & 0 \\
0 & 1 &0 \\
 0 & 0 & h^{-2} \end{array} \right),\slabel{gijup}
\eeas}
{\be\label{gij}
g_{ij}\equiv \gr{e}_i \cdot \gr{e}_j=\left( \begin{array}{ccc}
1 & 0 & 0 \\
0 & 1 &0 \\
 0 & 0 & h^2 \end{array} \right)\qquad
g^{ij}\equiv\gr{e}^i \cdot \gr{e}^j=\left( \begin{array}{ccc}
1 & 0 & 0 \\
0 & 1 &0 \\
 0 & 0 & h^{-2} \end{array} \right)
\ee}
We notice that it looks very much like a metric of cylindrical
coordinates. The main difference lies in the fact
that $h$ depends on all coordinates $(y^1,y^2,s)$ whereas for cylindrical coordinates $h$ is replaced by the radial coordinate [$h \to
r$ in \eqref{gij}].

Finally, we define the unit directional vector $\yu^a$, and the unit
orthoradial vector $\tyu^a$ in the section plane by
\be
y^a \equiv r \yu^a\,,\qquad\tyu^a \equiv-\eps^{ab} \yu_b\,,\qquad r^2 \equiv y_a y^a\,.
\ee
Since $\yu^a$ and $\tyu^a$ are unit vectors which are mutually
orthogonal and projected, then we find the identities
\bea\label{MegaPratique}
{P_\perp}^a_b=\delta^a_b &=& \yu^a \yu_b +\tyu^a \tyu_b\\
\epsilon^{ab}&=&\yu^a \tyu^b -\tyu^a \yu^b\,. 
\eea

Using $(r,\yu^a)$ rather than $(y^1,y^2)$ in the section plane amounts simply
to using polar coordinates instead of Cartesian coordinates. With the
variables $(r,\yu^a,s)$, the FA coordinates can be understood as a
cylindrical coordinates system that one would have deformed so that the
$z$ axis is curved and onto the FCL. The factor $h$ defined in \eqref{Defh} which is larger
that unity in the exterior of curvature and smaller than unity in the
interior ($\tkappa^a$ points toward the exterior of curvature) is the
local amount of
stretching which had to be applied to perform this deformation.

\subsection{$2+1$ decomposition}\label{Defunplusdeux}

As mentioned in~\S~\ref{SecCentralLine}, any vector can be decomposed
into its part along $\gr{T}$ (its longitudinal part) and a
sectional part according to
\be\label{ProjVectors}
X^\mu = \para{X} T^\mu+ {X_\perp}^\mu\,\myquad {X_\perp}^\mu \equiv
{P_\perp}^\mu_\nu X^\nu,\myquad \para{X} \equiv \gr{X} \cdot \gr{T}
\ee
or in components
\be
\para{X} = X^\tetr{3}\,,\qquad {X_\perp}^a = X^a \,.
\ee
Given this last property, we will omit the $\perp$ subscript when
dealing with the components of the projection of a tensor.
This decomposition is easily extended to tensors as each index needs
to be decomposed into a longitudinal and a sectional part. 

From now on, we thus use the notation $\para{U} = U^\tetr{3}$ and
$\para{\omega}=\omega^\tetr{3}$ for the longitudinal components of the
fiber velocity and of the rotation rate.

\subsection{Spatial derivatives in fiber adapted coordinates}\label{SecCovD}

The components of the derivative of a vector in the canonical basis
of the FA coordinates are given by~\footnote{For simplicity we have
  chosen to use the Cartesian coordinates $x^\mu$ in the ambient space. If we
were to choose general curvilinear coordinates, e.g., spherical
coordinates, then in all our equations we should promote the partial
derivative to a covariant derivative and perform the replacement
$\partial_\mu \to \nabla_\mu,\quad \partial_i \to \nabla_i$ and $\delta^{\mu\nu}\to g^{\mu\nu},\quad
\delta_{\mu\nu}\to g_{\mu\nu}$ everywhere. We would also have
to use the rule $\partial_s \to {e_3}^\mu \nabla_\mu$. The indices
$\mu,\nu\dots$ could even be given an abstract meaning and not refer to a particular system of coordinates. This is a standard
notation for general relativity~\citep{Wald1984} in which the background space is
unavoidably curved, but for classical physics this extra layer of
abstraction is not necessary. We have thus chosen the simplest and
most transparent notation based on an ambient set of Cartesian
coordinates with its associated partial derivative, but we must bear
in mind that the results are more general.}
\beas\label{ChristoUseful}
{e_i}^\mu {e_j}^\nu (\partial_\mu v_\nu) &=& \partial_i v_j-\Gamma^{k}_{ij} v_k,\\
{e_i}^\mu {e^j}_\nu (\partial_\mu v^\nu) &=& \partial_i v^j+\Gamma^{j}_{ik} v^k,
\eeas
where the Christoffel symbols are defined as
\be
\Gamma^{j}_{ik}\equiv {e_i}^\mu{e^j}_\nu\left(\partial_\mu {e_k}^\nu\right)\,.
\ee
They are related to the components of the metric in the canonical basis of
the FA coordinates by
\be
\Gamma^{i}_{jk} = \frac{1}{2}g^{il}\left(\partial_{j}
  g_{lk}+\partial_k g_{lj}-\partial_l g_{jk}\right)\,.
\ee
The only non-vanishing Christoffel symbols are
\be\label{Christoffels}
\Gamma^{3}_{33}=\frac{\partial_s h}{h},\myquad
\Gamma^{3}_{a3}=\frac{\widetilde{\kappa}_a}{h},\myquad \Gamma^{a}_{33}=
-h\widetilde{\kappa}_a\,.
\ee

The derivative of the orthonormal basis is then deduced from \eqref{ChristoUseful} and the relations~\eqref{etod}. We find
\be\label{Riccirotation}
{d^{\tetr{j}}}_\nu\left(\partial_\mu {d_\tetr{i}}^\nu\right)=
{d^{\tetr{j}}}_\nu {e^3}_\mu \eps^{\alpha \beta
\nu} \kappa_\alpha (d^\tetr{i})_\beta={e^3}_\mu \eps^{a \tetr{i}\tetr{j}}\kappa_a\,,
\ee
where from~\eqref{etod} we must use ${e^3}_\mu=T_\mu/h$.
Finally, the divergence of a vector has a simple expression in FA
coordinates. For a general vector $X^\mu$ it reads as
\prebool{
\bea\label{EqDiv}
 \partial_\mu X ^\mu =\frac{1}{h}\partial_i(h
 X^i)&=&\frac{1}{h}\left[\partial_a(h X^a) + \partial_s X^\tetr{3}
 \right] \\
&=&\frac{1}{h}\left[\tkappa_a X^a + \partial_s X^\tetr{3}\right]
+ \partial_a X^a\,.\nonumber
\eea}
{\be\label{EqDiv}
 \partial_\mu X ^\mu =\frac{1}{h}\partial_i(h
 X^i)=\frac{1}{h}\left[\partial_a(h X^a) + \partial_s X^\tetr{3}
 \right] 
=\frac{1}{h}\left[\tkappa_a X^a + \partial_s X^\tetr{3}\right]
+ \partial_a X^a\,.
\ee}
In particular, this allows to obtain the Laplacian of a scalar
function $S$ by using $X^\mu = \partial^\mu S $, and we get
\be\label{LaplacianOpened}
\Delta S = \partial^a \partial_a S +\frac{\partial_s^2
  S}{h^2}-\frac{(\partial_s S) y^a \partial_s
  \tkappa_a}{h^3}+\frac{\tkappa^a \partial_a S}{h}\,.
\ee

\section{Fields expansion on sections}\label{SecTaylors}

With the FA coordinates we have an appropriate method to describe the
position of a fluid particle inside the viscous jet. However in order to describe the dynamics of the viscous fluid inside the
fiber, we also need to find an appropriate description for the fluid
velocity itself, and this is the goal of this section. Clearly, in the slender approximation the velocity field is necessarily very
close to the velocity  $U^\mu$ on the FCL. It is thus natural to Taylor expand
the velocity field around the velocity on the FCL. However, we cannot keep all orders
of this expansion and we need general principles to guide us in
truncating such expansion. Since the fiber sections are
two-dimensional, it is natural to classify the various orders of the
Taylor expansion according to their transformation property under
${\rm SO}(2)$, that is the local group of rotations around the tangential
direction. In this section we give the essential steps to find the
irreducible representations of ${\rm SO}(2)$ and their tensorial
expressions. Further details for this method are reported in Appendix~\ref{STF}.

\subsection{Taylor expansion}

Any $2$-field, that is a field on the fiber section with only
sectional indices, can be Taylor
expanded in the variables $(y^1,y^2)$ for each $(s,t)$. For a scalar
field, this Taylor expansion is of the form
\be\label{DefSexpand}
S(y^a,s,t) = \sum_{\ell=0}^\infty S_L(s,t) y^L
\ee
where we use the multi-index notation $L=a_1\dots a_\ell$ on which
the Einstein summation convention applies. The $S_L$
are necessarily symmetric rank-$\ell$ tensors since the $y^L$ are symmetric. They
are obtained by successive applications of $\partial/\partial y^a$ on
$S$. If there is no index, that is, for $\ell=0$ in \eqref{DefSexpand},
then we use the notation $S_\myes$.
For a $2$-vector the expansion is instead of the form
\be\label{DefVaExpand}
V^a(y^i,t) = \sum_{\ell=0}^\infty {V^a}_L(s,t) y^L\,,
\ee
and this is obviously extended to higher order tensors.
The tensors ${V^a}_L$ are symmetric in the $\ell$ indices $L$ but
there is no particular symmetry involving the index $a$.

If we decompose the total fluid velocity $V^\mu$ into its longitudinal
part $\para{V}$ and its projected part $V^a$ as
in~\S~\ref{Defunplusdeux}, then the former is a scalar function
whereas the latter is a $2$-vector and they are Taylor expanded,
respectively, as in~\eqref{DefSexpand} and \eqref{DefVaExpand}.


\subsection{Irreducible representations of ${\rm SO}(2)$}\label{SecIrreps}

It proves useful to decompose the $y^a$ dependence in the Taylor
expansion of \eqref{DefSexpand} by separating the dependence in
the radial coordinate $r$ and the direction $\yu^a$. The dependence in
the direction $\yu^a$ can be further decomposed onto the irreducible
representations (irreps hereafter) of ${\rm SO}(2)$. This method follows a method long known in three dimensions
where the direction vector lies in the two-sphere, and the irreps
considered are those of ${\rm SO}(3)$~\citep{CourantHilbert,Thorne1980,Blanchet1985,Blanchet1998}. 

The irreps of ${\rm SO}(2)$ are
given by the functions ${\rm e}^{\ii n\theta}$. More precisely the
irreps are two-dimensional as they are represented by ${\rm
  e}^{\pm \ii n\theta }$, except for $D_0$ which is one dimensional
and which is just the set of constants. We note these irreps $D_n$. Any function depending on
$\yu^a=(\cos \theta,\sin \theta)$ is indeed expanded in Fourier series as
\be\label{Usualirreps}
f(\theta) = \sum_{n=-\infty}^\infty f_n {\rm e}^{\ii n
  \theta}=f_0+\sum_{n=1}^\infty f_n {\rm e}^{\ii n
  \theta} + f_{-n} {\rm e}^{-\ii n\theta}\,.
\ee
The {\it symmetric trace-free} (STF) $2$-tensors are also irreps
of the rotation group ${\rm SO}(2)$ just like STF 3-tensors are irreps of
the group ${\rm SO}(3)$. The corresponding expansion of $f$ reads as simply
\be
f(\yu^i) = \sum_{\ell=0}^\infty f_L \yu^L\,.
\ee
where this time the $f_L$ are symmetric but also traceless tensors of
rank $\ell$ (we recall the multi-index notation $L = a_1\dots a_\ell$). The two
expansions are related thanks to (for $m>0$)
\beas\label{DefYmL}
{\cal Y}^m_{a_1\dots a_m} &\equiv& (d_1+\ii d_2)^{a_1}\dots  (d_1+\ii d_2)^{a_m}\\
{\cal Y}^{-m}_{a_1\dots a_m} &\equiv& (d_1-\ii d_2)^{a_1}\dots  (d_1-\ii d_2)^{a_m} \,. 
\eeas
Indeed it is easily found that
\be
f_L =\sum_{m=\pm|\ell|}{\cal Y}_L^m f_m\,,\qquad f_m =\frac{1}{2^\ell} f^L {\cal Y}_L^m\,.
\ee
The STF tensors of rank $\ell$ have indeed only two degrees of
freedom. For instance for a symmetric rank-$4$ tensor, the only independent
degrees of freedom are $f_{1111}$ and $f_{1112}$, since from the traceless condition the
other components are related through $f_{1222}=-f_{1112}$, $f_{2222}=-f_{1122}=f_{1111}$.
We are led to decompose all tensors appearing in Taylor
expansions in STF tensors so as to obtain a decomposition in terms of irreps.

\subsection{Irreps of scalar functions}\label{SecIrrepScalar}

For scalar functions in the fiber section plane, the expansion~\eqref{DefSexpand} is already made in
terms of symmetric tensors and we only need to remove the traces.  For
any symmetric tensor $S_L$, the traceless part can be extracted as
\be
S_{\langle a_1\dots a_\ell\rangle} = \sum_{n=0}^{[\ell/2]} a_\ell^n \delta_{(a_1 a_2}
\dots \delta_{a_{2n-1} a_{2n}} {S_{a_{2n+1}\dots a_\ell) b_1
   \dots b_n}}^{b_1\dots b_n}
\ee
\be
a_\ell^n \equiv (-1)^n \frac{(2\ell-2n-2)!!(2n-1)!!}{(2\ell-2)!!} {\ell \choose 2n}\,.
\ee
In the expression above, the symmetric part of a tensor $T_{a_1\dots
  a_\ell}$ which is not
initially symmetric is denoted $T_{(a_1 \dots a_\ell)}$, and the STF
part is denoted with angle brackets $T_{\langle a_1
  \dots a_\ell \rangle}$. If the symmetrization ranges on the indices
of a product of tensors, this product has to be considered as a single
tensor for which the indices are symmetrized. We also use the notation $n!!=n(n-2)(n-4)\dots$.

From the Taylor expansion~\eqref{DefSexpand}, we see that by removing
the traces in the tensors $S_L$, we get factors of $r^2 \equiv  y_a
y^a$. The expansion in terms of STF tensors only is thus of the form
\be\label{ScalarSTF}
S(y^i,t) = \sum_{\ell=0}^\infty \sum_{n=0}^\infty  S^{(n)}_{L}(s,t) y^L r^{2n}\,,
\ee
where the $S^{(n)}_{L}$ are STF. The dependence in the direction and the radial coordinates have now
been clearly separated. The directional dependence is decomposed onto
STF tensors, and the dependence in the radial distance is an even
polynomial as it is a polynomial in $r^2$. If no ambiguity can arise,
we can omit the sums over $\ell,n$ so as to alleviate the notation.
The STF multipoles of the decomposition~\eqref{ScalarSTF} can be obtained from angular integrals
as explained in Appendix~\ref{STF}. 

Finally, in the case $\ell=0$, that is for the monopole of the directional
dependence, the coefficients are noted $S^{(n)}_\myes$. For
instance, if we consider the longitudinal part of the velocity field
$\para{V}$, then $\para{V}^{(0)}_\myes$ corresponds to a uniform
flow, and $\para{V}^{(1)}_\myes$ corresponds to a parabolic velocity
profile, also known as a Hagen-Poiseuille (HP) flow.

\subsection{Irreps of $2$-vector fields}\label{SecIrrepVector}

It is shown in Appendix~\ref{AppOrdSTF} that the expansion in
irreps of a 2-vector is necessarily of the form

\be\label{IrrepsVa}
V_a = \stf{V}^{(n)}_{aL} y^L r^{2n} +\frac{y_a}{2}
\mysh{V}^{(n)}_L y^L r^{2n}-\eps_{ab}y^b \omeg{V}^{(n)}r^{2n}\,,
\ee
where the $\stf{V}^{(n)}_{L}$ and $\mysh{V}^{(n)}_L$ are STF tensors and
where we recall that there is an implicit sum on $\ell$ and $n$. At first sight, this form is very cumbersome if we are familiar with
the irreps of 3-vectors [that is, irreps of ${\rm SO}(3)$]. Indeed for 3-vectors, the directional dependence can always be expanded in terms of electric
type and magnetic type multipoles~\citep{Thorne1980,Pitrou2008}. The result for 2-vectors is
necessarily very different because the antisymmetric tensor has only
two indices in two dimensions, hence the expressions from the
three-dimensional results, which also involve the antisymmetric tensor
in three dimensions, cannot be exported directly to two dimensions.

Note also that the tensors $\stf{V}_L^{(n)}$ have no monopole since
they must have at least one index. Conversely, the functions $\omeg{V}^{(n)}$ are
purely monopolar since they do not have any tensorial index. So, we can
interpret~\eqref{IrrepsVa} by saying that we have also two sets of STF
tensors. The first set is made of the $\mysh{V}^{(n)}_L$, $\ell\geq 0$,
and the second set consists in the $\omeg{V}^{(n)}$ and the
$\stf{V}^{(n)}_L$, $\ell \geq 1$.
  
Physically, $\omeg{V}^{(0)}$ corresponds to a solid rotation around
the fiber tangential axis (axial rotation hereafter) and $\mysh{V}_\myes^{(0)}$ corresponds to a
radial infall of the fluid. 

Finally, note that we could also decompose a vector field by projecting
its free index so as to obtain a scalar function. For instance, we can
select the radial and orthoradial components by projecting along
$\yu^a$ and $\tyu^a$. Then each scalar function can be expanded as in
\eqref{ScalarSTF}. The relation between the two methods is (with an
implied sum on $\ell$ and $n$)
\beas
r V_\tetr{r}&\equiv& y^a V_a = \stf{V}^{(n)}_{L}y^L r^{2n} + \mytfrac{1}{2}\mysh{V}^{(n)}_L y^L r^{2(n+1)},\\
r V_\tetr{\theta}&\equiv&\widetilde{y}^b V_b = {\eps_{a_1}}^b
\stf{V}^{(n)}_{a_{L-1} b}y^{a_L} r^{2n}+\omeg{V}^{(n)}r^{2(n+1)}\,.
\eeas
This points to a simple method to extract the STF components of a
vector. We first extract the STF components of its radial and orthoradial
projections as for scalar functions (see Appendix~\ref{STF}) and then deduce the STF multipoles of the
decomposition~\eqref{IrrepsVa} from the relations
\beas
\left[r V_\tetr{r}\right]_L^{(n)}&=&\stf{V}^{(n)}_L + \mytfrac{1}{2} \mysh{V}^{(n-1)}_L\\
\left[r V_\tetr{\theta}\right]_L^{(n)}&=&{\eps_{\langle a_1}}^b
\stf{V}_{a_{L-1}\rangle b }^{(n)}+\delta_\ell^0 \omeg{V}^{(n-1)}\,,
\eeas
which are inverted as
\beas
\omeg{V}^{(n)} &=& \left[r V_\tetr{\theta}\right]_\myes^{(n+1)}\\
\stf{V}_L^{(n)}&=&- {\eps_{\langle a_1}}^b  \left[r
  V_\tetr{\theta}\right]^{(n)}_{a_{L-1}\rangle b}\\
\mysh{V}_L^{(n)}&=& -2 \stf{V}_L^{(n+1)}+2\left[r V_\tetr{r}\right]_L^{(n+1)}\,.
\eeas
As a final comment on STF tensors, we must stress that these are much
more adapted to abstract tensor manipulation than the usual
representation~\eqref{Usualirreps}, and we chose to perform all the
tensor manipulations of this article with {\it xAct}~\citep{xAct}.

\section{Kinematics}\label{SecKinematics}

We are now equipped with all the necessary formalism to study in
details the kinematics and dynamics of viscous fibers. In this section
we present all the relations needed for the kinematics, and
in the next section we shall focus on the description of dynamics, so
as to obtain the one-dimensional reduction of viscous fibers
from its intrinsic physical laws.

\subsection{Velocity parameterization}

We separate the total fluid velocity $\totV^\mu$ into the velocity of
the FCL ($U^\mu$) and the small difference $V^\mu$ as
\be\label{totVUV}
\totV^\mu = U^\mu +V^\mu\,.
\ee
$\gr{V}$ is decomposed into a longitudinal part $\para{V}$  and a
sectional part $V^a$ which are decomposed as in \eqref{ScalarSTF} and \eqref{IrrepsVa}.
In order to avoid cluttering of indices when the decomposition~\eqref{IrrepsVa} is
used for the velocity field, we shall use the short notation
\beas\label{SimpleNotation}
\Sv^{(n)}_L&\equiv& \para{V}^{(n)}_L\,\qquad \Sv^{(n)}\equiv
\Sv^{(n)}_\myes\,\qquad \Sv \equiv \Sv^{(0)}_\myes,\\
 &&\quad\qquad\quad\, \Sw^{(n)}\equiv{\omeg{V}}^{(n)}\,\qquad \Sw\equiv{\omeg{V}}^{(0)}\,\\
\shV^{(n)}_L&\equiv& \mysh{V}^{(n)}_L\,\qquad \shV^{(n)}\equiv
\shV^{(n)}_\myes\,\qquad \shV \equiv \shV^{(0)}_\myes\,.
\eeas

\subsection{Incompressibility}

The incompressibility translates into a condition on the velocity
field as it implies that its divergence vanishes. This
incompressibility condition is a scalar equation 
\be\label{EqC}
\EqC \equiv \left[\partial_\mu \totV^\mu = 0\right] \,,
\ee
which is a constraint for the velocity field. 
Since \eqref{dsUortho} or \eqref{dsU3} imply that the fiber central line velocity $U^\mu$ is also divergenceless ($\partial_\mu U^\mu=0$), we
deduce that $\partial_\mu V^\mu =0$ and from~\eqref{EqDiv} it reads as
in terms of the FA coordinates
\be
h \partial_a V^a + \tkappa_a V^a + \partial_s \para{V}=0\,.
\ee
As this constraint is scalar, it can be expanded into irreps just like
\eqref{ScalarSTF} and each STF tensor of this expansion must
vanish identically. Using the property~\eqref{AbyyMagic}, we get
\prebool{\bea\label{EqCLn}
&&0=\left[\EqC\right]_L^{(n)} = \partial_s {\Sv}^{(n)}_L + 2(n+1) \stf{V}_L^{(n+1)}+\delta^1_\ell \kappa_a
\Sw^{(n)}\nonumber\\
&&\qquad\qquad+ 2(n+1)\stf{V}^{(n+1)}_{\langle L-1} \kappa_{a_\ell \rangle} +( n+1) \widetilde{\kappa}^a \stf{V}^{(n)}_{a L} \\
&&+ \left(\frac{\ell}{2}+n+1\right)\left(\frac{1}{2}\widetilde{\kappa}^a \shV^{(n-1)}_{a L}+ \shV_L^{(n)}+\shV^{(n)}_{\langle L-1} \widetilde{\kappa}_{a_\ell \rangle} \right)\nonumber\,.
\eea}
{\bea\label{EqCLn}
0=\left[\EqC\right]_L^{(n)} &=& \partial_s {\Sv}^{(n)}_L + 2(n+1) \stf{V}_L^{(n+1)}+\delta^1_\ell \kappa_a
\Sw^{(n)}+ 2(n+1)\stf{V}^{(n+1)}_{\langle L-1} \kappa_{a_\ell \rangle} \\
&&+( n+1) \widetilde{\kappa}^a \stf{V}^{(n)}_{a L} + \left(\frac{\ell}{2}+n+1\right)\left(\frac{1}{2}\widetilde{\kappa}^a \shV^{(n-1)}_{a L}+ \shV_L^{(n)}+\shV^{(n)}_{\langle L-1} \widetilde{\kappa}_{a_\ell \rangle} \right)\nonumber\,.
\eea}
In general, the tensors $\shV^{(n)}_L$ can always be expressed in terms
of the other tensors using the incompressibility
conditions~\eqref{EqCLn}. In particular, from $[\EqC]^{(0)}_\myes$, we get\be
\shV^{(0)}_\myes = - {{\stf{V}}^{(0)a}} \widetilde{\kappa}_{a} -  \partial_s{{\Sv}}\label{uprime}\,.
\ee
This relation has a simple physical interpretation. Indeed, the right-hand side is (minus) the stretching rate of
the velocity field on the central line $\gr{T} \cdot \partial_s
\gr{V}$ given that on the central line the velocity is just $\gr{V} =
\stf{V}^{(0)a} \gr{d}_a + \Sv^{(0)} \gr{T}$. It thus states that a
radial infall necessarily appears when the fiber is stretching, so as to ensure volume conservation.

\subsection{Coordinate velocity}

Initially, the total velocity $\totV^\mu$ is defined as the rate of
coordinate change, that is as $\dd x^\mu/\dd t$ for the fluid elementary particles. However, if we take the components of the
velocity in the basis associated with FA coordinates, that is, the
$\totV^i$, we do not get the rate of change of the FA coordinates $\dd
y^i/\dd t$. We thus need to infer the non-trivial relation between $\totV^i$ and $\dd y^i /\dd t$.
For this we define the speed of the coincident point as
\be\label{VCDef}
\totV_C^\mu \equiv \left.\frac{\partial x^\mu}{\partial t}\right|_{y^i}\qquad
\totV_C^i \equiv -\left.\frac{\partial y^i}{\partial t}\right|_{x^\mu} =
{e^i}_\mu \totV_C^\mu\,,
\ee
that we gave in both the Cartesian and the FA system of coordinates
(see Appendix~\ref{AppGalilee} for details). The speed of the coincident point is the speed of a point that would
have constant FA coordinates $y^1,y^2,s$. 
The total velocity is related by
\be
\totV^\mu = \frac{\dd x^\mu}{\dd t} = \frac{\partial x^\mu}{\partial y^i
}\frac{\dd y^i}{\dd t} + \totV_C^\mu
\ee
that is, in FA coordinates by
\be\label{VRdef}
\totV^i = \totV_R^i +\totV_C^i\,,\qquad \totV_R^i \equiv  \frac{\dd y^i}{\dd t}\,.
\ee
$\totV_R^i$ is thus the coordinate velocity for the FA coordinates
since it equals the rate of change of FA coordinates.
From the above definition of $\totV_C^\mu$ and using the
parameterization~\eqref{xRy} together with the property of the rotation~\eqref{DefRotation}, we get the
expressions 
\prebool{\beas\label{VRVC}
\totV_C^\mu &=& U^\mu + [\omega \times (y^a d_a)]^\mu \nonumber\\
&=& U^\mu +
(\widetilde \omega_a y^a) T^\mu+\para{\omega}\widetilde{y}^a {d_a}^\mu\slabel{VCmu}\\
\totV_R^\mu &=&V^\mu - [\omega \times (y^a d_a)]^\mu\nonumber\\
&=& V^\mu- (\widetilde \omega_a y^a) T^\mu-\para{\omega} \widetilde{y}^a {d_a}^\mu\,.\slabel{VRVC2}
\eeas}
{\beas\label{VRVC}
\totV_C^\mu &=& U^\mu + [\omega \times (y^a d_a)]^\mu = U^\mu +
(\widetilde \omega_a y^a) T^\mu+\para{\omega}\widetilde{y}^a {d_a}^\mu\slabel{VCmu}\\
\totV_R^\mu &=&V^\mu - [\omega \times (y^a d_a)]^\mu= V^\mu- (\widetilde \omega_a y^a) T^\mu-\para{\omega} \widetilde{y}^a {d_a}^\mu\,.\slabel{VRVC2}
\eeas}
On the expression~\eqref{VCmu} it appears that the coincident point
velocity is the velocity of the FCL, on which is added the solid
rotation of the system of FA coordinates which does not vanish when
the point considered is not lying exactly on the FCL. Part of the
rotation is due to the projected part of the rotation rate $\omega_a$
and corresponds to the rotation of section planes, and the rest of the
rotation is due to the longitudinal part of rotation $\para{\omega}$
and corresponds to a rotation of the basis vectors $d_1$ and $d_2$
around the fiber tangential direction, that is, to a rotation inside
the section plane itself.

\subsection{Section shape description}

The section shape can be characterized by its radius as a function
of the direction in the section, that is, by its curve in polar coordinates
\be
\rb(s,t,\yu^a)\,.
\ee
As any function depending on the direction inside the fiber section,
it can be decomposed in STF multipoles ${\rb}_L$ as
\be\label{Defrb}
\rb(s,t,\yu^a)=\ra(s,t) \left(1+\sum_{\ell=1}^\infty {\rb}_L(s,t) \ra^\ell \yu^L \right)\,.
\ee
We call $\ra(s,t)$ the radius of the fiber and the lowest multipole $\rb_{ab}$ can be
interpreted as an elliptic elongation of the fiber section
shape. Instead of working with the multipoles $\rb_L$ which have a
dimension $1/L^\ell$, we can define dimensionless STF moments as
\be\label{Defhrb}
\hrb_L \equiv {\rb}_L R^\ell\,.
\ee
There exist alternate ways for describing the shape of the fiber section, and we give an example of another method in Appendix~\ref{AppAltShape} and relate it to the description~\eqref{Defrb}.

\subsection{Normal vector}

For a function depending on the direction $f(\yu^a)$, we can define an
orthoradial derivative by
\be
\frac{D }{D \yu^a}f(\yu^c) \equiv \left.\perp_a^b \frac{\partial
  f(y^c)}{\partial y^b}\right|_{y^c=\yu^c}
\ee
where the orthoradial projector is defined by
\be
\perp^a_b \equiv \delta^a_b -\yu^a \yu_b\,.
\ee
This projector satisfies $\perp_b^a \yu^b = 0$, $\perp^a_b
\widetilde{\yu}^b = \widetilde{\yu}^a$,  and $\perp_a^a=1$. Note that it can also be written as
\be
\perp_a^b = \widetilde{\yu}_a \widetilde{\yu}^b = r \partial_b \yu^a\,.
\ee
When applied on the fiber radius, this yields
\prebool{\bea
\frac{D\rb} {D \yu^a}=r \frac{\partial \rb}{\partial y^a} &=&  \ra\sum_{\ell=1}^\infty (\ell+1)\perp_a^b \hrb_{b L} \yu^L\\
&=&\ra \sum_{\ell=1}^\infty (\ell+1)\tyu_a \left(\hrb_{b L} \tyu^b \yu^L\right)\,.
\eea}
{\be
\frac{D\rb} {D \yu^a}=r \frac{\partial \rb}{\partial y^a} =  \ra\sum_{\ell=1}^\infty (\ell+1)\perp_a^b \hrb_{b L} \yu^L=\ra \sum_{\ell=1}^\infty (\ell+1)\tyu_a \left(\hrb_{b L} \tyu^b \yu^L\right)\,.
\ee}

A normal co-vector (not necessarily unity) to the boundary surface is $\gr{\Nv} = \Nv_\mu \dd
x^\mu$ with
\be\label{DefNormalvector}
\Nv_\mu = \left.\partial_\mu \Phi\right|_{\Phi=0}\,
\ee
where the boundary function $\Phi$ is defined as
\be\label{DefPhi}
\Phi(s,t,y^a)=r-\rb(s,t,\yu^a) = \sqrt{y^a y_a}-\rb(s,t,\yu^a) \,.
\ee
The components of the normal vector are in our case
\prebool{\beas\label{NormalVector}
\Nv_a &=& \yu_a - \frac{1}{\rb}\frac{D \rb}{D \yu^a} \\
&=& \yu_a -\widetilde{\yu}_a\frac{  (\ell+1)
  \hrb_{b L} \widetilde{\yu}^b\yu^L}{1+ \hrb_L\yu^L} \\
\Nv_3 &=& -\partial_s \rb \,,\qquad \Nv_\tetr{3}=\frac{1}{h}\Nv_3\,,
\eeas}
{\beas\label{NormalVector}
\Nv_a &=& \yu_a - \frac{1}{\rb}\frac{D \rb}{D \yu^a} = \yu_a -\widetilde{\yu}_a\frac{  (\ell+1) \hrb_{b L} \widetilde{\yu}^b\yu^L}{1+ \hrb_L\yu^L} \\
\Nv_3 &=& -\partial_s \rb \,,\qquad \Nv_\tetr{3}=\frac{1}{h}\Nv_3\,,
\eeas}
where sums over $\ell$ are implied.
This vector can be normalized and the unit normal vector is just
\be\label{DefUnitN}
\widehat{\Nv}^\mu \equiv \frac{\Nv^\mu}{\sqrt{\Nv_a \Nv^a + (\Nv_\tetr{3})^2}}\,\,.
\ee

\subsection{Scalar extrinsic curvature}

In order to use Young-Laplace law for surface tension, we need the
general expression for the scalar part of the extrinsic curvature of
the fiber boundary, which by definition is the divergence of the unit
normal vector. Using~\eqref{EqDiv}, it is thus obtained as
\be\label{calK}
{\cal K} \equiv \partial_\mu \widehat{\Nv}^\mu = \frac{\widetilde{\kappa}_a \widehat{\Nv}^a
}{h}+\partial_a \widehat{\Nv}^a +\frac{\partial_s \widehat{\Nv}^\tetr{3}}{h}\,.
\ee

\subsection{Boundary kinematics}

Since the boundary must follow the velocity field, the constraint $\Phi=0$ must
propagate with the velocity, that is,
\be
\left[\frac{\dd \Phi}{ \dd t}\right]_{\Phi=0} = \left[\frac{\partial
    \Phi}{ \partial t}+\frac{\dd y^i}{\dd t}\frac{\partial \Phi}{ \partial y^i}\right]_{\Phi=0}  =0\,.
\ee
Since $\partial_s r= \partial_t r = 0$, then $\partial_t \Phi =
-\partial_t \rb$ and given the definition~\eqref{DefNormalvector} for the non-unit normal vector and
the definition~\eqref{VRdef} of the coordinate velocity, this constraint is simply rewritten as
\be\label{EqPropaR}
\EqBD \equiv\left[\partial_t \rb=\totV_R^i N_i =  \totV_R^3 N_3 + \totV_R^a N_a \right]_{\Phi=0}\,.
\ee
We must be careful with the fact that it is
$\totV_R^3 = \dd s/\dd t$ which appears and not $\totV_R^{\tetr{3}}$,
and we must thus use $\totV_R^3 = {d_\tetr{3}}^3 \totV_R^\tetr{3} =
h^{-1} \totV_R^\tetr{3}$ to relate them.

Despite its apparent simplicity, this equation is actually rather complicated.
First we stress that it is the coordinate velocity which appears since
it is the velocity which gives the rate of change for FA
coordinates.  But more importantly, all quantities must be evaluated on the
fiber side, meaning that every occurrence of $y^a$ must be replaced by
$\rb \yu^a$ where $\rb$ itself has a directional dependence given by \eqref{Defrb}. This equation is thus in general extremely non-linear in the
STF multipoles $\rb_L$. We are forced to realize that it is hopeless to solve the general problem of curved viscous jets without
major simplifications, which leads to consider a perturbative scheme
on which we should perform a consistent truncation.

\subsection{Slenderness perturbative expansion}

If the typical length for variations in the velocity field is $L$ and
is much larger than the radius $\ra$, then the approximation that the
body considered is elongated holds and we can hope to find a coherent
one-dimensional reduction. The small parameter of our perturbative
expansion is thus
\be
\epsilon_\ra \equiv \ra/L\,\,.
\ee
For instance, the moment $\Sv^{(1)}$ comes originally from a Taylor
expansion of the velocity, as can be checked from the dimension $[\Sv^{(1)}]=[\Sv^{(0)}] /
L^2$. A term like $\Sv^{(1)}\ra^2$ is thus of order $\epsilon_\ra^2$
compared to the lowest order velocity $\Sv^{(0)}$, meaning that the
former is really a correction to the latter. In general, higher order
multipoles correspond to higher orders in $\epsilon_\ra$ because they
primarily come from gradients of the velocity fields which bring
inverse powers of $L$. As an example, $\Sv^{(0)}_{ab} y^a y^b$ is of order
$\epsilon_\ra^2$ compared to $\Sv=\Sv^{(0)}_\myes$.

When including the effect of curvature, we also assume that the scale
$1/|\gr{\kappa}|$ (the curvature radius) is also of the order of
$L$ at most. Indeed if there is curvature, the velocity flow must adapt on
scales which are commensurate with the curvature radius. As a
consequence, terms of the type $\kappa_a y^a$ must also be of order
$\epsilon_\ra$. In any case it would be impossible to consider
sections which have a section radius larger than the fiber curvature
radius. Indeed, in that case fiber sections would intersect, hence
$|\gr{\kappa}| \ra \propto \epsilon_\ra$ must be small to obtain a satisfactory one-dimensional approximation.

\subsection{Gauge fixing}\label{GaugeFix}

Since there are three degrees of freedom in the position of the FCL inside the
viscous jet, we can fix two of these by asking that there is no dipole
in $\rb$.  This corresponds to the intuitive requirement that the
fiber central line should be {\it in the middle} of the fiber section.
Formally, this means that we fix the gauge by setting 
\be\label{Gauge}
\rb_a =0\,,
\ee 
and it leads to a constraint equation when considering the dipole of \eqref{EqPropaR}. 

The  gauge restriction~\eqref{Gauge} is an order $\epsilon_\ra^2$ expression as it is automatically satisfied at lowest order. It
reads as indeed
\prebool{\bea\label{GaugeConstraint}
\stf{V}^{(0)}_a&=&\mytfrac{1}{6} {\ra}^2 \left(-2 {{\stf{V}}^{(1)}_{a}} + 6 {\cal H} {{\Sv}^{(0)}_{a}} - 6 {\cal H} {\Sv} \widetilde{\kappa}_{a} + 2 \kappa_{a} {\Sw} \right.\nonumber \\ 
&&\left.\quad- 6 {\cal H} \widetilde{\omega}_{a}  - 3 \widetilde{\kappa}_{a} \partial_s{{\Sv}} + 2 \partial_s{{{\Sv}^{(0)}_{a}}}\right)+\calO{\epsilon_\ra^4}
\eea}
{\be\label{GaugeConstraint}
\stf{V}^{(0)}_a=\mytfrac{1}{6} {\ra}^2 \left(-2 {{\stf{V}}^{(1)}_{a}} + 6 {\cal H} {{\Sv}^{(0)}_{a}} - 6 {\cal H} {\Sv} \widetilde{\kappa}_{a} + 2 \kappa_{a} {\Sw} - 6 {\cal H} \widetilde{\omega}_{a}  - 3 \widetilde{\kappa}_{a} \partial_s{{\Sv}} + 2 \partial_s{{{\Sv}^{(0)}_{a}}}\right)+\calO{\epsilon_\ra^4}
\ee}
where we have defined the stretching factor
\be\label{DefH}
{\cal H}\equiv \partial_s \ln \rb \,.
\ee
The gauge restriction fixes the global velocity shift with respect to the fiber
central line velocity $U^a$ which is encoded by the velocity moment
$\stf{V}_a^{(0)}$. And, since this shift is of order $\epsilon_\ra^2$, we can state that the
projected component of the velocity on the central line is nearly the
projected velocity of the central line itself, that is, $[\totV^a \simeq U^a]_{y^1=y^2=0}$.

With~\eqref{Gauge}, we have fixed only two of the three gauge degrees
of freedom which arise from the fact that we are free to choose any
curve as the FCL. The third degree of freedom corresponds to the
possible reparameterization of the fiber inside the same curve, that
is to the replacement $s\to s+f(s,t)$. Given the choice of normalization for
the tangent vector in \eqref{TTun}, this freedom is only a global but time-dependent
reparameterization freedom $s \to s+f(t)$. For every problem considered, there is a
natural way to fix unambiguously the affine parameter $s$, for
instance setting it to $s=0$ at the boundary.

\subsection{Shape restriction} \label{SecShapeRestrict}

With the gauge choice of the previous section, we have managed to
cancel the shape dipole. However, we cannot assume that in general higher order shape multipoles
vanish. Indeed, if we have curvature, then terms of the type $\propto \ra^2 \kappa_{\langle a} \kappa_{b
  \rangle}$ which are of order $\epsilon_\ra^2$ would source the
section shape quadrupole $\rb_{ab}$. Typically, we expect to find that
shape terms like $\rb_L y^L$ or $\ra^\ell \rb_L$ are of order
$\epsilon_\ra^\ell$. In~\S~\ref{SecCurvedJets}, we discuss this
scaling and show that circular sections are compatible with the string
description of curved fibers, which is the lowest order description in which order $\epsilon_\ra^2$ effects are ignored.

Of course, if we restrict to axisymmetric jets as we shall do in~\S~\ref{SecApplications}, then $\kappa_a=0$ and circular
sections are consistent throughout even though one might still want to
consider straight jets with non-circular sections.

\subsection{Velocity shear rate}

In order to describe the dynamics of the viscous fluid inside the
fiber, we will need to consider the gradient of the velocity field. Let us define the non-symmetric tensor
\be
S_{\mu\nu}\equiv\partial_\mu \totV_\nu\,.
\ee
Its components are given by
\beas\label{Sabincomponents}
S_{ab} &=& \partial_a \totV_b\\
S_{a\tetr{3}} &=& \partial_a \totV_\tetr{3}\\
h S_{\tetr{3}a} &=& \partial_s \totV_a-\tkappa_a \totV_\tetr{3}\\
h S_{\tetr{3}\tetr{3}} &=&  \partial_s \totV_\tetr{3} +\tkappa^a \totV_a\,.
\eeas
These components are obtained either from the components in the
canonical basis of the FA coordinates, that is the $S_{ab}$, $S_{3a}$,
$S_{a3}$ and $S_{33}$ which we compute from~\eqref{ChristoUseful}, that we then
project on the orthonormal basis, or using
directly~\eqref{Riccirotation} in the velocity decomposition
$\gr{\totV} = \totV^\tetr{i} \gr{d}_\tetr{i}$.  This tensor can be decomposed as
\be\label{Ssigmavorty}
S_{\mu\nu} = \frac{1}{2}\sigma_{\mu\nu}+\omega_{\mu\nu}\,.
\ee
where we used that the velocity shear rate is (twice) the symmetric part
\be
\sigma_{\mu\nu} \equiv 2S_{(\mu\nu)}=S_{\mu \nu}+S_{\nu\mu}\,,
\ee
and the vorticity is the antisymmetric part
\be
\myvort_{\mu\nu} \equiv S_{[\mu\nu]} = \frac{1}{2}(S_{\mu\nu}-S_{\nu\mu})\,.
\ee
We also define the vorticity (Hodge) dual vector by
\be
\myvort_\alpha \equiv \frac{1}{2}\eps_{\alpha \mu\nu} \myvort^{\mu\nu}\myquad\Rightarrow\myquad\myvort_{\mu\nu} = \eps_{\mu\nu\alpha} \myvort^\alpha \,.
\ee

From~\eqref{Sabincomponents}, we find that the components of the shear
in the orthonormal basis are then given by
\beas\label{Sigincomponents}
\sigma_{ab} &=& \partial_a \totV_b + \partial_b \totV_a,\\
h \sigma_{a\tetr{3}} &=& h \partial_a \totV_\tetr{3}+\partial_s
\totV_a-\tkappa_a \totV_\tetr{3},\\
h \sigma_{\tetr{3}\tetr{3}} &=& 2 (\partial_s \totV_\tetr{3} +\tkappa^a \totV_a)\,.\slabel{Eqs33}
\eeas
Similarly, the components of the vorticity vector are
\prebool{\beas\label{varpiComponents}
\myvort^a&=&\mytfrac{1}{2}\eps^{ab}\left(S_{b\tetr{3}}-S_{\tetr{3}b}\right)\\
&=&\mytfrac{1}{2}\eps^{ab}\left[\partial_b \totV_\tetr{3}
  -\frac{1}{h}\left(\partial_s \totV_b-\tkappa_b \totV_\tetr{3}\right)\right]\nonumber\\
\myvort^\tetr{3}&=&\mytfrac{1}{2}\eps^{ab}\partial_{[a}\totV_{b]}=\mytfrac{1}{4}\eps^{ab} (\partial_a \totV_b-\partial_b \totV_a)\,.
\eeas}
{\beas\label{varpiComponents}
\myvort^a&=&\mytfrac{1}{2}\eps^{ab}\left(S_{b\tetr{3}}-S_{\tetr{3}b}\right)=\mytfrac{1}{2}\eps^{ab}\left[\partial_b \totV_\tetr{3}
  -\frac{1}{h}\left(\partial_s \totV_b-\tkappa_b \totV_\tetr{3}\right)\right]\\
\myvort^\tetr{3}&=&\mytfrac{1}{2}\eps^{ab}\partial_{[a}\totV_{b]}=\mytfrac{1}{4}\eps^{ab} (\partial_a \totV_b-\partial_b \totV_a)\,.
\eeas}

\section{Dynamics}\label{SecDynamics}

The dynamics of viscous fluids is well known and arises from the
Navier-Stokes equation. However, in order to achieve a one-dimensional
reduction for viscous jets, we must find a way to get rid of the
physics on the fiber boundary. Enforcing the boundary conditions on
the stress tensor on the fiber boundary leads to a set of three
constraints which can be conveniently used to reduce the number of
free fields in our one-dimensional reduction. This section is
dedicated to the general construction of this method and we then apply
it in the subsequent sections for axisymmetric and curved fibers.

\subsection{Total stress tensor and viscous forces}

The total stress tensor is decomposed as
\be\label{taucontribs}
\tau_{\mu\nu} = \tau^{(P)}_{\mu\nu}+\tau^{(\visc)}_{\mu\nu}\,,
\ee
where the two components arise from pressure forces and shear
viscosity. For a Newtonian fluid, they are simply given by
\be
\tau^{(P)}_{\mu\nu}=-P g_{\mu\nu}\,,\qquad \tau^{(\visc)}_{\mu\nu}=\visc \sigma_{\mu\nu}\,.
\ee
The pressure is a scalar and it is thus decomposed as in \eqref{ScalarSTF}
\be
P = \sum_{\ell=0}^\infty \sum_{n=0}^\infty  P^{(n)}_{L} y^L r^{2n}\,\,,
\ee
where the $P^{(n)}_{L}$ are STF tensors. 

From the total stress tensor~\eqref{taucontribs}, we can define a
viscous force per unit area on the fiber sections
\be\label{DefVF}
\VF_\mu \equiv \tau_{\mu \tetr{3}}\,.
\ee
As any vector it has a longitudinal part $\para{\VF}$ which corresponds to viscous
traction or compression on fiber sections, and a sectional part $\VF^a$.

\subsection{Boundary conditions}\label{SecBoundary}

The boundary conditions for the stress tensor is the vector constraint
\bea\label{VectorBoundaryConstraint}
\EqBC_\mu &\equiv& \left[\tau_{\mu\nu} \widehat{\Nv}^\nu + \ST {\cal K}
  \widehat{\Nv}_\mu=0\right]\\
&\equiv&\left[\tau^{(\visc)}_{\mu\nu}\widehat{\Nv}^\nu+\ST {\cal K}\widehat{\Nv}_\mu-P \widehat{\Nv}_\mu=0\right]\,.\nonumber
\eea
As any vector, it can be decomposed into its longitudinal component
$\para{\EqBC}$ and its sectional component $\EqBC^a$. The latter can be
further decomposed into a radial contribution and an orthoradial
contributions as
\be\label{DefCrCt}
\EqCr\equiv \EqBC_a \yu^a \,,\qquad \EqCtheta \equiv \EqBC_a \tyu^a \,.
\ee
However, these are not fields on the fiber sections since they are defined only at the
boundary.  They depend on the position on the FCL $s$, on time $t$, but
their sectional dependence is only a dependence in the direction
vector $\yu^a$. They also depend on the fiber radius $\ra$ and on the
shape multipoles $\ra_L$. So, in general they can be expanded in STF
components as
\beas\label{ScalarSTFSide}
\para{\EqBC}(s,t,\yu^a)&=&\sum_{\ell=0}^\infty \para{\EqBC}_L(s,t,\ra)\ra^\ell\yu^{L}\slabel{ScalarSTFSide1}\\
&=&\sum_{\ell=0}^\infty \sum_n
\para{\EqBC}_L^{(n)}(s,t)\ra^{2n+\ell}\yu^{L}\,,\slabel{ScalarSTFSide2}
\eeas
where in the second line we have also expanded the dependence of the
STF multipoles in powers of $\ra$, and with similar expansions for the
radial and orthoradial boundary constraints $\EqCtheta$ and
$\EqCr$. To be precise, for the latter, the expansion takes the
form~\eqref{ScalarSTFSide} for $\EqCr/\ra$. Note again the difference
with \eqref{ScalarSTF} as the powers of $r$ are replaced by powers of
$\ra$ because constraints are only defined on the boundary. In
practice, it is simpler to consider the constraint
\eqref{VectorBoundaryConstraint} with the non-normalized normal vector
$\Nv^\mu$ instead of $\widehat{\Nv}^\mu$ since they are equivalent.

We realize that the total sum involves all orders of the form $\epsilon_\ra^{2m}$. Since we are eventually interested in results
which are valid up to a given order in $\epsilon_\ra$, we define the
moments of the constraints {\it up to a given order} by
\be\label{DefUpTon}
\para{\EqBC}^{(\leq n)}_L = \sum_{m=0}^{m=n} \para{\EqBC}^{(m)}_L \ra^{2m}\,,
\ee
with similar definitions for the radial and orthoradial constraints.

As we shall detail in two examples in \S~\ref{SecApplications} and~\S\ref{SecCurvedJets}, we
will deduce general relations from the vector
constraint~\eqref{VectorBoundaryConstraint}, or, more precisely, from
its three scalar components (longitudinal, radial, and
orthoradial). 

\subsection{Volumic forces}

Once the stress tensor is computed, it is straightforward to get the
volumic forces $f^\mu$ since they are expressed as
\be\label{rawfmu}
f^\mu=\partial_\nu \tau^{\nu\mu}+g^\mu=\mu\Delta \totV^\mu - \partial^\mu P+g^\mu,
\ee
where $g^\mu$ are long range volumic forces such as gravity. Its
components are related to the components of the stress tensor thanks to
\prebool{\bea
f^\tetr{3} &=& \partial_a \tau^{a \tetr{3}} +
\frac{1}{h}\left(\partial_s \tau^{\tetr{3}\tetr{3}} +2\tkappa_a
\tau^{a\tetr{3}}\right)+g^{\tetr{3}}\\
f^a&=& \partial_b \tau^{b a}+
\frac{1}{h}\left(\partial_s \tau^{\tetr{3}a} +\tkappa_b
\tau^{ba}-\tkappa^a \tau^{\tetr{3}\tetr{3}}\right)+g^a.\nonumber
\eea}
{\beas
f^\tetr{3} &=& \partial_a \tau^{a \tetr{3}} +
\frac{1}{h}\left(\partial_s \tau^{\tetr{3}\tetr{3}} +2\tkappa_a
\tau^{a\tetr{3}}\right)+g^{\tetr{3}},\\
f^a&=& \partial_b \tau^{b a}+
\frac{1}{h}\left(\partial_s \tau^{\tetr{3}a} +\tkappa_b
\tau^{ba}-\tkappa^a \tau^{\tetr{3}\tetr{3}}\right)+g^a.
\eeas}

\subsection{Navier-Stokes equation}

The dynamics of a viscous fluid is governed by the conservation
equation and the Navier-Stokes equation. The conservation equation has
already been used since from incompressibility it implied the
divergenceless condition~\eqref{EqC}. Having developed all the tools to express
the volumic forces, we are now in position to write the Navier-Stokes equation. It is of the form
\be\label{SecNS}
\EqD^\mu\equiv \left[A^\mu=f^\mu\right]
\ee
where the acceleration vector is
\be\label{DefAmu}
A_\mu \equiv \left.\partial_t \totV_\mu\right|_{y^i} + {\totV_R}^\nu S_{\nu \mu} + G_\mu\,.
\ee
If the dynamics is considered in a (constantly) rotating frame, the
fictitious or geometrized forces (inertial and Coriolis forces)
gathered in $G^\mu$ are expressed as
\be\label{Imu}
G^\mu \equiv 2 \eps^{\mu\alpha\beta}\Omega_\alpha
\totV_\beta+(\Omega_\nu x^\nu) \Omega^\mu - \Omega^2 x^\mu
\ee
where $\Omega^\mu$ is the rotation of the frame with respect to a
Galilean (inertial) frame.

Just as any vector, the Navier-Stokes equation~\eqref{SecNS} is
decomposed into a longitudinal part $\para{\EqD}$ and a sectional part
$\EqD^a$ which we can further decompose in irreps as
in~\S~\ref{SecIrrepVector}. The components of the first term of
\eqref{DefAmu} are simply obtained from
\beas\label{dtVcomp}
(\partial_t \totV)^\tetr{3}&\equiv&{d^\tetr{3}}_\mu \partial_t \totV^\mu=\partial_t \para{\totV}+\widetilde{\omega}_a \totV^a\\
(\partial_t {\totV})^a &\equiv&{d^a}_\mu \partial_t \totV^\mu=\partial_t \totV^a-\para{\totV} \widetilde{\omega}^a+\widetilde{\totV}^a \para{\omega} \,,
\eeas
where we recall the notation $\para{\totV} \equiv \totV^\tetr{3}$ for
the longitudinal component of the velocity.

\subsection{Secondary incompressibility constraint}

The incompressibility implies the divergenceless
condition~\eqref{EqC}, and as such it can be considered as a primary
constraint. Then by ensuring that this constraint remains satisfied
when time evolves, we obtain a secondary constraint. Indeed taking the
divergence of the Navier-Stokes equation~\eqref{SecNS} we obtain an incompressibility {\it constraint}
\be\label{EqCD}
\partial_\mu A^\mu = \partial_\mu f^\mu = -\Delta  P\,,
\ee
where we have assumed that the long range forces are constant
(e.g., for gravity) or have at least no divergence ($\partial_\mu
g^\mu=0$). If the long range forces have a divergence, then this
should be included in \eqref{EqCD}.

From the expression~\eqref{DefAmu} of the acceleration, this constraint is
\be
\EqCD\equiv \left[S_{\mu \nu}S^{\nu \mu}+2 \eps^{\alpha \beta \mu} \Omega_\beta
S_{\alpha \mu} -2 \Omega^2 = -\Delta  P\right]\,.
\ee
Using the decomposition~\eqref{Ssigmavorty} of the velocity gradient tensor into
velocity shear rate and vorticity, it can be recast nicely as
\be
\frac{1}{4}\sigma_{\mu \nu}\sigma^{\nu \mu}-2
  (\Omega_\mu +\myvort_\mu)(\Omega^\mu+\myvort^\mu) = -\Delta  P.
\ee

The right-hand-side of this equation can be computed by using \eqref{LaplacianOpened}. We will use this secondary incompressibility constraint to get constraints on the various moment
of the pressure field. 

\subsection{Dimensionless reduction}

So far, all physical quantities have a physical dimension. It is, however,
possible to build dimensionless quantities. Usually for viscous
fluids, this is done by noting that the dimensions of viscosity and surface tension [recalling that we have divided them by
the mass density in~\eqref{mutomurho}] are
\be
[\visc] = L^2 T^{-1}\,\qquad [\ST] = L^3 T^{-2}\,.
\ee 
It is thus possible to define a length scale $\visc^2/\ST$ and a time
scale $\visc^3/\ST^2$ from which we can define dimensionless
quantities for all physical quantities in the problem at hand. However,
we want to allow for the possibility of having no surface tension, and
we do not wish to use $\ST$ to define dimensionless
quantities. Instead, we decide that there is a natural length scale
$\Ls$ in our problem which corresponds to the typical length of
velocity variations. The time scale is then obtained from
$\Ls^2/\visc$. The main quantities in the problem are simply
adimensionalized using these length scales and timescales. For instance, we
define dimensionless quantities with
\prebool{\bea
&&t = \frac{\Ls^2}{\visc}\,\dimless{t}\,,\myquad s = \Ls
\,\dimless{s}\,,\myquad \Sv = \frac{\visc}{\Ls}\,\dimless{\Sv}\,,\myquad
g=\frac{\mu^2}{\Ls^3}\dimless{g}\,,
\nonumber\\
&&\Sw = \frac{\visc}{\Ls^2}\,\dimless{\Sw}\,,\myquad R=\Ls
\dimless{R}\,,\myquad \nu = \frac{\mu^2}{\Ls}\dimless{\nu}\,,\nonumber\\
&&\kappa^a = \frac{1}{\Ls}\dimless{\kappa}^a \,,\myquad \omega^a = \frac{\visc}{\Ls^2}\,\dimless{\omega}^a\,,
\eea}
{\bea
&&t = \frac{\Ls^2}{\visc}\,\dimless{t}\,,\myquad s = \Ls
\,\dimless{s}\,,\myquad \Sv = \frac{\visc}{\Ls}\,\dimless{\Sv}\,,\myquad
g=\frac{\mu^2}{\Ls^3}\dimless{g}\,,\Sw = \frac{\visc}{\Ls^2}\,\dimless{\Sw}\,,\myquad R=\Ls
\dimless{R}\,,\nonumber\\
&&\nu = \frac{\mu^2}{\Ls}\dimless{\nu}\,,\myquad\kappa^a = \frac{1}{\Ls}\dimless{\kappa}^a \,,\myquad \omega^a = \frac{\visc}{\Ls^2}\,\dimless{\omega}^a\,,
\eea} 
where we recall the notation $\Sv=\Sv_\myes^{(0)}$ and
$\Sw=\Sw_\myes^{(0)}$. For higher order multipoles this construction
of dimensionless variables is straightforwardly performed as the
dimension of the multipoles is read from the expansion from which it
is defined. For instance from~\eqref{ScalarSTF} we get $\Sv_L^{(n)} =
\mu/L^{1+\ell+2n} \dimless{\Sv}_L^{(n)}$. 
Note also that by construction $\epsilon_\ra =
\dimless{R}$ so that in practice it is by expanding in powers of
$\dimless{\ra}$ that we identify for any expression the various orders in powers of $\epsilon_\ra$.

The dimensionless ratios of fluid mechanics which are relevant for viscous jets in a
rotating frame are the Reynolds number, the Froude number, the Rossby
number, and the Weber number. If we consider a typical reference reduced velocity
${\dimless{\Sv}}^{\rm r}$, then they are simply expressed as
\be\label{Numbers}
{\rm Re} = \dimless{\Sv}^{\rm r},\myquad {\rm Fr}^2 =
\frac{(\dimless{\Sv}^{\rm r})^2}{\dimless{g}},\myquad {\rm Rb} =
\frac{\dimless{\Sv}^{\rm r}}{\dimless{\Omega}},\myquad {\rm We} = \frac{(\dimless{\Sv}^{\rm r})^2}{\dimless{\ST}}\,.
\ee
In the remainder of this article, we use the dimensionless
physical quantities rather than the dimensionless numbers, but using the dictionary~\eqref{Numbers}, all
expressions can be recast with these dimensionless numbers.

In the next section, we present two main applications of our formalism
and we shall assume that we have performed such a dimensionless
reduction for all quantities. In order to avoid cluttering of notation, we will omit in the remainder of this article to specify that the quantities are
dimensionless. In practice, the dimensionless reduction amounts simply
to replacing $\mu \to 1$ in all our equations but keeping $\ST$, and it
is thus used as a consistency check.

Note that several other schemes would have been possible to build dimensionless
quantities, just by using other physical quantities that might be
present in the problem. For instance if we use gravity, then we can
build a length scale and a time scale without resorting to a choice
of $\Ls$ simply by
\be
L = \visc^{2/5} g^{1/5}\,,\qquad T = \mu^{2/5} g^{-4/5}\,.
\ee 
After building dimensionless quantities with these scales, the gravity vector $g^\mu$ is replaced by a unit
vector while $\visc \to 1$ in all equations.

Similarly, if we are working in a rotating frame we can use
the rotation vector magnitude to build a time scale and then from the
viscosity we can build a length scale as
\be
T = \Omega^{-1}\,,\qquad L = \sqrt{\mu/\Omega}\,.
\ee
After building dimensionless quantities, the rotation vector
$\Omega^\mu$ is replaced by a unit vector while $\visc \to 1$ in all equations.

\section{Application to axisymmetric jets} \label{SecApplications}

Using the formalism developed so far in the particular case of a
straight viscous jet ($\kappa^a=0$) would be equivalent to kill a fly with a
sledgehammer, especially if we also require axisymmetry of the fiber
around the FCL.  Indeed, in that case the FA coordinates are just Cartesian coordinates, with the
third coordinate $s$ corresponding to the axis of symmetry of the
problem. This problem has been studied already in the
literature~\citep{GarciaCastellanos,EggersDupont} and
we shall rederive, recover and extend these standard results to
include higher order corrections, and a possible rotation of the
viscous fluid around the axis of symmetry (called torsion by~\citet{BBCF}).

From the assumed rotational symmetry of the problem, only the
monopolar moments are non-vanishing and we need only to consider the
$\Sv^{(n)}=\Sv_\myes^{(n)}$,  $\Sw^{(n)}=\Sw_\myes^{(n)}$  and $\shV^{(n)}=\shV_\myes^{(n)}$. We recall that we use the short notation introduced in the
definitions~\eqref{SimpleNotation}, and in particular we emphasize the
notation $ \Sv = \Sv^{(0)}$ for the fundamental lowest order component
of the longitudinal velocity. It is also possible to further restrict the problem to non-rotational
flows, and require that all the $\Sw^{(n)}$ vanish, as
done in e.g. \citet{GarciaCastellanos,EggersDupont} but we will not
perform such simplification and allow for a rotation of the fluid around the
axis of symmetry as in \citet{BBCF}.

From the symmetry of the problem we also deduce that $U^a=0$ and the
rotation rate of the central line satisfies necessarily
$\omega^a=0$. Then from~\eqref{Reldtdsko2}, the longitudinal component
of rotation $\para{\omega}$ is constant along the fiber so it is
reasonable to choose $\para{\omega}=0$.  Eventually, the only possible velocity components for the central line is
$\para{U}=U_\tetr{3}$. However it is natural to also set $U_\tetr{3}=0$ which amounts to taking a non-moving FCL. As a
consequence from \eqref{VRVC} the total velocity $\totV^\mu$ is also the coordinate
velocity $\totV_R^\mu$. The sledgehammer comes from the fact that the
FA coordinates $(y^1,y^2,y^3=s)$ are just Cartesian coordinates so they can be chosen to
be the Cartesian coordinates $(x^1,x^2,x^3)$ and all the machinery of
FA coordinates is not used in this case.

Finally, the long range volumic forces need also to respect the
symmetry and $g^a=0$, so if we consider the effect of gravity the fiber needs to
be vertical and we will write simply  $\para{g}=g^\tetr{3}=g$. 
Let us also mention that we discard the possibility of considering a
rotating frame ($\Omega^a =\para{\Omega}=\Omega^\mu=0$). Indeed, even though
it is possible in principle to also consider the problem in a  frame
rotating around the axis of symmetry (that is $\para{\Omega} \neq 0$),
this would be of very limited interest as it can be obtained from the
replacement $\Sw^{(0)}\to \Sw^{(0)}+\para{\Omega}$ (see the discussion
in \S~\ref{SecAltRot}).

\subsection{The lowest order viscous string model}\label{SecStringStraight}

The divergenceless condition~\eqref{EqCLn} leads simply to the set of relations
\be\label{IncompAxial}
\shV^{(n)}=-\frac{1}{n+1}\partial_s \Sv^{(n)}\,,
\ee
so that we need only  to consider the $\Sv^{(n)}$ and the $\Sw^{(n)}$.
At lowest order ($n=0$), \eqref{IncompAxial} has a very simple interpretation. A gradient in the longitudinal
velocity $\partial_s \Sv^{(0)}$ leads to a radial inflow $\shV^{(0)}$ because of
incompressibility. Indeed, if the flow stretches, the radius shrinks to ensure volume conservation and thus incompressibility.


The lowest contributions of the Navier-Stokes
equation are $\para{\EqD}^{(0)}$ for the longitudinal part and
$\omeg{\EqD}^{(0)}$ for the rotational part. At lowest order they
lead to
\beas\label{BasicSvSwEqs}
&&\partial_t \Sv + \Sv \partial_s \Sv = - \partial_s P^{(0)}+ 4
\Sv^{(1)}+\partial_s^2 \Sv +g\slabel{BasicSvSwEqs1}\\
&&\partial_t \Sw+\Sv \partial_s \Sw- \Sw\partial_s \Sv= \partial_s^2
\Sw + \Sw^{(1)}\,.\slabel{BasicSvSwEqs2}
\eeas
The evolution of the radius is easily obtained from~\eqref{EqPropaR} and it
reads as, at lowest order,
\be\label{EvolRCircLow}
\partial_t \ln \ra  = -{\cal H}\Sv -\mytfrac{1}{2}\partial_s \Sv+\calO{\epsilon_\ra^2}\,.
\ee

In order to find a closed system of equations from~\eqref{BasicSvSwEqs}, we need to find
$P^{(0)}$, $\Sv^{(1)}$, and $\Sw^{(1)}$ from the boundary constraint
equation~\eqref{VectorBoundaryConstraint}. First, it turns out that in the
axisymmetric case, the contribution of surface tension can only be in
$P^{(0)}$, and we can separate the pressure field as
\be\label{PPST}
P \equiv  P_\ST +p\,,\qquad P_\ST \equiv  \ST {\cal K}\,,
\ee
where $p$ is the contribution coming from viscous forces and where the
extrinsic scalar curvature ${\cal K}$ does not depend on $r$. It is
also more convenient to combine the longitudinal
part and  the radial part of the boundary constraint~\eqref{VectorBoundaryConstraint} to obtain a constraint
which gives directly $p$ and remove the pressure dependence in
the longitudinal constraint. Indeed, the components of the normal vector
take exactly the form 
\be
N^a=\yu^a\,\qquad N^\tetr{3}=-{\cal H}\ra\,,
\ee 
and we find that the three scalar constraints can be expressed in the form
\beas\label{ConsAllAxisym}
\EqCP&\equiv& \left[\left(1-\myalpha^2\right)p = \visc \left(\sigma_{ab}\yu^a\yu^b -\myalpha^2\sigma_{\tetr{3}\tetr{3}}\right)\right]\slabel{ConsP}\\
\para{\EqBC}&\equiv& \left[\left(1-\myalpha^2\right)\sigma_{\tetr{3}a}\yu^a = \myalpha(\sigma_{\tetr{3}\tetr{3}}-\sigma_{ab}\yu^a\yu^b)\right]\slabel{Consr}\\
\EqCtheta&\equiv& \left[\sigma_{ab}\yu^a\tyu^b =\alpha \sigma_{\tetr{3}a}\tyu^a\right]\,,\slabel{Constheta}
\eeas
where $\myalpha\equiv {\cal H} \ra$.

Using that constraints are expanded according to
\eqref{ScalarSTFSide}, then from the lowest order of the pressure
constraint ($ \EqCP_{\myes}^{(0)}$) we obtain the pressure monopole at
lowest order
\be\label{ConsaxiP0}
p^{(0)} = \shV^{(0)}+\calO{\epsilon_R^2}=-\partial_s v+\calO{\epsilon_\ra^2}\,.
\ee
The longitudinal constraint $\para{\EqBC}$ at lowest order  (that is $
\para{\EqBC}_{\myes}^{(0)} $) gives the HP profile encoded by $v^{(1)}$
\be\label{ConsaxiV1}
v^{(1)}=\mytfrac{3}{2} {\cal H} \partial_s v + \mytfrac{1}{4} \partial_s^2 v+\calO{\epsilon_\ra^2}\,.
\ee

Finally the orthoradial constraint $\EqCtheta$ does not vanish
identically if we have rotation. Instead, if we consider $\EqCtheta_{\myes}^{(1)} $, we get
\be\label{ConsaxiW1}
\Sw^{(1)} =\mytfrac{1}{2}{\cal H} \partial_s \Sw +\calO{\epsilon_\ra^2}\,.
\ee
The lowest order of the incompressibility constraint $\EqCD$ is not needed for the lowest
order dynamics and is only required when considering higher order
corrections as we shall see in~\S~\ref{SecFirstCorrectionsAxi}.

Using \eqref{ConsaxiP0}, \eqref{ConsaxiV1}, and \eqref{ConsaxiW1}
replaced in \eqref{BasicSvSwEqs} we are now able to obtain
\prebool{\beas\label{EqF1DLow}
\partial_t \Sv &=&\mathit{g}
 -  \ST \partial_s {\cal K}
 -  \Sv \partial_s\Sv
\nonumber\\
&&+ 6 \mathcal{H} \partial_s\Sv
 + 3 \partial_s^2\Sv +\calO{\epsilon_\ra^2}\slabel{dtvlowAxi}\\
\partial_t \Sw&=& \Sw \partial_s\Sv
 + 4 \mathcal{H} \partial_s\Sw
 -  \Sv \partial_s\Sw
 + \partial_s^2\Sw +\calO{\epsilon_\ra^2}.\slabel{dtwlowAxi}
\eeas}
{\beas\label{EqF1DLow}
\partial_t \Sv &=&\mathit{g}
 -  \ST \partial_s {\cal K}
 -  \Sv \partial_s\Sv
+ 6 \mathcal{H} \partial_s\Sv
 + 3 \partial_s^2\Sv +\calO{\epsilon_\ra^2}\slabel{dtvlowAxi}\\
\partial_t \Sw&=& \Sw \partial_s\Sv
 + 4 \mathcal{H} \partial_s\Sw
 -  \Sv \partial_s\Sw
 + \partial_s^2\Sw +\calO{\epsilon_\ra^2}.\slabel{dtwlowAxi}
\eeas}
Together with~\eqref{EvolRCircLow}, it forms a closed set of equations. 
Since rotation does not couple to the longitudinal velocity
in~\eqref{dtvlowAxi} it is reasonable to consider that the
dynamical equation~\eqref{dtwlowAxi} should be considered only when
including the first corrections. In fact in order to obtain it we had to
consider $\EqCtheta_{\myes}^{(1)}$ and not $\EqCtheta_{\myes}^{(0)} $
which vanishes identically, so we realize that we have been using a
higher order constraint to be able to close~\eqref{BasicSvSwEqs2}.

We note finally that if surface tension is ignored, the evolution of velocity and rotation does not depend on the
radius. However, as soon as we consider surface tension, the
dependence in $\ra$ appears of course through ${\cal K}$, and we thus need \eqref{EvolRCircLow} to complement the dynamical equations for $\Sv$. We intentionally did not replace in \eqref{EqF1DLow}
the expression of the extrinsic scalar curvature ${\cal K}$ since even though we might use
our perturbative expansion in powers of $\ra$, it proves often useful
to keep its most general expression when considering the axisymmetric geometry. Instead of using ${\cal K}=1/R$, which is the
lowest order expression obtained from~\eqref{calK}, we obtain a much better description if we use instead the exact expression~\citep{Eggers2008}
\be\label{PSTexact}
{\cal K} = \frac{1}{\ra\sqrt{1+(\partial_s\ra)^2}} -\frac{\partial_s^2 \ra}{[1+(\partial_s\ra)^2]^{3/2}}\,,
\ee
which is easily obtained from the general expression of the extrinsic
scalar curvature~\eqref{calK}. In \eqref{dtvlowAxi}, one should thus
use~\eqref{PSTexact}. It amounts to resumming all higher order
contributions from surface tension effects, and this is made possible
thanks to the decoupling property~\eqref{PPST}.

Equation~\eqref{dtvlowAxi} together with~\eqref{PSTexact} for surface
tension induced pressure, and the boundary kinematic
relation~\eqref{EvolRCircLow} constitute the lowest order model for an
axisymmetric jet.
Note that the expression for $\partial_t \Sv$ involves second order derivative with respect to the affine parameter $s$.
The factor $3$ in front of the last term of \eqref{dtvlowAxi} is the famous Trouton
factor~\citep{Trouton} and we review its origin in \S~\ref{SecMonopolePressure}. As for the factor $6{\cal H}$ in the previous
term, it is easily understood from the longitudinal component of
viscous forces per unit area~\eqref{DefVF}
$\para{\VF}_\myes^{(0)}=3 \visc \partial_s \Sv$. Indeed, this implies that
the longitudinal viscous forces integrated on a circular section are
\be\label{paraF}
\para{F} \simeq 3 \mu \pi \ra^2 \partial_s \Sv\,,
\ee
which implies that the lineic density of longitudinal forces is $\mu \partial_s
(3 \pi \ra^2 \partial_s \Sv)$.

In the next section we detail the general method to obtain corrections
up to any order in the small parameter $\epsilon_\ra$ and report the
detailed expressions of the first set of corrections (corrections up
to order $\epsilon_\ra^2$). The second set of corrections (that is up to order $\epsilon_\ra^4$) is reported in Appendix~\ref{AppOrder2}.

\subsection{General method for higher order corrections}\label{SecFirstCorrectionsAxi}

The equations~\eqref{BasicSvSwEqs} are formally unchanged when
considering higher orders because they are exact. However they involve
quantities that we have determined from the side constraints up to
order $\epsilon_\ra^0$ contributions. We thus need to determine these
quantities ($\Sv^{(1)}$, $\Sw^{(1)}$, $p^{(0)}$) with greater
precision, that is also taking into account contributions of order
$\epsilon_\ra^2$. To this end, we need to consider $\EqCP^{(\leq 1)}$,
$\para{\EqBC}^{(\leq 1)}$, and $\EqCtheta^{(\leq 1)}$. After a straightforward computation, we can show that the
longitudinal constraint $\para{\EqBC}$ reads as
\prebool{\bea\label{EqC1}
&&\sum_{n=0}^\infty\ra^{2n}\left[\partial_s \shV^{(n)} + 4 (n+1)
  \Sv^{(n+1)}-{\cal H}^2 \partial_s \shV^{(n-1)} \right.\nonumber\\
&&\qquad\qquad \left.- 4 n {\cal H}^2 \Sv^{(n)}+(8n+6){\cal H}\shV^{(n)}\right]=0\,,
\eea}
{\be\label{EqC1}
\sum_{n=0}^\infty\ra^{2n}\left[\partial_s \shV^{(n)} + 4 (n+1)
  \Sv^{(n+1)}-{\cal H}^2 \partial_s \shV^{(n-1)} - 4 n {\cal H}^2 \Sv^{(n)}+(8n+6){\cal H}\shV^{(n)}\right]=0\,,
\ee}
and the truncated constraint $\para{\EqBC}^{(\leq 1)}$ is found by keeping only $n=0$ and $n=1$ in this sum. However we notice that if we want to deduce $\Sv^{(1)}$ from
$\para{\EqBC}^{(\leq 1)}$, that is, keeping corrections of order
$\epsilon_\ra^{2}$, then we need an expression for $\Sv^{(2)}$ at lowest order. Similarly, if we want corrections up to order $\epsilon_\ra^{6}$, we
need $\Sv^{(3)}$ at lowest order and $\Sv^{(2)}$ up to corrections of
order $\epsilon_\ra^2$ and $\Sv^{(1)}$ up to corrections of order $\epsilon_\ra^4$.

It is straightforward to show that the pressure constraint $\EqCP$ has the general form
\be\label{EqC2}
\sum_n p^{(n)} \ra^{2n} = \sum_n \ra^{2n}\left[\frac{(2n+1)(1+{\cal H}^2 \ra^2 )}{1-{\cal H}^2 \ra^2}\right]\shV^{(n)}
\ee
and if we want to deduce $p^{(0)}$ up to order corrections of order
$\epsilon_\ra^2$ we need the truncation  $\EqCP^{(\leq 1)}$ in which
we need to replace an expression for $p^{(1)}$ at lowest
order. Similarly, if we want $p^{(0)}$ up to corrections
$\epsilon_\ra^4$, then from  $\EqCP^{(\leq 2)}$  we need $p^{(2)}$
at lowest order and $p^{(1)}$ up to corrections $\epsilon_\ra^2$, and
so on.

Finally, the orthoradial constraint $\EqCtheta$ reads as in full generality
\be\label{EqC3}
\sum_{n=1}^\infty \ra^{2n}\left[2n \Sw^{(n)} - {\cal H}\partial_s
\Sw^{(n-1)}\right]=0\,,
\ee
and in particular, from $\EqCtheta^{(\leq 2)}$ we deduce that if we
need $\Sw^{(1)}$ up to corrections of order $\epsilon_\ra^2$, then we need an expression for $\Sw^{(2)}$ at lowest
order. The structure of these recursive dependencies in the three boundary
constraints is summarized in Table~\ref{Tab1}.

\begin{table}
\ifpre
\begin{tabular}{|l|c|c|}
\else
\begin{center}
\begin{tabular}{lcc}
\fi
\hline
Equation & Variable & Dependence\\[0.1cm]
\hline
   $\displaystyle {\EqCP}\,\,$  \eqref{EqC1}& $\displaystyle p^{(0)}$&$\displaystyle\Sv^{(0)}$ and $\displaystyle p^{(1)}\ra^2$,  $\displaystyle p^{(2)}\ra^4\dots$\\[0.1cm]
\hline
   $\,\,\displaystyle\para{\EqBC}\,\,\,$ \eqref{EqC2}& $\displaystyle\Sv^{(1)}$& $\displaystyle\Sv^{(0)}$ and $\displaystyle\Sv^{(2)}\ra^2$,  $\displaystyle\Sv^{(3)}\ra^4\dots$ \\[0.1cm]
\hline
   $\displaystyle{\EqCtheta}\,\,\,$ \eqref{EqC3}&  $\displaystyle\Sw^{(1)}$& $\displaystyle\Sw^{(0)}$ and $\displaystyle\Sw^{(2)}\ra^2$,  $\displaystyle\Sw^{(3)}\ra^4\dots$  \\[0.1cm]
\hline
\end{tabular}
\caption{Structure of dependencies from the pressure, radial, and
  orthoradial boundary constraints.}
\label{Tab1}
\ifpre
\else
\end{center}
\fi
\end{table}

This problem is solved if we now also consider higher moments of the Navier-Stokes
equation~\eqref{SecNS} together with the incompressibility constraint~\eqref{EqCD} so as to find expressions for the missing
moments. The key is to notice that these equations contain a
Laplacian, e.g., $\Delta P$ for the incompressibility constraint or $\Delta
\totV^\mu$ for the Navier-Stokes equation. Since for any
scalar $S$ expanded as~\eqref{ScalarSTF}, the coefficients of the
expansion of $\Delta S$ are
\be\label{CoeffLaplacian}
[\Delta S]^{(n)}_L = 4(n+1)(n+1+\ell)S_L^{(n+1)}+\partial_s^2 S_L^{(n)}\,,
\ee
then in the axisymmetric case we can use this property for $\ell=0$,
and from the incompressibility constraint~\eqref{EqCD} we can express $P^{(n+1)}$ in terms of $\partial_s^2
P^{(n)}$. Similarly from the longitudinal part of the Navier-Stokes
equation ~\eqref{SecNS} we can express $\Sv^{(n+1)}$ in terms of  $\partial_s^2
\Sv^{(n)}$ since it contains $\Delta \para{V}$.

Indeed, the general expansion of the longitudinal part of the Navier-Stokes
equation in powers of $r^{2n}$ (the $\para{\EqD}^{(n)}$) is
\prebool{\bea\label{EqT1}
&&\partial_t \Sv^{(n)} + \sum_m \shV^{(n-m)}\partial_s
\Sv^{(m)}+\sum_m m \Sv^{(m)}\shV^{(n-m)}\\
&&= 4(n+1)^2 \Sv^{(n+1)}+\partial_s^2 \Sv^{(n)}-\partial_s p^{(n)}+\delta_n^0 \left(\para{g}-\partial_s P_\ST\right)\nonumber\,.
\eea}
{\bea\label{EqT1}
\partial_t \Sv^{(n)} + \sum_m \shV^{(n-m)}\partial_s
\Sv^{(m)}+\sum_m m \Sv^{(m)}\shV^{(n-m)}&=& 4(n+1)^2
\Sv^{(n+1)}+\partial_s^2 \Sv^{(n)}-\partial_s p^{(n)}\nonumber\\
&&+\delta_n^0 \left(\para{g}-\partial_s P_\ST\right)\,.
\eea}
For instance, using $\para{\EqD}^{(1)}$, we can obtain $\Sv^{(2)}$ at
lowest order, in function of $\partial_t \Sv^{(1)}$ and also $\partial_s
p^{(1)}$. Let us ignore this latter dependence for the sake of
simplicity. Given that we already know the lowest order expression of
$\Sv^{(1)}$ in terms of $\Sv=\Sv^{(0)}$ from~\eqref{ConsaxiV1}, then
using it we obtain $\Sv^{(2)}$ as a function of $\Sv^{(0)}$. The
time derivatives on $\Sv^{(0)}$ can be further replaced with the lowest
order dynamical equation of $\Sv^{(0)}$~\eqref{dtvlowAxi}. In the end,
we have obtained $\Sv^{(2)}$ in terms of $\Sv^{(0)}$ and its
derivatives with respect to $s$, thus having the structure of a
constraint equation.

As for the rotational part, its expansion in powers of $r^{2n}$, that
is $\omeg{\EqD}^{(n)}$, leads to
\prebool{\bea\label{EqT2}
&&\partial_t \Sw^{(n)} + \sum_m \left(\Sv^{(n-m)}\partial_s \Sw^{(m)}+
(m+1)\Sw^{(m)}\shV^{(n-m)}\right)\nonumber\\
&&=\partial_s^2 \Sw^{(n)}+ 4 (n+1)(n+2)\Sw^{(n+1)}\,.
\eea}
{\be\label{EqT2}
\partial_t \Sw^{(n)} + \sum_m \left(\Sv^{(n-m)}\partial_s \Sw^{(m)}+
(m+1)\Sw^{(m)}\shV^{(n-m)}\right)=\partial_s^2 \Sw^{(n)}+ 4 (n+1)(n+2)\Sw^{(n+1)}\,.
\ee}
Similarly, from $\omeg{\EqD}^{(n)}$ we see that we can obtain $\Sw^{(n+1)}$ as a
function of $\partial_s^2 \Sw^{(n)}$, and the time derivatives are
eventually eliminated in the same manner once all replacements with
lowest order relations are performed. And finally, using the expansion in powers of
$r^{2n}$ of the incompressibility constraint, that is using the
$\EqCD^{(n)}$, we get
\prebool{\bea\label{EqT3}
&&\sum_m\left[\frac{3n+5}{2}+2m(n-m)\right]\shV^{(m)}\shV^{(n-m)}\nonumber\\
&&+\sum_m\left[-2(n+1)\Sw^{(m)}\Sw^{(n-m)}+2m \Sv^{(m)}\partial_s \shV^{(n-m)}\right]\nonumber\\
&&=-\partial_s^2 p^{(n)}-4(n+1)^2 p^{(n+1)}\,.
\eea}
{\bea\label{EqT3}
0&=&\sum_m\left[\frac{3n+5}{2}+2m(n-m)\right]\shV^{(m)}\shV^{(n-m)}\\
&+&\sum_m\left[-2(n+1)\Sw^{(m)}\Sw^{(n-m)}+2m \Sv^{(m)}\partial_s \shV^{(n-m)}\right]+\partial_s^2 p^{(n)}+4(n+1)^2 p^{(n+1)}\,.\nonumber
\eea}
We see that from $\EqCD^{(n)}$ we can obtain $p^{(n+1)}$ in terms of
$\partial_s^2 p^{(n)}$ and this time it is directly in form of a
constraint since it does not involve any time derivatives. The structure of these recursive dependencies deduced from the
Navier-Stokes equation and the incompressibility constraint is summarized in Table~\ref{Tab2}.

\begin{table}
\ifpre
\begin{tabular}{|l|c|c|}
\else
\begin{center}
\begin{tabular}{lcc}
\fi
\hline
Equation & Variable & Dependence\\[0.1cm]
\hline
   $\para{\EqD}^{(n)}\,\,\,$  \eqref{EqT1}& $\Sv^{(n+1)}$& $p^{(n)}$, $\Sv^{(m)}$ $m\leq n$ \\[0.1cm]
\hline
   $\omeg{\EqD}^{(n)}\,\,\,$ \eqref{EqT2}& $\Sw^{(n+1)}$&
   $\Sw^{(m)}$, $\Sv^{(m)}$,   $m\leq n$ \\[0.1cm]
\hline
   ${\EqCD}^{(n)}$ \eqref{EqT3}& $p^{(n+1)}$& $p^{(n)}$,
   $\Sv^{(m)}$, $\Sw^{(m)}$, $m\leq n$ \\[0.1cm]
\hline
\end{tabular}
\caption{Structure of dependencies for the constraints deduced from the
  Navier-Stokes equation and the incompressibility constraint. }
\label{Tab2}
\ifpre
\else
\end{center}
\fi
\end{table}

We understand that the general procedure is very recursive but simple,
and this motivates the use of abstract calculus packages (such as {\it
Mathematica}) to circumvent the complexity of these tedious abstract
computations.
\begin{enumerate}
\item Initially from the lowest order of the constraints $\EqCP$,
  $\para{\EqBC}$ and $\EqCtheta$ (\ref{EqC1},\ref{EqC2},\ref{EqC3}) we
  get $p^{(0)}$, $\Sv^{(1)}$ and $\Sw^{(1)}$ at lowest order in
  function of $\Sv^{(0)}$ and $\Sw^{(0)}$, namely \eqref{ConsaxiP0}, \eqref{ConsaxiV1} and \eqref{ConsaxiW1}
\item Then, if we know the $p^{(q)}$, $\Sv^{(q)}$ and $\Sw^{(q)}$ with
$0\leq q \leq n$ up to order $\epsilon_\ra^{2m}$, then from (\ref{EqT1},\ref{EqT2},\ref{EqT3}) we can deduce $p^{(n+1)}$, $\Sv^{(n+1)}$ and $\Sw^{(n+1)}$ up to order
$\epsilon_\ra^{2m}$ as summarized in Table~\ref{Tab2}. 
\item By using the constraints $\EqCP$, $\para{\EqBC}$ and $\EqCtheta$  (\ref{EqC1},\ref{EqC2},\ref{EqC3}) we see that we can find
  $p^{(0)}$, $\Sv^{(1)}$ and $\Sw^{(1)}$ up to order
  $\epsilon_\ra^{2m}$ if we know the $p^{(q)}$, $\Sv^{(q+1)}$ and $\Sw^{(q+1)}$
up to order $\epsilon_\ra^{2(m-q)}$, and this is summarized in Table~\ref{Tab1}. 
\item From the second point, we see that if we know the $p^{(0)}$,
  $\Sv^{(1)}$ and $\Sw^{(1)}$ up to order $\epsilon_\ra^{2m}$ then we
  also know all $p^{(q)}$,
  $\Sv^{(q+1)}$ and $\Sw^{(q+1)}$  to the same order
  $\epsilon_\ra^{2m}$ and from the third point this is what we need to
  know the $p^{(0)}$, $\Sv^{(1)}$ and $\Sw^{(1)}$ up to order
  $\epsilon_\ra^{2(m+1)}$, which validates this recursive method.
\item Eventually, time derivatives on the fundamental
variables $\Sv$ and $\Sw$ which appear in corrective terms, can be
replaced by using their dynamical equations at a lower order and it is thus possible to obtain the corrections to the
lowest order model up to any order only in terms of derivatives with
respect to $s$.
\end{enumerate}

Finally, the evolution of the radius is easily found up to any
given order in $\epsilon_\ra^{2n}$. Indeed given that
\be\label{GenKinematicR}
\partial_t \ln \ra  =\sum_m \left(
  \mytfrac{1}{2}\shV^{(m)}-{\cal H} \Sv^{(m)}\right) \ra^{2m},
\ee
then in order to obtain the evolution up to  $\epsilon_\ra^{2n}$ corrections, we need the $\shV^{(m)}$ and $\Sv^{(m)}$  (with $m\leq n$) up to
$\epsilon_\ra^{2(n-m)}$ corrections.

\subsection{First corrections}\label{SecStraightFirst}

Implementing the procedure described in the previous section, we first get
the constraints
\prebool{\beas
p^{(1)} &=& \mytfrac{1}{2} \Sw^2
 -  \mytfrac{3}{8} (\partial_s\Sv)^2
 -  \ST\mytfrac{1}{4} \partial_s^2{\cal K}
 \nonumber\\
&&+ \mytfrac{1}{4} \partial_s^3\Sv+\calO{\epsilon_\ra^2}\\
\Sv^{(2)} &=& \mytfrac{9}{16} \mathcal{H} \partial_s\mathcal{H} \partial_s\Sv
 -  \mytfrac{3}{16} \mathcal{H} (\partial_s\Sv)^2
 + \mytfrac{1}{16} \Sw \partial_s\Sw\nonumber \\
&& -  \ST\mytfrac{3}{32} \mathcal{H} \partial_s^2{\cal K}
 + \mytfrac{9}{16} \mathcal{H}^2 \partial_s^2\Sv
 -  \mytfrac{9}{64} \partial_s\Sv \partial_s^2\Sv\nonumber \\
&& -  \ST\mytfrac{1}{32} \partial_s^3{\cal K}
 + \mytfrac{9}{32} \mathcal{H} \partial_s^3\Sv
 + \mytfrac{3}{64} \partial_s^4\Sv+\calO{\epsilon_\ra^2}\slabel{Sv2}\\
\Sw^{(2)} &=& - \mytfrac{1}{32} \Sw \partial_s\mathcal{H} \partial_s\Sv
 + \mytfrac{1}{12} \mathcal{H} \partial_s\mathcal{H} \partial_s\Sw
 -  \mytfrac{1}{48} \partial_s\Sw \partial_s^2\mathcal{H}\nonumber \\
&& -  \mytfrac{1}{96} \mathcal{H} \Sw \partial_s^2\Sv
 + \mytfrac{1}{12} \mathcal{H}^2 \partial_s^2\Sw
 \nonumber \\
&& -  \mytfrac{1}{24} \partial_s\mathcal{H} \partial_s^2\Sw-  \mytfrac{1}{192} \Sw \partial_s^3\Sv+\calO{\epsilon_\ra^2}\,.
\eeas}
{\beas
p^{(1)} &=& \mytfrac{1}{2} \Sw^2
 -  \mytfrac{3}{8} (\partial_s\Sv)^2
 -  \ST\mytfrac{1}{4} \partial_s^2{\cal K}
 + \mytfrac{1}{4} \partial_s^3\Sv+\calO{\epsilon_\ra^2}\\
\Sv^{(2)} &=& \mytfrac{9}{16} \mathcal{H} \partial_s\mathcal{H} \partial_s\Sv
 -  \mytfrac{3}{16} \mathcal{H} (\partial_s\Sv)^2
 + \mytfrac{1}{16} \Sw \partial_s\Sw 
 - \ST\mytfrac{3}{32} \mathcal{H} \partial_s^2{\cal K}
 + \mytfrac{9}{16} \mathcal{H}^2 \partial_s^2\Sv
 -  \mytfrac{9}{64} \partial_s\Sv \partial_s^2\Sv\nonumber \\
&& -  \ST\mytfrac{1}{32} \partial_s^3{\cal K}
 + \mytfrac{9}{32} \mathcal{H} \partial_s^3\Sv
 + \mytfrac{3}{64} \partial_s^4\Sv+\calO{\epsilon_\ra^2}\slabel{Sv2}\\
\Sw^{(2)} &=& - \mytfrac{1}{32} \Sw \partial_s\mathcal{H} \partial_s\Sv
 + \mytfrac{1}{12} \mathcal{H} \partial_s\mathcal{H} \partial_s\Sw
 -  \mytfrac{1}{48} \partial_s\Sw \partial_s^2\mathcal{H}
 -  \mytfrac{1}{96} \mathcal{H} \Sw \partial_s^2\Sv
 + \mytfrac{1}{12} \mathcal{H}^2 \partial_s^2\Sw
 -  \mytfrac{1}{24} \partial_s\mathcal{H} \partial_s^2\Sw\nonumber \\
&& -  \mytfrac{1}{192} \Sw \partial_s^3\Sv+\calO{\epsilon_\ra^2}\,.
\eeas}
We also find the corrections to the constrained quantities which were already
computed at lowest order when deriving the lowest order string model. Indeed, the expressions~\eqref{ConsaxiP0}, \eqref{ConsaxiV1}, and
\eqref{ConsaxiW1} need to be corrected with the terms
\prebool{\beas
p^{(0)} &\supset&\ra^2\left[- \mytfrac{1}{2} \Sw^2
 - 3 \mathcal{H}^2 \partial_s\Sv
 -  \mytfrac{9}{4} \partial_s\mathcal{H} \partial_s\Sv
 + \mytfrac{3}{8} (\partial_s\Sv)^2\right.\nonumber \\
&& \qquad \left.+\ST \mytfrac{1}{4} \partial_s^2{\cal K}
 -  \mytfrac{9}{4} \mathcal{H} \partial_s^2\Sv
 -  \mytfrac{5}{8} \partial_s^3\Sv\right]\\
\Sv^{(1)} &\supset&\ra^2\left[\mytfrac{3}{2} \mathcal{H}^3 \partial_s\Sv
 + \mytfrac{3}{2} \mathcal{H} \partial_s\mathcal{H} \partial_s\Sv
 + \mytfrac{3}{8} \mathcal{H} (\partial_s\Sv)^2\nonumber \right.\\
&&  \qquad-  \mytfrac{1}{8} \Sw \partial_s\Sw
 + \mytfrac{3}{16} \partial_s\Sv \partial_s^2\mathcal{H}
 +\ST \mytfrac{3}{16} \mathcal{H} \partial_s^2{\cal K}\nonumber \\
&&  \qquad+ \mytfrac{3}{2} \mathcal{H}^2 \partial_s^2\Sv
 + \mytfrac{3}{8} \partial_s\mathcal{H} \partial_s^2\Sv
 + \mytfrac{9}{32} \partial_s\Sv \partial_s^2\Sv\nonumber \\
&&  \qquad\left.+\ST \mytfrac{1}{16} \partial_s^3{\cal K}
 + \mytfrac{1}{16} \mathcal{H} \partial_s^3\Sv
 -  \mytfrac{1}{16} \partial_s^4\Sv\right]\slabel{Sv1corrected}\\
\Sw^{(1)} &\supset&\ra^2\left[\mytfrac{1}{16} \Sw \partial_s\mathcal{H} \partial_s\Sv
 + \mytfrac{1}{12} \mathcal{H} \partial_s\mathcal{H} \partial_s\Sw
 + \mytfrac{1}{24} \partial_s\Sw \partial_s^2\mathcal{H}\right.\nonumber \\
&& \qquad + \mytfrac{1}{48} \mathcal{H} \Sw \partial_s^2\Sv
 + \mytfrac{1}{12} \mathcal{H}^2 \partial_s^2\Sw
 + \mytfrac{1}{12} \partial_s\mathcal{H} \partial_s^2\Sw\nonumber \\
&& \qquad \left.+ \mytfrac{1}{96} \Sw \partial_s^3\Sv\right]\,.
\eeas}
{\beas
p^{(0)} &\supset&\ra^2\left[- \mytfrac{1}{2} \Sw^2
 - 3 \mathcal{H}^2 \partial_s\Sv
 -  \mytfrac{9}{4} \partial_s\mathcal{H} \partial_s\Sv
 + \mytfrac{3}{8} (\partial_s\Sv)^2
 +\ST \mytfrac{1}{4} \partial_s^2{\cal K}
 -  \mytfrac{9}{4} \mathcal{H} \partial_s^2\Sv
 -  \mytfrac{5}{8} \partial_s^3\Sv\right]\\
\Sv^{(1)} &\supset&\ra^2\left[\mytfrac{3}{2} \mathcal{H}^3 \partial_s\Sv
 + \mytfrac{3}{2} \mathcal{H} \partial_s\mathcal{H} \partial_s\Sv
 + \mytfrac{3}{8} \mathcal{H} (\partial_s\Sv)^2-  \mytfrac{1}{8} \Sw \partial_s\Sw
 + \mytfrac{3}{16} \partial_s\Sv \partial_s^2\mathcal{H}
 +\ST \mytfrac{3}{16} \mathcal{H} \partial_s^2{\cal K}\right.\nonumber \\
&&  \qquad+ \mytfrac{3}{2} \mathcal{H}^2 \partial_s^2\Sv
 + \mytfrac{3}{8} \partial_s\mathcal{H} \partial_s^2\Sv
 + \mytfrac{9}{32} \partial_s\Sv \partial_s^2\Sv\left.+\ST \mytfrac{1}{16} \partial_s^3{\cal K}
 + \mytfrac{1}{16} \mathcal{H} \partial_s^3\Sv
 -  \mytfrac{1}{16} \partial_s^4\Sv\right]\slabel{Sv1corrected}\\
\Sw^{(1)} &\supset&\ra^2\left[\mytfrac{1}{16} \Sw \partial_s\mathcal{H} \partial_s\Sv
 + \mytfrac{1}{12} \mathcal{H} \partial_s\mathcal{H} \partial_s\Sw
 + \mytfrac{1}{24} \partial_s\Sw \partial_s^2\mathcal{H} +
 \mytfrac{1}{48} \mathcal{H} \Sw \partial_s^2\Sv\right.\nonumber\\
&&\left.\qquad
 + \mytfrac{1}{12} \mathcal{H}^2 \partial_s^2\Sw
 + \mytfrac{1}{12} \partial_s\mathcal{H} \partial_s^2\Sw+ \mytfrac{1}{96} \Sw \partial_s^3\Sv\right]\,.
\eeas}
Once replaced in \eqref{BasicSvSwEqs}, we finally get the
corrections to \eqref{EqF1DLow} which read as
\prebool{\beas \label{dtAllCorrectionsAxi}
\partial_t \Sv &\supset&\ra^2\left[\mathcal{H} \Sw^2
 + 12 \mathcal{H}^3 \partial_s\Sv
 + \mytfrac{33}{2} \mathcal{H} \partial_s\mathcal{H} \partial_s\Sv\right.\nonumber \\
&&  \qquad+ \mytfrac{3}{4} \mathcal{H} (\partial_s\Sv)^2
 + \mytfrac{1}{2} \Sw \partial_s\Sw
 + 3 \partial_s\Sv \partial_s^2\mathcal{H}\nonumber \\
&& \qquad +\ST \mytfrac{1}{4} \mathcal{H} \partial_s^2{\cal K}
 + \mytfrac{27}{2} \mathcal{H}^2 \partial_s^2\Sv
 + 6 \partial_s\mathcal{H} \partial_s^2\Sv\nonumber \\
&& \qquad \left.+ \mytfrac{3}{8} \partial_s\Sv \partial_s^2\Sv
 + \mytfrac{15}{4} \mathcal{H} \partial_s^3\Sv
 + \mytfrac{3}{8} \partial_s^4\Sv\right]\slabel{dtvCorrectionsAxi}\\
\partial_t \Sw &\supset&\ra^2\left[\mytfrac{1}{2} \Sw \partial_s\mathcal{H} \partial_s\Sv
 + \mytfrac{2}{3} \mathcal{H} \partial_s\mathcal{H} \partial_s\Sw
 + \mytfrac{1}{3} \partial_s\Sw \partial_s^2\mathcal{H}\right.\nonumber \\
&&  \qquad+ \mytfrac{1}{6} \mathcal{H} \Sw \partial_s^2\Sv
 + \mytfrac{2}{3} \mathcal{H}^2 \partial_s^2\Sw
 + \mytfrac{2}{3} \partial_s\mathcal{H} \partial_s^2\Sw\nonumber \\
&&  \qquad\left.+ \mytfrac{1}{12} \Sw \partial_s^3\Sv\right]\slabel{dtwCorrectionsAxi}\\
\partial_t \ln \ra  &\supset&\ra^2\left(- \mytfrac{3}{2} \mathcal{H}^2 \partial_s\Sv
 -
 \mytfrac{3}{8} \partial_s\mathcal{H} \partial_s\Sv\right.\nonumber\\
&&\qquad \left.
 -  \mytfrac{5}{8} \mathcal{H} \partial_s^2\Sv
 -  \mytfrac{1}{16} \partial_s^3\Sv\right)\,. \slabel{dtRCorrectionsAxi}
\eeas}
{\beas \label{dtAllCorrectionsAxi}
\partial_t \Sv &\supset&\ra^2\left[\mathcal{H} \Sw^2
 + 12 \mathcal{H}^3 \partial_s\Sv
 + \mytfrac{33}{2} \mathcal{H} \partial_s\mathcal{H} \partial_s\Sv\right.+ \mytfrac{3}{4} \mathcal{H} (\partial_s\Sv)^2
 + \mytfrac{1}{2} \Sw \partial_s\Sw
 + 3 \partial_s\Sv \partial_s^2\mathcal{H} \nonumber \\
&& \,\, \left. +\ST \mytfrac{1}{4} \mathcal{H} \partial_s^2{\cal K}
 + \mytfrac{27}{2} \mathcal{H}^2 \partial_s^2\Sv
 + 6 \partial_s\mathcal{H} \partial_s^2\Sv+ \mytfrac{3}{8} \partial_s\Sv \partial_s^2\Sv
 + \mytfrac{15}{4} \mathcal{H} \partial_s^3\Sv
 + \mytfrac{3}{8} \partial_s^4\Sv\right]\slabel{dtvCorrectionsAxi}\\
\partial_t \Sw &\supset&\ra^2\left[\mytfrac{1}{2} \Sw \partial_s\mathcal{H} \partial_s\Sv
 + \mytfrac{2}{3} \mathcal{H} \partial_s\mathcal{H} \partial_s\Sw
 + \mytfrac{1}{3} \partial_s\Sw \partial_s^2\mathcal{H}
+ \mytfrac{1}{6} \mathcal{H} \Sw \partial_s^2\Sv\right.\nonumber\\
&&\left.\qquad
 + \mytfrac{2}{3} \mathcal{H}^2 \partial_s^2\Sw
 + \mytfrac{2}{3} \partial_s\mathcal{H} \partial_s^2\Sw
+ \mytfrac{1}{12} \Sw \partial_s^3\Sv\right]\slabel{dtwCorrectionsAxi}\\
\partial_t \ln \ra  &\supset&\ra^2\left(- \mytfrac{3}{2} \mathcal{H}^2 \partial_s\Sv
 -  \mytfrac{3}{8} \partial_s\mathcal{H} \partial_s\Sv
 -  \mytfrac{5}{8} \mathcal{H} \partial_s^2\Sv
 -  \mytfrac{1}{16} \partial_s^3\Sv\right)\,. \slabel{dtRCorrectionsAxi}
\eeas}

We report in Appendix~\ref{AppOrder2} the next order corrections
which are of order $\epsilon_\ra^4$. Note also that surface tension effects do not
enter explicitly the dynamical equation for axial rotation
\eqref{dtwCorrectionsAxi} nor the dynamical equation for the radius
\eqref{dtRCorrectionsAxi}, but they still matter due to the couplings between $\Sv$, $\Sw$ and $\ra$.

In general, if we keep terms of order $\epsilon_\ra^{2n}$ in the
expression giving $\partial_t\Sv$, that is, if we consider the $n$-th correction,
then it involves terms which have $2(n+1)$ order
derivatives in the affine parameter $s$, typically from terms of the
form $\ra^{2n}\partial_s^{2(n+1)} \Sv$, and the differential complexity
is increased. If we consider for instance a
steady regime in which all time derivatives vanish, this can lead to a
rather stiff differential system as the coefficients in front of the
highest derivatives are typically the smallest.

\subsection{Comparison with the Cosserat model}\label{SecCosserat}

As shown by~\citet{GarciaCastellanos} the Cosserat model can be viewed as an
averaged model. Indeed, given that the longitudinal velocity is given
by $\Sv^{(0)}+r^2\Sv^{(1)}+r^4\Sv^{(2)}+\dots$ it could be natural to
consider an averaged velocity and derive the dynamical equation for this variable
and not for $\Sv^{(0)}$ which should be considered as a derived
variable. The average longitudinal velocity is simply defined as
\prebool{\bea
\aver{\Sv} \equiv \frac{2}{\ra^2}\int_0^\ra \para{V} r \dd r &=&
\frac{2}{\ra^2}\int_0^\ra \sum_n \Sv^{(n)} r^{2n} r \dd r \,\nonumber\\
&=&\sum_n \frac{\ra^{2n}}{n+1} \Sv^{(n)} \label{averagev}\,\,.
\eea}
{\be
\aver{\Sv} \equiv \frac{2}{\ra^2}\int_0^\ra \para{V} r \dd r =
\frac{2}{\ra^2}\int_0^\ra \sum_n \Sv^{(n)} r^{2n} r \dd r =\sum_n \frac{\ra^{2n}}{n+1} \Sv^{(n)} \label{averagev}\,\,.
\ee}
We obtain that from~\eqref{GenKinematicR}
and~\eqref{IncompAxial} the kinematic equation for the time evolution
of the fiber radius reads exactly~\citep{GarciaCastellanos}
\be
\partial_t \ln \ra  = -{\cal H}\aver{\Sv} -\mytfrac{1}{2}\partial_s \aver{\Sv}\,.
\ee
Obviously the expression~\eqref{averagev} needs to be truncated at a
given power of $\ra$ which is the order at which the equations are considered. Once truncated, we can invert it because from our
algorithm, the $\Sv^{(n)}$ are expressed in terms of (the derivatives
of) $\Sv^{(0)}$ and $\Sw^{(0)}$. After inverting, we obtain
$\Sv=\Sv^{(0)}$ as a power series in $\para{\Sv}$ and its derivatives. For instance up to the first corrections we get
\beas
\aver{\Sv} &=&
\Sv+\frac{\ra^2}{2}\left(\mytfrac{3}{2}{\cal H}\partial_s
  \Sv+\mytfrac{1}{4}\partial_s^2 \Sv\right)+\calO{\epsilon_\ra^4}\,\\
 \Sv &=& \aver{\Sv}-\frac{\ra^2}{2}\left(\mytfrac{3}{2}{\cal H}\partial_s
  \aver{\Sv}+\mytfrac{1}{4}\partial_s^2 \aver{\Sv}\right)+\calO{\epsilon_\ra^4}\,.
\eeas
At lowest order, the form of \eqref{dtvlowAxi} is the same if it
is expressed with $\aver{\Sv}$ or with $\Sv$ because both
velocities are equal at lowest order. Differences appear only when
including the first set of corrections. Eventually, we find
\prebool{\bea
\partial_t \aver{\Sv}
&+&\ra^2\left(-\frac{1}{2}\mathcal{H}\partial_{t} \partial_s \aver{\Sv}-\frac{1}{8}\partial_{t} \partial_s^2 \aver{\Sv}\right)\\
&=& \mathit{g}
 -  \partial_s{P_{\ST}^{(0)}}
 + 6 \mathcal{H} \partial_s\aver{\Sv}
 -  \aver{\Sv} \partial_s\aver{\Sv}
 + 3 \partial_s^2\aver{\Sv}\nonumber\\
&&+\ra^2\left(\mathcal{H} \Sw^2
 - 6 \mathcal{H}^3  \partial_s\aver{\Sv}
 -  \mytfrac{1}{4} \mathcal{H}  (\partial_s\aver{\Sv})^2\right.\nonumber \\
&&\,\,\qquad + \mytfrac{1}{2}  \Sw \partial_s\Sw
 + \mytfrac{3}{4} \partial_s\aver{\Sv} \partial_s^2\mathcal{H}
 -  \mytfrac{3}{2} \mathcal{H}^2 \partial_s^2\aver{\Sv}\nonumber \\
&& \,\,\qquad\left.+ \mytfrac{1}{2} \mathcal{H} \aver{\Sv} \partial_s^2\aver{\Sv}
 + \mytfrac{3}{4} \partial_s\mathcal{H} \partial_s^2\aver{\Sv}
 + \mytfrac{1}{8}  \aver{\Sv} \partial_s^3\aver{\Sv}\right)\nonumber
\eea}
{\bea
&&\partial_t \aver{\Sv}
+\ra^2\left(-\frac{1}{2}\mathcal{H}\partial_{t} \partial_s
  \aver{\Sv}-\frac{1}{8}\partial_{t} \partial_s^2
  \aver{\Sv}\right)= \mathit{g}
 -  \partial_s{P_{\ST}^{(0)}}
 + 6 \mathcal{H} \partial_s\aver{\Sv}
 -  \aver{\Sv} \partial_s\aver{\Sv}
 + 3 \partial_s^2\aver{\Sv}\\
&&+\ra^2\left(\mathcal{H} \Sw^2
 - 6 \mathcal{H}^3  \partial_s\aver{\Sv}
 -  \mytfrac{1}{4} \mathcal{H}  (\partial_s\aver{\Sv})^2
 + \mytfrac{1}{2}  \Sw \partial_s\Sw
 + \mytfrac{3}{4} \partial_s\aver{\Sv} \partial_s^2\mathcal{H}
 -  \mytfrac{3}{2}
 \mathcal{H}^2 \partial_s^2\aver{\Sv}\right.\nonumber\\
&&\left.\qquad
 + \mytfrac{1}{2} \mathcal{H} \aver{\Sv} \partial_s^2\aver{\Sv}
 + \mytfrac{3}{4} \partial_s\mathcal{H} \partial_s^2\aver{\Sv}
 + \mytfrac{1}{8}  \aver{\Sv} \partial_s^3\aver{\Sv}\right)\nonumber
\eea}
which we have written in a form which matches Eq. (55) of
\citet{GarciaCastellanos}. However it does not match the Cosserat equation Eq. (53) of \citet{GarciaCastellanos} [which is also  Eq. (74) of
\citet{EggersRMP} derived with a Galerkin approximation method]. As explained in details by
\citet{GarciaCastellanos}, this is because some terms involving $\Sv^{(1)}$
are removed. In the case of  the Cosserat model of \citet{EggersRMP},
these terms would probably be recovered when including higher orders
of the Galerkin method, given that at lowest order in the
Galerkin approximation the information that the next corrections are
parabolic in nature has not been put in. In a sense, the lowest order
of the Galerkin approximation also amounts to ignoring some parabolic terms of
the type $\Sv^{(1)}r^2$. In our method, the contributions $\Sv^{(1)}$ and $\Sv^{(2)}$ are taken into account from the
constraints~\eqref{ConsaxiV1} [corrected by~\eqref{Sv1corrected} and
\eqref{Sv2}].

Furthermore, our results are extended to include a possible axial
rotation from the inclusion of the $\Sw^{(n)}$. We find that when including the
first corrections to the viscous string approximations, the corrected
dynamics of $\Sv$ couples with the lowest order axial rotation
$\Sw=\Sw^{(0)}$ whose evolution needs then to be computed from the lowest
order dynamical equation~\eqref{dtwlowAxi}. This interplay between
longitudinal velocity and axial rotation which takes place
when the first corrections are included has already been
described in \citet{BBCF}, but our formalism allows already for more compact
and geometrically more meaningful expressions. However, it is only
when considering curved fibers that our formalism based on STF
multipoles appears to be powerful as we shall now see in details in the following section.

\section{Application to curved fibers}\label{SecCurvedJets}

\subsection{Overview of curved fiber specificities}

Whenever we consider curved fibers, we must take into account the
property that the FCL curvature $\kappa^a$ does not vanish anymore, nor does the rotation rate
$\omega^\tetr{i}$ of the orthonormal basis. 
As stressed in~\S~\ref{SecShapeRestrict},  even if we restrict to a
stationary regime for which all time derivatives vanish and thus $\omega^a=0$, we can still
generically form STF products of the type 
\be
\kappa_{\langle a_1}\dots \kappa_{a_\ell \rangle}
\ee 
which might source the STF moments of order $\ell$. For instance, terms proportional to $\kappa_{\langle a}
\kappa_{b \rangle}$ would source the shear part of the velocity field
$\stf{V}_{ab}$ from the Navier-Stokes equation, and this would in turn induce a deformation of the fiber
shape of the type $\rb_{ab}$ from the kinetic condition on the fiber
side~\eqref{EqPropaR}. Since the lowest order description corresponds
to a string approximation for which the section size and shape are
irrelevant, these terms are expected to arise only when including the
first corrections. For instance, the combinations  
\be\label{kaka}
\kappa_{\langle a} \kappa_{b \rangle} y^a y^b\,, \myquad{\rm
  or}\myquad  \kappa_{\langle a} \kappa_{b \rangle} \ra^2
\ee
are of order $\epsilon_\ra^2$. For the lowest order description, one might consider circular sections, but as
soon as we consider refinements to this description, we must abandon
the circular shape assumption.  As we shall find in the remainder of
this section, the fundamental dynamical variables are the same as for
the axisymmetric case ($\Sv=\Sv_\myes^{(0)}$, $\Sw=\Sw_\myes^{(0)}$,
and $\ra$) on which we add the various shape multipoles $\rb_L$ and
also the position and velocity of the FCL. If we
consider a description up to order $n$ in powers of $\epsilon_\ra$,
then we shall find that we must include at least the multipoles $\rb_L$ with $\ell \leq
n$. As soon as we leave the realm of straight fibers,
we open Pandora's box and we cannot obtain all STF devils from
constraints, as the shape multipoles become dynamical.\\

This departure from circular fiber sections is even more obvious when
considering the motion in a steady rotating frame. Indeed, the
inertial forces will bring typical contributions of the form
$\Omega_{\langle a}\Omega_{b \rangle}$, which are similar to tidal forces and induce an elliptic elongation. Such contribution arises naturally in the
Navier-Stokes equation~\eqref{SecNS} as can be seen from the general
expression of the inertial forces~\eqref{Imu}, and they typically
source the velocity shear moments $\stf{V}^{(n)}_{ab}$, which in turns
source the elliptic deformation $\rb_{ab}$ from the boundary kinematics~\eqref{EqPropaR}.

We first derive the lowest order viscous string model in the curved
case in~\S~\ref{SecCurvedString} and discuss corrections in~\S~\ref{SecCurvedExtended}.
In~\S~\ref{SecElliptic} we also consider the special case of straight but
non axisymmetric fibers with mild elliptic shapes.

\subsection{Viscous string model}\label{SecCurvedString}

\subsubsection{Incompressibility conditions}

The incompressibility conditions on moments \eqref{EqCLn} are used to
replace the moments $\shV_L^{(n)}$. In particular, from
$[\EqC]^{(0)}_\myes$, $[\EqC]^{(1)}_\myes$, $[\EqC]^{(0)}_a$,
$[\EqC]^{(1)}_a$, and $[\EqC]^{(0)}_{ab}$ we obtain the general
conditions
\prebool{\beas\label{EqsIncompressibilityCurved}
\shV^{(1)}_\myes &=&- {{\stf{V}}^{(1)a}} \widetilde{\kappa}_{a} -  \mytfrac{1}{2} \widetilde{\kappa}^{a} {{\mysh{V}}^{(0)}_{a}} -  \mytfrac{1}{2} \partial_s\Sv^{(1)}_\myes,\\ 
\shV^{(0)}_a&=&- \mytfrac{4}{3} {{\stf{V}}^{(1)}_{a}} -  \mytfrac{2}{3} {{\stf{V}}^{(0)}_{ab}} \widetilde{\kappa}^{b} -  \widetilde{\kappa}_{a} {\shV}_\myes^{(0)} -  \mytfrac{2}{3} \kappa_{a} {\Sw} \nonumber \\ 
&& -  \mytfrac{2}{3} \partial_s{{{\Sv}^{(0)}_{a}}}\slabel{EqsIncompressibility3},\\
\shV^{(1)}_a&=&- \mytfrac{8}{5} {{\stf{V}}^{(2)}_{a}} -  \mytfrac{4}{5} {{\stf{V}}^{(1)}_{ab}} \widetilde{\kappa}^{b} -  \widetilde{\kappa}_{a} \shV^{(1)}_\myes -  \mytfrac{1}{2} \widetilde{\kappa}^{b} {\mysh{V}}^{(0)}_{ab} \nonumber \\ 
&& -  \mytfrac{2}{5} \kappa_{a} {\Sw^{(1)}} -  \mytfrac{2}{5} \partial_s{{{\Sv}^{(1)}_{a}}},\\
\shV^{(0)}_{ab}&=&- {{\stf{V}}^{(1)}_{ab}}-{{\stf{V}}^{(1)}_{\langle
    a}} \widetilde{\kappa}_{b \rangle}- \widetilde{\kappa}_{\langle a}
{{\mysh{V}}^{(0)}_{b \rangle}} -
\mytfrac{1}{2} \partial_s{{\Sv}^{(0)}_{ab}}\nonumber\\
&&-\mytfrac{1}{2} \stf{V}_{abc}^{(0)}\tkappa^c\,,
\eeas}
{\beas\label{EqsIncompressibilityCurved}
\shV^{(1)}_\myes &=&- {{\stf{V}}^{(1)a}} \widetilde{\kappa}_{a} -  \mytfrac{1}{2} \widetilde{\kappa}^{a} {{\mysh{V}}^{(0)}_{a}} -  \mytfrac{1}{2} \partial_s\Sv^{(1)}_\myes\\ 
\shV^{(0)}_a&=&- \mytfrac{4}{3} {{\stf{V}}^{(1)}_{a}} -  \mytfrac{2}{3} {{\stf{V}}^{(0)}_{ab}} \widetilde{\kappa}^{b} -  \widetilde{\kappa}_{a} {\shV}_\myes^{(0)} -  \mytfrac{2}{3} \kappa_{a} {\Sw} -  \mytfrac{2}{3} \partial_s{{{\Sv}^{(0)}_{a}}}\slabel{EqsIncompressibility3}\\
\shV^{(1)}_a&=&- \mytfrac{8}{5} {{\stf{V}}^{(2)}_{a}} -  \mytfrac{4}{5} {{\stf{V}}^{(1)}_{ab}} \widetilde{\kappa}^{b} -  \widetilde{\kappa}_{a} \shV^{(1)}_\myes -  \mytfrac{1}{2} \widetilde{\kappa}^{b} {\mysh{V}}^{(0)}_{ab} -  \mytfrac{2}{5} \kappa_{a} {\Sw^{(1)}} -  \mytfrac{2}{5} \partial_s{{{\Sv}^{(1)}_{a}}}\\
\shV^{(0)}_{ab}&=&- {{\stf{V}}^{(1)}_{ab}}-{{\stf{V}}^{(1)}_{\langle
    a}} \widetilde{\kappa}_{b \rangle}- \widetilde{\kappa}_{\langle a}
{{\mysh{V}}^{(0)}_{b \rangle}} -
\mytfrac{1}{2} \partial_s{{\Sv}^{(0)}_{ab}}-\mytfrac{1}{2} \stf{V}_{abc}^{(0)}\tkappa^c\,,
\eeas}
that are used extensively together with \eqref{uprime}  throughout \S~\ref{SecCurvedJets}.

\subsubsection{Normal vector and curvature}

At lowest order, the unit normal vector components are simply
\beas
\widehat{\Nv}^\tetr{3}&=&-{\cal H} \ra+\calO{\epsilon_\ra^2}\\
\widehat{\Nv}^a&=&\yu^a +\calO{\epsilon_\ra^2}
\eeas
and from \eqref{calK} the scalar extrinsic curvature reads as
\be
{\cal K} = \frac{1}{\ra}+\tkappa_a \yu^a +\calO{\epsilon_\ra}\,.
\ee
If we expand $\ra {\cal H}$ as in \eqref{ScalarSTFSide2}, then ${[\ra
  {\cal K}]}_\myes^{(0)}=1$ and ${[\ra  {\cal
    K}]}_a^{(0)}=\tkappa_a$. Note that we cannot separate the pressure contribution from viscous
and surface tension effects as in~\eqref{PPST}. The scalar extrinsic curvature must be used directly inside the boundary constraint~\eqref{VectorBoundaryConstraint}.

\subsubsection{Navier-Stokes equation}

Using the components expression for the velocity
gradient~\eqref{Sabincomponents}, the velocity shear
\eqref{Sigincomponents}, and the velocity derivatives \eqref{dtVcomp}, the lowest multipoles of the longitudinal and
sectional components of acceleration and of the volumic forces are [reminding that we are using the notation~\eqref{SimpleNotation}
throughout]
\prebool{\beas\label{ComponentsNSpara}
\para{A}_\myes^{(0)} &=&\left(\partial_t + {\Sv} \partial_s\right){\Sv}+ \partial_t \bar{U}+ U^{a} \widetilde{\omega}_{a}\nonumber\\
&&+\overline{I} + 2 U^{a} \widetilde{\Omega}_{a} ,\\
\para{f}_\myes^{(0)} &=&\para{g} + 4 {\Sv^{(1)}} + {{\Sv}^{(0)}_a}
\widetilde{\kappa}^{a} -  \kappa^{a} {\omega}_{a}
-  \partial_s{{P^{(0)}_\myes}} \\
&&-\Sv \kappa_a \kappa^a+2\tkappa^a \partial_s
\stf{V}_a^{(0)}+\stf{V}_a^{(0)}\partial_s \tkappa^a+ \partial^2_s{{\Sv}},\nonumber
\eeas
\beas\label{ComponentsNSsectional}
\stf{A}_a^{(0)} &=&- {\Sv}^2 \widetilde{\kappa}_{a} +2 \widetilde{U}_{a} \para{\Omega} + I_{a} - 2 \bar{U} \widetilde{\Omega}_{a} - 2 {\Sv} \widetilde{\Omega}_{a} \nonumber \\ 
&& -  \bar{U} \widetilde{\omega}_{a} - 2 {\Sv} \widetilde{\omega}_{a}
+ \partial_t U_{a}+\widetilde{U}_a \para{\omega},\\
\stf{f}_a^{(0)} &=&g_{a} -  {P^{(0)}_a} + \mytfrac{8}{3} {{\stf{V}}^{(1)}_a} -  \mytfrac{5}{3} \kappa_{a} {\Sw} \nonumber \\ 
&& -  \mytfrac{3}{2} \widetilde{\kappa}_{a} \partial_s{{\Sv}} -  \mytfrac{2}{3} \partial_s{{{\Sv}^{(0)}_a}}-  {\Sv} \partial_s{\widetilde{\kappa}_{a}}
-  \partial_s{\widetilde{\omega}_{a}},
\eeas
\beas\label{ComponentsNSomeg}
\omeg{A}^{(0)} &=&- {{\Sv}^{(0)}_a} \left({\Omega}^{a} +{\omega}^{a} +\kappa^{a} \Sv\right) -  \left({\Sw} +\para{\Omega} \right)\partial_s{{\Sv}} \nonumber \\ 
&&+ \left(\partial_t+{\Sv} \partial_s\right){{\Sw}} ,\\
\omeg{f}^{(0)} &=&- \mytfrac{4}{3} {{\stf{V}}^{(1)}_a} \kappa^{a}-
\mytfrac{7}{6} \kappa_{a} \kappa^{a} {\Sw} + 8 {\Sw^{(1)}}  \nonumber\\
&&-  \mytfrac{7}{6} \kappa^{a} \partial_s{{{\Sv}^{(0)}_{a}}} -  \mytfrac{1}{2} {{\Sv}^{(0)a}} \partial_s{\kappa_{a}} + \mytfrac{1}{2} {\Sv} \widetilde{\kappa}_{a} \partial_s{\kappa^{a}} 
 \nonumber\\
&&+ \mytfrac{1}{2} {\omega}^{a} \partial_s{\widetilde{\kappa}_{a}} + \widetilde{\kappa}_{a} \partial_s{{\omega}^{a}} + \partial^2_s{{\Sw}}\,.
\eeas}
{\beas\label{ComponentsNSpara}
\para{A}_\myes^{(0)} &=&\left(\partial_t + {\Sv} \partial_s\right){\Sv}+ \partial_t \bar{U}+ U^{a} \widetilde{\omega}_{a}+\overline{I} + 2 U^{a} \widetilde{\Omega}_{a} \\
\para{f}_\myes^{(0)} &=&\para{g} + 4 {\Sv^{(1)}} + {{\Sv}^{(0)}_a}
\widetilde{\kappa}^{a} -  \kappa^{a} {\omega}_{a}
-  \partial_s{{P^{(0)}_\myes}} -\Sv \kappa_a \kappa^a+2\tkappa^a \partial_s
\stf{V}_a^{(0)}\nonumber\\
&&+\stf{V}_a^{(0)}\partial_s \tkappa^a+ \partial^2_s{{\Sv}}
\eeas
\beas\label{ComponentsNSsectional}
\stf{A}_a^{(0)} &=&- {\Sv}^2 \widetilde{\kappa}_{a} +2 \widetilde{U}_{a} \para{\Omega} + I_{a} - 2 \bar{U} \widetilde{\Omega}_{a} - 2 {\Sv} \widetilde{\Omega}_{a} -  \bar{U} \widetilde{\omega}_{a} - 2 {\Sv} \widetilde{\omega}_{a}
+ \partial_t U_{a}+\widetilde{U}_a \para{\omega}\\
\stf{f}_a^{(0)} &=&g_{a} -  {P^{(0)}_a} + \mytfrac{8}{3} {{\stf{V}}^{(1)}_a} -  \mytfrac{5}{3} \kappa_{a} {\Sw} -  \mytfrac{3}{2} \widetilde{\kappa}_{a} \partial_s{{\Sv}} -  \mytfrac{2}{3} \partial_s{{{\Sv}^{(0)}_a}}-  {\Sv} \partial_s{\widetilde{\kappa}_{a}}
-  \partial_s{\widetilde{\omega}_{a}}
\eeas
\beas\label{ComponentsNSomeg}
\omeg{A}^{(0)} &=&- {{\Sv}^{(0)}_a} \left({\Omega}^{a} +{\omega}^{a} +\kappa^{a} \Sv\right) -  \left({\Sw} +\para{\Omega} \right)\partial_s{{\Sv}} + \left(\partial_t+{\Sv} \partial_s\right){{\Sw}} \\
\omeg{f}^{(0)} &=&- \mytfrac{4}{3} {{\stf{V}}^{(1)}_a} \kappa^{a}-
\mytfrac{7}{6} \kappa_{a} \kappa^{a} {\Sw} + 8 {\Sw^{(1)}}  -
\mytfrac{7}{6} \kappa^{a} \partial_s{{{\Sv}^{(0)}_{a}}} -
\mytfrac{1}{2} {{\Sv}^{(0)a}} \partial_s{\kappa_{a}} + \mytfrac{1}{2}
{\Sv} \widetilde{\kappa}_{a} \partial_s{\kappa^{a}} \nonumber\\
&&+ \mytfrac{1}{2} {\omega}^{a} \partial_s{\widetilde{\kappa}_{a}} + \widetilde{\kappa}_{a} \partial_s{{\omega}^{a}} + \partial^2_s{{\Sw}}\,.
\eeas}
In these expressions, we have defined the inertial force on the FCL by
\be
\gr{I} \equiv \gr{\Omega}\times \left(\gr{\Omega}\times\gr{R}\right)\,. 
\ee 
$\para{I}$ and $I^a$ are as usual its longitudinal and sectional
components. Similarly, we recall that $\Omega^a$ and $g^a$ are the sectional components of the steady frame
rotation (if any) and of long range forces, whereas $\para{\Omega}$
and $\para{g}$ are their longitudinal projections. Since $\gr{\Omega}$
and $\gr{g}$ are constant vectors, then from \eqref{Propkappa} their components vary along the FCL
according to
\be
\partial_s \para{g}= -\tkappa_a g^a\,,\qquad \partial_s g^a = \kappa^a \para{g}\,,
\ee
with similar expressions for the components of $\gr{\Omega}$. However
$\gr{I}$ is not constant and we must use
\be
\partial_s \para{I} = -\Omega^2-\tkappa_a I^a\,,\qquad \partial_s I^a
= \para{\Omega} \Omega^a+\tkappa_a \para{I}\,.
\ee

In reality, the expressions~\eqref{ComponentsNSpara}, \eqref{ComponentsNSsectional} and \eqref{ComponentsNSomeg} are formally more complex,
as they also involve terms which contain $\stf{V}^{(0)}_a$ and
$\stf{V}^{(0)}_{ab}$. The former vanishes at lowest order from the
gauge condition~\eqref{GaugeConstraint}, and when studying the
structure of corrections in~\S~\ref{SecCurvedExtended}, we will show that from the
orthoradial boundary condition the latter vanishes as well at lowest
order. The missing terms are gathered in~\S~\ref{SecFundDynCorrection}. 

From \eqref{ComponentsNSpara}  we can infer the
lowest moment of the longitudinal part of Navier-Stokes equation, that
is $\para{\EqD}_\myes^{(0)}$, which could also be obtained from the
longitudinal projection of a momentum balance equation. From
\eqref{ComponentsNSsectional} we can infer the lowest order
multipoles of the sectional part of the Navier-Stokes equations
$\EqD_a^{(0)}$. It could also be obtained from the sectional
projection of the momentum balance equation. Finally,
from~\eqref{ComponentsNSomeg}, we infer $\omeg{\EqD}^{(0)}$ which governs the dynamics of the axial
rotation rate $\Sw=\Sw^{(0)}$. In the axisymmetric case, it was not strictly speaking part of the lowest order string description since
axial rotation decoupled at lowest order. For curved fibers, it seems
at first sight when examining \eqref{ComponentsNSsectional} that axial rotation retroacts on the dynamical equations for
$U_a$. However, when including the boundary constraints it will appear
that this is not the case. As in the axisymmetric case, the
dynamical equation for the axial rotation is in fact already a first
correction and not part of the viscous string model.

In order to obtain closed dynamical equations from
\eqref{ComponentsNSpara}, \eqref{ComponentsNSsectional}, and
\eqref{ComponentsNSomeg}, we need to find the lowest order expressions for $\Sv_a^{(0)}$,
$P^{(0)}=P_\myes^{(0)}$, $\Sv^{(1)}=\Sv_\myes^{(1)}$, $\stf{V}^{(1)}_a$, and $\Sw^{(1)}$, which as in the axisymmetric case
are going to be obtained from the lowest order of the boundary condition~\eqref{VectorBoundaryConstraint}.

\subsubsection{Constraint equations}\label{SecCons1}

We recall the notation introduced in~\S~\ref{SecBoundary} for the various
constraints obtained on the boundary and their multipoles
expansions. 

\begin{itemize}
\item First, from the lowest order monopole and dipole of the longitudinal constraint, that is $\para{\EqBC}^{(0)}_\myes$ and $\para{\EqBC}^{(0)}_a$ we get
\beas\label{RadialConstraintLowCurved}
\Sv^{(1)}&=& \mytfrac{3}{2} {\cal H} \partial_s{{\Sv}} + \mytfrac{1}{4}\partial^2_s{{\Sv}} +\calO{\epsilon_\ra^2} \slabel{RadialConstraintLowCurved1}\\
\Sv^{(0)}_a&=&{\Sv} \widetilde{\kappa}_{a} + \widetilde{\omega}_{a}+\calO{\epsilon_\ra^2}\,. \slabel{RadialConstraintLowCurved2}
\eeas
The constraint \eqref{RadialConstraintLowCurved1} is exactly the same as the one obtained in the
axisymmetric case~\eqref{ConsaxiV1} and we will find that the effect of curvature
appears only at higher orders. 
However, the latter constraint \eqref{RadialConstraintLowCurved2} deserves a thorough comment. From the expression
of the total velocity~\eqref{totVUV} and the
decomposition~\eqref{IrrepsVa} for the relative velocity $\gr{V}$, the
velocity field on the FCL (that is when $y^1=y^2=0$) is
\be\label{TotVCenter}
\qquad\,\gr{\totV}_{\rm Cen}=\gr{U} + \Sv \gr{T}+\stf{V}_a^{(0)} \gr{d}_a =\gr{U} + \Sv \gr{T}+\calO{\epsilon_\ra^2}
\ee 
since from the gauge condition~\eqref{GaugeConstraint} $\stf{V}_a^{(0)}$ is only an order
$\epsilon_\ra^2$ quantity. The rotation rate of the fluid on the central line is thus approximately
\be\label{omegaomega}
\qquad \omega_{\rm Cen}^a \equiv [\gr{T}\times \partial_s(\gr{U}+\Sv\gr{T})]^a=\omega^a+\kappa^a \Sv\,.
\ee
Furthermore, the dipolar component $\Sv_a^{(0)}$ corresponds to the
sectional part of the solid rotation of the fluid contained in a fiber
section. Indeed, if we focus on terms which are linear in $y^a$ in the
decomposition~\eqref{IrrepsVa}, and if we ignore radial infall
($\shV_\myes^{(0)}$) and axial rotation ($\Sw^{(0)}$), we realize that
the relative velocity $\gr{V}$ contains
\be
\qquad\gr{V} \supset \Sv_a^{(0)} y^a \gr{T} = [{\omega}_{\rm Sec}^a \gr{d}_a
\times y^b \gr{d}_b]\,,\myquad {\omega}^{\rm Sec}_a \equiv -\widetilde{\Sv}_a^{(0)}\,.\prebool{\nonumber}{}
\ee
The constraint~\eqref{RadialConstraintLowCurved2} thus states that
the solid rotation of the fiber section $\omega_{\rm Sec}^a$  is
equal to the rotation of the fluid on the FCL $\omega_{\rm
  Cen}^a$. It means that once the central line velocity $U^a$ is
determined (together with the curvature and the longitudinal lowest
order velocity $\Sv$), then the rotation of the fluid on the FCL is
determined, and the fiber section rotation must follow exactly the
same rotation rate. This fact can be rephrased more rigorously by replacing the
constraint~\eqref{RadialConstraintLowCurved2} in the
expression~\eqref{varpiComponents} for the vorticity, as we obtain
\be\label{CoRotations}
\qquad\myvort^a=\omega_{\rm Cen}^a +\calO{\epsilon_\ra^2}=\omega^a+\kappa^a \Sv +\calO{\epsilon_\ra^2}\,.
\ee
It is yet another way to see that the rotation of the fluid in
fiber sections (vorticity) is guided by the rotation of the fluid on the FCL.

The consequence is that the fluid contained in a given section, which by
construction is orthogonal to the tangential direction $\gr{T}$,
remains always in a geometrically defined fiber section. Or, said
differently, the fluid particles belonging to different fiber sections
are not mixed by the fluid velocity. We must stress again that this result is only valid at
lowest order in $\epsilon_\ra$. 

Hence, just by contemplating of \eqref{RadialConstraintLowCurved2} we
can understand why the lowest order approximation is called a string approximation. It is because
the fiber sections are slaves of the FCL, as they are determined from
it without retroaction at lowest order. Furthermore, and this is even
more important, this type of behavior also corresponds to a form of flexible rod model since the fiber sections are not
mixed and remain orthogonal to the FCL. If there was no longitudinal velocity ($\Sv=0$), a good analogy would be the spinal column with the
vertebra being the sections. If we want to consider a longitudinal
velocity, a good analogy would be a collar made of beads. The beads
have a cylindrical hole through which the collar string is passed, and if
the beads can slide along the string of the collar, their orientation
with respect to the string tangential direction is necessarily fixed
thanks to the cylindrical hole. We understand already at that point that it is hopeless
to try to find consistently the corrections of the string model if we start from a
flexible rod model, since the lowest order description is already a form of flexible rod model.

\item Second, from the lowest order monopole and dipole of the orthoradial constraint, that is ${\EqCtheta}^{(1)}_\myes$ and
${\EqCtheta}^{(0)}_a$, we obtain
\prebool{\beas
\Sw^{(1)}&=&\mytfrac{1}{2} ({\cal H} \widetilde{\kappa}_{a} {\omega}^{a} + {\cal H} \partial_s{{\Sw}}) +\calO{\epsilon_\ra^2} ,\\
\stf{V}^{(1)}_a&=&\mytfrac{1}{8} (2 \kappa_{a} {\Sw} -
\widetilde{\kappa}_{a} \partial_s{{\Sv}} + 2
{\Sv} \partial_s{\widetilde{\kappa}_{a}} +
2 \partial_s{\widetilde{\omega}_{a}}) \nonumber\\
&&+\calO{\epsilon_\ra^2} \,.
\eeas}
{\beas
\Sw^{(1)}&=&\mytfrac{1}{2} ({\cal H} \widetilde{\kappa}_{a} {\omega}^{a} + {\cal H} \partial_s{{\Sw}}) +\calO{\epsilon_\ra^2} \\
\stf{V}^{(1)}_a&=&\mytfrac{1}{8} (2 \kappa_{a} {\Sw} -
\widetilde{\kappa}_{a} \partial_s{{\Sv}} + 2
{\Sv} \partial_s{\widetilde{\kappa}_{a}} +
2 \partial_s{\widetilde{\omega}_{a}}) +\calO{\epsilon_\ra^2} \,.
\eeas}

\item And third, from the  lowest order monopole and dipole of the
  radial boundary constraint, that is from ${\EqCr}^{(0)}_\myes$ and
${\EqCr}^{(0)}_a$, and using the gauge condition~\eqref{GaugeConstraint} and the previous constraints we get
\prebool{\beas
P^{(0)}_\myes&=&\frac{\ST}{{\ra}} -  \partial_s{{\Sv}}
+\calO{\epsilon_\ra}\\
P^{(0)}_a&=& -  \kappa_{a} {\Sw} + \mytfrac{1}{2}
\widetilde{\kappa}_{a} \partial_s{{\Sv}} -
{\Sv} \partial_s{\widetilde{\kappa}_{a}}
-  \partial_s{\widetilde{\omega}_{a}}\nonumber\\
&&+\frac{\ST \widetilde{\kappa}_{a}}{{\ra}} +\calO{\epsilon_\ra} \,.\slabel{P0alow}
\eeas}
{\beas
P^{(0)}_\myes&=&\frac{\ST}{{\ra}} -  \partial_s{{\Sv}}
+\calO{\epsilon_\ra}\\
P^{(0)}_a&=& -  \kappa_{a} {\Sw} + \mytfrac{1}{2}
\widetilde{\kappa}_{a} \partial_s{{\Sv}} -
{\Sv} \partial_s{\widetilde{\kappa}_{a}}
-  \partial_s{\widetilde{\omega}_{a}}+\frac{\ST \widetilde{\kappa}_{a}}{{\ra}} +\calO{\epsilon_\ra} \,.\slabel{P0alow}
\eeas}
\end{itemize}

\subsubsection{Dynamics of the string model}

Inserting the constraints of the previous section in the expressions~\eqref{ComponentsNSpara}, \eqref{ComponentsNSsectional}, and
\eqref{ComponentsNSomeg}, we finally obtain the system of equations
\prebool{\beas\label{AllEqskappalow}
\partial_t \Sv&=&-  \partial_t \bar{U}-  U^{a} \widetilde{\omega}_{a}+\para{g}
 + \frac{\ST {\cal H}}{{\ra}} \nonumber\\
&& -  {\Sv} \partial_s{{\Sv}} + 6 {\cal H} \partial_s{{\Sv}}+ 3 \partial^2_s{{\Sv}}+\calO{\epsilon_\ra}\slabel{Eqkappalowdtv},\\
(\partial_t U)_a &=&g_{a} + {\Sv}^2 \widetilde{\kappa}_{a}  + 2 {\Sv} \widetilde{\omega}_{a}- 3 \widetilde{\kappa}_ {a} \partial_s{{\Sv}}
 \nonumber\\
&&-  \frac{\ST
  \widetilde{\kappa}_{a}}{{\ra}}+\calO{\epsilon_\ra}\slabel{EqkappalowdtUa},\\
\partial_t \Sw&=&  4 {\cal H} \widetilde{\kappa}^{a} {\omega}_{a}
+ {\Sw} \partial_s{{\Sv}} + {\omega}^{a} \partial_s{\widetilde{\kappa}_{a}}
 + 4 {\cal H} \partial_s{{\Sw}}
 -  {\Sv} \partial_s{{\Sw}}\nonumber \\
&&  + \partial^2_s{{\Sw}}+\calO{\epsilon_\ra^2}\slabel{Eqkappalowdtw}\,,
\eeas}
{\beas\label{AllEqskappalow}
\partial_t \Sv&=&-  \partial_t \bar{U}-  U^{a} \widetilde{\omega}_{a}+\para{g}
 + \frac{\ST {\cal H}}{{\ra}}  -  {\Sv} \partial_s{{\Sv}} + 6 {\cal H} \partial_s{{\Sv}}+ 3 \partial^2_s{{\Sv}}+\calO{\epsilon_\ra}\slabel{Eqkappalowdtv}\\
(\partial_t U)_a &=&g_{a} + {\Sv}^2 \widetilde{\kappa}_{a}  + 2 {\Sv} \widetilde{\omega}_{a}- 3 \widetilde{\kappa}_ {a} \partial_s{{\Sv}}
 -  \frac{\ST
  \widetilde{\kappa}_{a}}{{\ra}}+\calO{\epsilon_\ra}\slabel{EqkappalowdtUa}\\
\partial_t \Sw&=&  4 {\cal H} \widetilde{\kappa}^{a} {\omega}_{a}
+ {\Sw} \partial_s{{\Sv}} + {\omega}^{a} \partial_s{\widetilde{\kappa}_{a}}
 + 4 {\cal H} \partial_s{{\Sw}}
 -  {\Sv} \partial_s{{\Sw}}+ \partial^2_s{{\Sw}}+\calO{\epsilon_\ra^2}\slabel{Eqkappalowdtw}\,,
\eeas}
where using~\eqref{dtXdtX1} we used the compact expression
\be
(\partial_t U)_a = \partial_t U_a+\para{\omega}\widetilde{U}_a-\bar{U} \widetilde{\omega}_{a}\,.
\ee
Note also that from~\eqref{dtXidtXi1} we also get
$(\partial_t U)^\tetr{3}= \partial_t \para{U}+\widetilde{\omega}_a U^a$ allowing to
rewrite~\eqref{Eqkappalowdtv} in a slightly more compact form if
desired. Finally, when surface tension effects are included, we also need to
determine the dynamics of the fiber radius as it couples
to~\eqref{EqkappalowdtUa} and ~\eqref{Eqkappalowdtv}, and at lowest
order it reads exactly as in the axisymmetric case, that is
\be\label{LowRadCurvedString}
\partial_t \ln \ra = - {\cal H} {\Sv} - \mytfrac{1}{2} \partial_s{{\Sv}}\,.
\ee

Several comments are in order for this viscous string model.
\begin{itemize}
\item We can check that there is no retroaction of $\Sw$ on the lowest
order dynamical equations for $\Sv$ and $U_a$. The axial rotation dynamical equation~\eqref{Eqkappalowdtw} is in fact part of the first corrections and not part of the lowest order
string model.
\item The equations are at most linear in the curvature $\kappa_a$
  even though we have not linearized in this variable.
\item To compare \eqref{Eqkappalowdtv} and \eqref{EqkappalowdtUa}
  with the result obtained from a momentum balance equation by~\citet{Fraunhofer1,Fraunhofer2,Ribe2004,Ribe2006}, we must
 first use that the average velocity inside a fiber is approximately the velocity on the FCL given at lowest order
  by~\eqref{TotVCenter}. Then, from the properties
\beas\label{ConvectiveDerivativeComponents}
(\partial_t \totV_{\rm    Cen})^\tetr{3} &=& \partial_t \Sv + (\partial_t U)^\tetr{3}\\
(\partial_t \totV_{\rm  Cen})_a &=& (\partial_t U)_a-\Sv \widetilde{\omega}_a\\
\Sv (\partial_s \totV_{\rm Cen})^\tetr{3}&=&\Sv \partial_s \Sv\\
\Sv (\partial_s \totV_{\rm Cen})_a&=&-\Sv^2 \tkappa_a-\Sv \widetilde{\omega}_a\,,
\eeas
and defining a convective derivative by ${\cal
  D}_t\equiv \partial_t+\Sv \partial_s$, the string equations are
simply recast as
\beas\label{Convective}
({\cal D}_t \totV_{\rm    Cen})^\tetr{3}&\simeq& \para{g}+ \frac{\ST
  {\cal H}}{{\ra}} + 6 {\cal H} \partial_s{{\Sv}}+3 \partial^2_s{{\Sv}}\\
({\cal D}_t \totV_{\rm    Cen})_a&\simeq&g_{a} - 3 \widetilde{\kappa}_ {a} \partial_s{{\Sv}}- \frac{\ST \widetilde{\kappa}_{a}}{{\ra}}\,.
\eeas
\item We can in particular check that the contributions from the surface tension are
  exactly those found in Eqs. (37) (49) and (57) of
  \citet{Ribe2006}. Since the vector $\tkappa^a$ points toward the exterior of the FCL curvature, the contribution $- \ST
  \widetilde{\kappa}_{a}/\ra$ tends to unfold the viscous jet as
  strongly curved region will be accelerated toward the center of curvature.
\item If we ignore the surface tension contributions, and using the
  covariantization relations \eqref{Covariantization}, the
  equation~\eqref{Convective} can be recast in the compact form
\prebool{\bea\label{MOmentumBalance}
\qquad{\cal D}_t \gr{\totV}_{\rm    Cen}&\simeq&\gr{g}+\frac{1}{\pi
  \ra^2}\partial_s \gr{F}\,,\\\
 \gr{F}&\equiv& \pi
 \ra^2 \para{\VF}_\myes^{(0)}\gr{T}\,,\myquad \,\para{\VF}_\myes^{(0)}  =
 3(\partial_s\Sv)
\eea}
{\be\label{MOmentumBalance}
\qquad{\cal D}_t \gr{\totV}_{\rm    Cen}\simeq\gr{g}+\frac{1}{\pi
  \ra^2}\partial_s \gr{F}\,,\qquad\gr{F}\equiv\pi
\ra^2 \para{\VF}_\myes^{(0)}\gr{T}\,,\myquad \,\para{\VF}_\myes^{(0)}
= 3(\partial_s\Sv)
\ee}
where the link with the momentum balance has now been made
obvious. The expression of the viscous force $\gr{F}$ which is purely longitudinal has
the same origin as in the axisymmetric case [see Eq.~\eqref{paraF}].
\end{itemize}

\subsubsection{Rotating frame}

If we now consider that the problem is studied in a steady rotating
frame, then Eqs.~\eqref{AllEqskappalow} need to be supplemented with the contributions
\beas\label{InertialTerms}
\partial_t \Sv &\supset&-  \overline{I} - 2 U^{a} \widetilde{\Omega}_{a}\\
(\partial_t U)_a &\supset& -2 \widetilde{U}_a \para{\Omega}
 -  I_{a}+ 2 (\para{U}+\Sv) \widetilde{\Omega}_{a}\\
\partial_t \Sw &\supset&-\widetilde{\Omega}^{a} \left({\omega}_{a}+{\Sv} \kappa_{a}\right)+ \para{\Omega} \partial_s{{\Sv}}\,.
\eeas
The first two equations arise naturally from the longitudinal and
sectional components of the inertial and Coriolis forces.

\subsubsection{Longitudinal central line velocity}

We notice that the dynamical equation for $v$ \eqref{Eqkappalowdtv} contains $\partial_t
\para{U}$. However, we must remember that the longitudinal velocity of
the central line contains a remaining gauge degree of freedom. Indeed,
we have fixed the position of the central line inside the fiber by
asking that there should be no shape dipole. However, as argued
in~\S~\ref{GaugeFix}, we can still displace the fiber inside that curved
central line, equivalent to a reparameterization $s\to s+f(t)$ which
changes the velocities as $\para{U} \to \para{U}-\partial_t
f$ and $\Sv \to \Sv + \partial_t f$.

Eq.~\eqref{Eqkappalowdtv} is in fact an equation for
$\partial_t(\Sv+\para{U})$ and we must find a unique way to determine
$\para{U}$ independently. Let us fix the value of the central line
longitudinal velocity $\para{U}$ for a given affine parameter (say
$s=0$ at a fiber boundary) at all times. Typically the fiber is
attached at the boundary so we choose simply $\para{U}(s=0)=0$. Then from the condition~\eqref{dsUortho} we obtain
\be\label{FixGaugethird}
\partial_s \para{U} = \tkappa_a U^a
\ee
and thus $\para{U}$ must be determined
everywhere on the fiber at all times, effectively breaking the degeneracy in \eqref{Eqkappalowdtv}.

\subsubsection{Full-set of equations for the string model}\label{secFullSystem}

We are now in position to gather all the equations which are required
for the curved viscous string model. First there is a set of equations
which are constraint equations and which must be solved at a given
initial time, since they are ordinary differential equations in $s$.
\begin{enumerate}
\item Once $\gr{R}(s,t)$ is known at a given time, e.g. at initial
  time, then the tangential direction of the FCL is obtained from~\eqref{TTun}.
\item The curvature $\gr{\kappa}$ of the FCL is then determined at that same given
  time from its definition~\eqref{DefCurvature}.
\item The orthonormal basis $\gr{d}_\tetr{i}$, can be determined everywhere on the fiber
  at that same given time from~\eqref{Propkappa}, provided some choice
  is made on a fiber boundary.
\item With this orthonormal basis we can extract the components
  $\kappa^a$ of the curvature.
\item When the FCL velocity components $U^a$ and $\para{U}$ are known at a given time, and the
  curvature components $\kappa^a$ as well, then the rotation components
  $\omega^\tetr{i}$ can be found from~\eqref{omegacomponent} for the
  sectional component and from~\eqref{Reldtdsko2} for the longitudinal
  component at the same given time.
\item The longitudinal part of the FCL velocity $\para{U}$ is not dynamical but
  it is instead constrained by~\eqref{FixGaugethird}.
\end{enumerate}
Then we have dynamical equations which give the time evolution of variables
from initial conditions, but they depend on partial derivatives
$\partial_s$ so they are partial differential equations.
\begin{enumerate}
\item The position in space of the FCL $\gr{R}(s,t)$ is modified due
  to the FCL velocity $\gr{U}$, as seen on~\eqref{DefU}.
\item The curvature components evolve in time thanks to~\eqref{Reldtdsko1}.
\item The orthonormal basis evolves in time with~\eqref{DefRotation}.
\item The longitudinal velocity $\Sv$ of the fluid inside the fiber evolves in time according to Eq.~\eqref{Eqkappalowdtv}.
\item The sectional part of the FCL velocity $U^a$ evolves according to
  Eq.~\eqref{EqkappalowdtUa}.
\item The fiber radius $\ra$ evolves according to~\eqref{LowRadCurvedString}.
\end{enumerate}
When considering corrections to the viscous string model, this
structure between constraints and dynamical equations is preserved.

\subsubsection{Covariant expressions}\label{SecCovariantMethod}

If we prefer to work fully in the Cartesian canonical basis, that is,
if for a vector we prefer using the covariant form $\gr{X}$ than the $X^\tetr{i}$, then
from \eqref{XtoXi} this is immediate. However, we need to use the covariantization relations~\eqref{Covariantization} to
recast the derivatives in the desired form $\partial_t \gr{X}$ or
$\partial_s \gr{X}$. The resulting equations take a more transparent
form if we separate all vectors in sectional and longitudinal parts
according to  \eqref{ProjVectors} and write equations for these
components. Since the longitudinal and sectional projections involve only the tangential
direction $\gr{T}$, then the dependence in the
section basis $\gr{d}_a$ disappears. The FA coordinates and the
orthonormal basis were introduced to perform all
intermediary computations, but the final results need not be expressed
in this language. This justifies {\it a posteriori}  why we have chosen the special choice \eqref{DefCurvature} for the
curvature and discarded the possibility of having a non-vanishing
longitudinal component for the FCL curvature. Indeed, this would lead to the same final equations when expressed in
a covariant form. However, all intermediate computations
would be more involved because the expression of the metric \eqref{gij}
would be much more complicated and non-diagonal, and the Christoffels \eqref{Christoffels}
would also be more complex. In particular, as a consequence of this choice for curvature,
the partial derivatives with respect to $s$ are easily written in a
covariant form since for any vector $\gr{X}$ we deduce from \eqref{dtXdtX2} the property
\be\label{Magicds}
P_\perp(\partial_s \gr{X}_\perp) = (\partial_s X^a) \gr{d}_a\,.
\ee
This allows to read the covariant form of a given equation written in
terms of sectional components nearly instantly.

We gather in covariant form the complete set of equations described in
\S~\ref{secFullSystem}. First, the vectors $\gr{\omega}$ and $\gr{U}$ are
decomposed in sectional parts and longitudinal parts as in
\eqref{ProjVectors}. The constraint equations are
\beas\label{ConsCov}
\partial_s \gr{R} &=& \gr{T},\\
\partial_s \gr{T}&=& \gr{\kappa} \times \gr{T}=-\gr{\tkappa} \myquad\Leftrightarrow\myquad \gr{\kappa} = \gr{T} \times \partial_s \gr{T},\\
\gr{\omega}_\perp &=&\gr{T}\times \partial_s
\gr{U}=\partial_s \widetilde{\gr{U}}+\gr{\kappa}\para{U},\\
0&=&\gr{T}\cdot \partial_s
\gr{\omega}\qquad\myquad\Leftrightarrow\myquad \partial_s \para{\omega} =-
\gr{\omega} \cdot \gr{\tkappa},\\
0&=&\gr{T}\cdot \partial_s
\gr{U}\qquad\,\,\,\,\,\Leftrightarrow\myquad \partial_s \para{U} = -\gr{U}\cdot \gr{\tkappa}\,.
\eeas

As for the dynamical equations, they are recast as
\prebool{\beas\label{DynCov}
\partial_t \gr{R} &=& \gr{U},\\
\partial_t \gr{\kappa}&=&\partial_s \gr{\omega}+\gr{\omega}\times
\gr{\kappa}\,\,\Leftrightarrow \,\,P_\perp(\partial_t
\gr{\kappa})=P_\perp(\partial_s\gr{\omega}_\perp),\nonumber\\
\partial_t \gr{T}&=&\gr{\omega}\times \gr{T},\\
\partial_t \ln \ra& =&  - {\cal H} {\Sv} -
\mytfrac{1}{2} \partial_s{{\Sv}},\\
\partial_t \Sv &= &-\partial_t \para{U}-\gr{U} \cdot
\widetilde{\gr{\omega}}+\para{g}+\ST\frac{{\cal
    H}}{\ra}-\Sv \partial_s \Sv\\
&&+6 {\cal H}\partial_s \Sv+3\partial_s^2
\Sv-\para{I}-2(\gr{\Omega}\times \gr{U})\cdot \gr{T},\nonumber\\
P_\perp[\partial_t
\gr{U}_\perp]&=&\gr{T}\times\left[(\para{U}+2\Sv)\gr{\omega} +\left(\Sv^2-3\partial_s\Sv-\frac{\ST}{\ra}\right) \gr{\kappa}\right]\nonumber\\
&+&[\gr{g}-\gr{I}]_\perp+2 \gr{T}\times \left[(\Sv+\para{U}) \gr{\Omega} -\para{\Omega}\gr{U}\right],
\eeas}
{\beas\label{DynCov}
\partial_t \gr{R} &=& \gr{U},\\
\partial_t \gr{\kappa}&=&\partial_s \gr{\omega}+\gr{\omega}\times
\gr{\kappa}\,\,\Leftrightarrow \,\,P_\perp(\partial_t
\gr{\kappa})=P_\perp(\partial_s\gr{\omega}_\perp),\\
\partial_t \gr{T}&=&\gr{\omega}\times \gr{T},\\
\partial_t \ln \ra& =&  - {\cal H} {\Sv} -
\mytfrac{1}{2} \partial_s{{\Sv}},\\
\partial_t \Sv &= &-\partial_t \para{U}-\gr{U} \cdot
\widetilde{\gr{\omega}}+\para{g}+\ST\frac{{\cal
    H}}{\ra}-\Sv \partial_s \Sv+6 {\cal H}\partial_s \Sv+3\partial_s^2
\Sv-\para{I}\nonumber\\
&&-2(\gr{\Omega}\times \gr{U})\cdot \gr{T} ,\\
P_\perp[\partial_t
\gr{U}_\perp]&=&\gr{T}\times\left[(\para{U}+2\Sv)\gr{\omega}
  +\left(\Sv^2-3\partial_s\Sv-\frac{\ST}{\ra}\right)
  \gr{\kappa}\right]+[\gr{g}-\gr{I}]_\perp\nonumber\\
&&+2 \gr{T}\times \left[(\Sv+\para{U}) \gr{\Omega} -\para{\Omega}\gr{U}\right]
\eeas}
where we should use that for any vector
\beas
\partial_t \gr{X}_\perp &=& P_\perp[\partial_t\gr{X}_\perp] + 
(\gr{X}_\perp\cdot \widetilde{\gr{\omega}}) \gr{T}\\
\partial_s \gr{X}_\perp &=& P_\perp[\partial_s\gr{X}_\perp] + 
(\gr{X}_\perp\cdot \widetilde{\gr{\kappa}}) \gr{T}\,.
\eeas

\subsubsection{Stationary regime}

In the stationary regime, all partial time derivatives vanish and the
viscous fiber model takes a simpler form. The velocity of the fiber
center also necessarily vanishes and $U^a=\para{U}=0$ as well as the
rotation ($\omega^\tetr{i}=0$). In that case, it becomes an ordinary
differential equation in the FCL parameter $s$. For
completeness, we report here the set of stationary equations, and they
read as
\beas\label{FullSteadyString}
\partial_s \gr{R} &=& \gr{T},\\
\partial_s \ln \ra &\simeq& - \mytfrac{1}{2} \partial_s{\ln
  {\Sv}}\slabel{Debit},\\
3 \partial^2_s{{\Sv}}+6 {\cal H} \partial_s{{\Sv}}-{\Sv} \partial_s{{\Sv}}&\simeq& -  \frac{\ST {\cal H}}{{\ra}}
 + \overline{I}- \para{g}\slabel{EqkappalowdtvSteady},\\
\widetilde{\kappa}_{a}\left(
  {\Sv}^2-\frac{\ST}{\ra}-3 \partial_s{{\Sv}}\right) &\simeq&I_{a} - g_{a}- 2 {\Sv}
\widetilde{\Omega}_{a} \slabel{EqkappalowdtUaSteady},\\
\partial_s \gr{d}_\tetr{i} &=& \gr{\kappa} \times \gr{d}_\tetr{i}\,.
\eeas
Eq.~\eqref{EqkappalowdtUaSteady} is used to determine the curvature
$\kappa^a$ (but it can become singular) and \eqref{EqkappalowdtvSteady} is used to integrate $\Sv$
along the FCL. Eq.~\eqref{Debit} is the statement that $\Sv \ra^2$ is
constant in a stationary regime, due to incompressibility.

If written in covariant form, the last three equations
of~\eqref{FullSteadyString} read as
\beas
3 \partial^2_s{{\Sv}}+6 {\cal H} \partial_s{{\Sv}}-{\Sv} \partial_s{{\Sv}}&\simeq& -  \frac{\ST {\cal H}}{{\ra}}
 + \gr{T}\cdot \left(\gr{I}- \gr{g}\right),\\
\gr{\kappa}\left(
  {\Sv}^2-\frac{\ST}{\ra}-3 \partial_s{{\Sv}}\right) &\simeq& \gr{T}\times\left[{\gr{g}}-
\gr{I}\right] - 2 {\Sv} \gr{\Omega},\\
\partial_s \gr{T} &=& \gr{\kappa} \times \gr{T}=-\gr{\tkappa}\,,
\eeas
which is the standard form for the stationary curved string model in the literature.

\subsection{Beyond the string model}\label{SecCurvedExtended}

\subsubsection{Limitations of the string model}

Apart for surface tension effects, the spatial extension of sections, that is the radius
$\ra$, does not play any role in the dynamical
equations, meaning that the internal structure of the fiber has no
impact on the dynamics. Furthermore, at lowest order the sectional
part of the viscous forces which reduces to the components $\stf{\VF}_a^{(0)}$ vanishes. The viscous forces are purely
longitudinal, as would be the case in a string, thus justifying the
name of the approximation. If we want to consider a model for viscous
fibers in which the size of spatial sections plays a role, we must
necessarily consider higher order terms in the parameter $\epsilon_\ra$.

Since fiber sections rotate at the same angular velocity as the fluid located on the FCL [see
Eq.~\eqref{CoRotations}], it means also that a rod model in which sections are
not mixed and remain orthogonal to the fiber tangential direction
cannot be part of this higher order model. Hence, when considering the lowest order of the angular momentum balance
equation, we do not obtain a dynamical equation which gives the time
evolution of the fiber section rotation as a function of viscous
forces, but rather obtain a constraint on the viscous forces (more
precisely on their sectional component) given that the fiber
section rotation is already determined by the string model.

For higher order models, the corrections of order $\epsilon_\ra^2$ in Eq.~\eqref{CoRotations}
will imply that fluid particles of different sections will be mixed as
a result of time evolution. Higher order models must also necessarily take into account that the
velocity of the fluid {\it on the central line}, is not exactly the
velocity {\it of the central line}. Indeed, Eq.~\eqref{GaugeConstraint} implies that there is a tiny shift between
the two which is an order $\epsilon_\ra^2$ correction.

Finally, terms of the type~\eqref{kaka} typically source the shape quadrupole $\rb_{ab}$ and a model restricted to circular sections cannot be
sufficient when considering corrections to the string limit. As we
shall explain in this section, when including order $\epsilon_\ra^n$
effect, we must include all multipoles $\rb_{L}$ with $\ell \leq
n$. Since we are interested in the first corrections which are of
order $\epsilon_\ra^2$, we consider quadrupolar shape moments thereafter.

\subsubsection{Normal vector and extrinsic curvature}

When including a first set of corrections, the components of the normal vector and the unit normal vector are approximately given by
\prebool{\beas
\Nv_\tetr{3}&=&- {\cal H} {\ra} + {\cal H} {\ra}^2 ({\yu}_{a} \widetilde{\kappa}^{a})+\calO{\epsilon_\ra^3}\\
\Nv_a&=&\yu_a-2 \ra^2 \perp_a^b \rb_{bc} \yu^c \nonumber\\
&&-3 \ra^3 \perp_a^b
\rb_{bcd} \yu^c \yu^d +\calO{\epsilon_\ra^4}\\
\widehat{\Nv}_\tetr{3} &=&- {\cal H} {\ra} + {\cal H} {\ra}^2 ({\yu}_{a} \widetilde{\kappa}^{a}) +\calO{\epsilon_\ra^3}\\
\widehat{\Nv}_a &=&{\yu}_{a}  -\ra^2\left(  \frac{ {\cal
      H}^2}{2}{\yu}_{a}  -2  \perp_a^b \rb_{bc} \yu^c
\right)\\
&&+\ra^3\left(-3 \perp_a^b \rb_{bcd} \yu^c \yu^d +{\cal H}^2 \yu_a
\yu_b \tkappa^b \right)+\calO{\epsilon_\ra^4}\,.\nonumber
\eeas}
{\beas
\Nv_\tetr{3}&=&- {\cal H} {\ra} + {\cal H} {\ra}^2 ({\yu}_{a} \widetilde{\kappa}^{a})+\calO{\epsilon_\ra^3}\\
\Nv_a&=&\yu_a-2 \ra^2 \perp_a^b \rb_{bc} \yu^c -3 \ra^3 \perp_a^b
\rb_{bcd} \yu^c \yu^d +\calO{\epsilon_\ra^4}\\
\widehat{\Nv}_\tetr{3} &=&- {\cal H} {\ra} + {\cal H} {\ra}^2 ({\yu}_{a} \widetilde{\kappa}^{a}) +\calO{\epsilon_\ra^3}\\
\widehat{\Nv}_a &=&{\yu}_{a}  -\ra^2\left(  \frac{ {\cal
      H}^2}{2}{\yu}_{a}  -2  \perp_a^b \rb_{bc} \yu^c
\right)\nonumber\\
&&+\ra^3\left(-3 \perp_a^b \rb_{bcd} \yu^c \yu^d +{\cal H}^2 \yu_a
\yu_b \tkappa^b \right)+\calO{\epsilon_\ra^4}\,.
\eeas}
From~\eqref{calK}, the extrinsic curvature can then be
obtained. Expanding $\ra {\cal K}$ in moments as in \eqref{ScalarSTFSide1}, the
first moments which are used to compute the first set of corrections
to the string model are
\prebool{\beas
{[\ra {\cal K}]}_\myes &=& 1-\ra^2\left(\mytfrac{3}{2}{\cal H}^2
  +\mytfrac{1}{2}\kappa_a \kappa^a +\partial_s {\cal
    H}\right)\nonumber\\
&&+\calO{\epsilon_\ra^4}\slabel{KR}\\
{[\ra {\cal K}]}_a &=&\tkappa_a\left[1+\ra^2\left(\mytfrac{7}{2}{\cal
      H}^2+\mytfrac{3}{4}\kappa_b \kappa^b+2\partial_s {\cal
      H}\right)\right]\nonumber\\
&&+\ra^2\left({\cal H}\partial_s \tkappa_a -\rb_{ab} \tkappa^b
\right)+\calO{\epsilon_\ra^4}\\
{[\ra {\cal K}]}_{ab} &=&3 \rb_{ab}+\kappa_{\langle a} \kappa_{b \rangle}+\calO{\epsilon_\ra^2}\,.
\eeas}
{\beas
{[\ra {\cal K}]}_\myes &=& 1-\ra^2\left(\mytfrac{3}{2}{\cal H}^2
  +\mytfrac{1}{2}\kappa_a \kappa^a +\partial_s {\cal
    H}\right)+\calO{\epsilon_\ra^4}\slabel{KR}\\
{[\ra {\cal K}]}_a &=&\tkappa_a\left[1+\ra^2\left(\mytfrac{7}{2}{\cal
      H}^2+\mytfrac{3}{4}\kappa_b \kappa^b+2\partial_s {\cal
      H}\right)\right]\nonumber\\
&&+\ra^2\left({\cal H}\partial_s \tkappa_a -\rb_{ab} \tkappa^b
\right)+\calO{\epsilon_\ra^4}\\
{[\ra {\cal K}]}_{ab} &=&3 \rb_{ab}+\kappa_{\langle a} \kappa_{b \rangle}+\calO{\epsilon_\ra^2}\,.
\eeas}

\subsubsection{Fundamental dynamical equations}\label{SecFundDynCorrection}

The general method to build higher order corrections is similar to the
axisymmetric case. The only difference is that now a given equation will not just give a constraint on the monopole but also on other
moments. As we restrict to second order, we shall need to consider the
monopole, dipole, and quadrupole of equations only. The
incompressibility conditions \eqref{EqsIncompressibilityCurved} are
also used throughout to express any dependence in multipoles
$\shV_L^{(n)}$ in terms of other types of multipoles.

We start from lowest moments of the Navier Stokes equation components
($\para{\EqD}_\myes^{(0)}$, $\stf{\EqD}_a^{(0)}$), which give the fundamental dynamical equations for the longitudinal
velocity $\Sv=\Sv_\myes^{(0)}$, and the FCL sectional velocity $U^a$. The evolution of the axial rotation $\Sw=\Sw^{(0)}$ has already
been found in~\eqref{Eqkappalowdtw} and we have argued that it should be considered
as part of the first corrections. Eqs~\eqref{ComponentsNSpara} and
\eqref{ComponentsNSsectional} were not given in full generality as we
had removed the contributions from $\stf{V}_a^{(0)}$ and
$\stf{V}_{ab}^{(0)}$ which are order $\epsilon_\ra^2$ quantities. The Navier-Stokes components \eqref{ComponentsNSpara} and \eqref{ComponentsNSsectional} must be supplemented by
the contributions
\beas\label{ComponentsNSparafirst}
\para{A}_\myes^{(0)} &\supset& {{\stf{V}}^{(0)}_a} \left(2\widetilde{\Omega}^{a}+\widetilde{\omega}^{a}+{\Sv} \widetilde{\kappa}^{a}+{{\Sv}^{(0)a}} \right) \\
\para{f}_\myes^{(0)} &\supset&2 \widetilde{\kappa}^{a} \partial_s{{{\stf{V}}^{(0)}_a}} + {{\stf{V}}^{(0)}_a} \partial_s{\widetilde{\kappa}^{a}} 
\eeas
\prebool{\beas\label{ComponentsNSsectionalfirst}
\stf{A}_a^{(0)} &\supset&{{\stf{V}}^{(0)b}} {\stf{V}}^{(0)}_{ab} -
\mytfrac{1}{2} {{\stf{V}}^{(0)}_a} \left(\partial_s{{\Sv}} +{{\stf{V}}^{(0)}_b}
\widetilde{\kappa}^{b} \right)+ {\Sv} \partial_s{{{\stf{V}}^{(0)}_a}}
\nonumber\\
&&-  {\varepsilon_{a}}^b {{\stf{V}}^{(0)}_b}\left( {\Sw}+ \bar{\omega} -2 \para{\Omega} \right) + \partial_t {{\stf{V}}^{(0)}_a}\\
\stf{f}_a^{(0)} &\supset&\mytfrac{1}{2} {{\stf{V}}^{(0)}_b} \kappa_{a}
\kappa^{b} -  \mytfrac{1}{2} {{\stf{V}}^{(0)}_a} \kappa_{b} \kappa^{b} +
\mytfrac{1}{3} {\stf{V}}^{(0)}_{ab} \widetilde{\kappa}^{b}+ \partial^2_s{{{\stf{V}}^{(0)}_a}}\nonumber
\eeas}
{\beas\label{ComponentsNSsectionalfirst}
\stf{A}_a^{(0)} &\supset&{{\stf{V}}^{(0)b}} {\stf{V}}^{(0)}_{ab} -
\mytfrac{1}{2} {{\stf{V}}^{(0)}_a} \left(\partial_s{{\Sv}} +{{\stf{V}}^{(0)}_b}
\widetilde{\kappa}^{b} \right)+
{\Sv} \partial_s{{{\stf{V}}^{(0)}_a}}\nonumber\\
&&
-  {\varepsilon_{a}}^b {{\stf{V}}^{(0)}_b}\left( {\Sw}+ \bar{\omega} -2 \para{\Omega} \right) + \partial_t {{\stf{V}}^{(0)}_a}\\
\stf{f}_a^{(0)} &\supset&\mytfrac{1}{2} {{\stf{V}}^{(0)}_b} \kappa_{a}
\kappa^{b} -  \mytfrac{1}{2} {{\stf{V}}^{(0)}_a} \kappa_{b} \kappa^{b} +
\mytfrac{1}{3} {\stf{V}}^{(0)}_{ab} \widetilde{\kappa}^{b}+ \partial^2_s{{{\stf{V}}^{(0)}_a}}
\eeas}
to be fully general. 

We also need to consider the dynamics of the shape. As in the string
model, we need to consider  the monopole of
the boundary kinematics~\eqref{EqPropaR} to determine the evolution of
the radius. The dipole of this equation has already been considered in
\eqref{GaugeConstraint} to fix the gauge and determine $\stf{V}_a^{(0)}$. Furthermore, we also
need to consider the quadrupole of the boundary kinematics
equation~\eqref{EqPropaR} so as to determine the evolution of
$\rb_{ab}$. These dynamical equations need to be truncated at the
required order. In Table~\ref{Table3} we summarize the essential dependencies of the
fundamental dynamical equations which need to be determined from constraints.

\begin{table}
\ifpre
\begin{tabular}{|l|c|l|}
\else
\begin{center}
\begin{tabular}{lcl}
\fi
\hline
Equation & Variable & Essential dependence\\[0.1cm]
\hline
   $\para{\EqD}^{(0)}_\myes\,\,\,$  \eqref{ComponentsNSpara} and
   \eqref{ComponentsNSparafirst} & $\partial_t \Sv^{(0)}$&
   $\Sv^{(1)}_\myes$, $P^{(0)}_\myes$, $\Sv^{(0)}_a$, $\stf{V}_a^{(0)}$ \\[0.1cm]
\hline
   $\stf{\EqD}^{(0)}_a\,\,\,$ \eqref{ComponentsNSsectional} and
   \eqref{ComponentsNSsectionalfirst} & $\partial_t U_a$& $\stf{V}^{(1)}_a$,
   $P^{(0)}_a$, $\Sv^{(0)}_a$, $\stf{V}_a^{(0)}$, $\stf{V}^{(0)}_{ab}$  \\[0.1cm]
\hline
$\EqBD^{(\leq 2)}_\myes$ \eqref{EqPropaR}& $\partial_t \ra$&$\Sv^{(0)}_a$, $\stf{V}^{(1)}_a$, $\Sv^{(0)}_\myes$,
$\Sv^{(1)}_\myes\ra^2$\\[0.1cm]
\hline
$\displaystyle\EqBD^{(\leq 2)}_{ab}$ \eqref{EqPropaR} &$\displaystyle\partial_t \rb_{ab}$&$\displaystyle\stf{V}_{ab}^{(0)}$, $\displaystyle\stf{V}_{ab}^{(1)}$\\[0.1cm]
\hline
\end{tabular}
\caption{Fundamental dynamical equations, with the corresponding
  variables and the main variables on which their evolution depends. The dependencies for the last
  two equations giving the evolution of radius and shape quadrupole are given only when order $\epsilon_\ra^2$ terms are included.}
\label{Table3}
\ifpre
\else
\end{center}
\fi
\end{table}

\subsubsection{General structure of constraint equations}

The method follows essentially the same steps as in the axisymmetric
case. The constraints which were obtained at lowest order in~\S~\eqref{SecCons1} need
to be extended to include order $\epsilon_\ra^2$ corrections, so as to
be replaced in the fundamental dynamical equations. We follow the same
procedure except that all constraints are considered up to a higher
order. For instance, when deriving the string model we considered the
monopole of the radial constraint at lowest order ${\EqCr}^{(0)}_\myes$ to constrain $P_\myes^{(0)}$, and now we must consider ${\EqCr}^{(1)}_\myes$.

However, just like when finding the corrections of the axisymmetric
case, the price to pay is that we introduce new dependencies. For
instance from ${\EqCr}^{(1)}_\myes$ we can obtain corrections for
the constraint which determines $P_\myes^{(0)}$, but it involves
$P_\myes^{(0)}\ra^2$. This set of dependencies is summarized in Table~\ref{Table4}.

The solution to this problem follows exactly the method used in the axisymmetric case. We
use the higher order moments of the Navier-Stokes, which we
consider as constraints and not dynamical equations, together with the
incompressibility constraint~\eqref{EqCD} to remove these newly
introduced variables. This is made possible since, as in the axisymmetric
case, these equations have Laplacians which allow us to use the property~\eqref{CoeffLaplacian}.

We thus need to follow a recursive algorithm which is very similar to the one used in the axisymmetric case, but
which is more involved since it involves the $\ell$-th order STF moments
when considering order $\epsilon_\ra^\ell$ corrections. Furthermore, the
structure of the recursive algorithm is slightly different for the
dipole components since {\it i)} the gauge constraint already
determines one dipolar moment ($\stf{V}_a^{(0)}$) and {\it ii)} one equation is used to
determine the external variable $U^a$. The corresponding set of dependencies is summarized in Table~\ref{Table5}.

\begin{table}
\ifpre
\begin{tabular}{|c|c|l|}
\else
\begin{center}
\begin{tabular}{ccl}
\fi
\hline
Equation & Variable & Essential dependence\\[0.1cm]
\hline
   ${\EqCr}_\myes\,\,$  & $P^{(0)}$&$\Sv^{(0)}$ and
   $P^{(1)}\ra^2$,  $P^{(2)}\ra^4\dots$\\[0.1cm]
\hline
   ${\EqCr}_a\,\,$  & $P^{(0)}_a$&$\Sv^{(0)}_a$ and
   $P^{(1)}_a\ra^2$,  $P^{(2)}_a\ra^4\dots$\\[0.1cm]
\hline
   ${\EqCr}_{ab}\,\,$  & $P^{(0)}_{ab}$&$\Sv^{(0)}_{ab}$ and $P^{(1)}_{ab}\ra^2$,  $P^{(2)}_{ab}\ra^4\dots$\\[0.1cm]
\hline
   $\para{\EqBC}_\myes\,\,\,$ & $\Sv^{(1)}$& $\Sv^{(0)}$ and
   $\Sv^{(2)}\ra^2$,  $\Sv^{(3)}\ra^4\dots$ \\[0.1cm]
\hline
   $\para{\EqBC}_a\,\,\,$ & $\Sv^{(0)}_a$& $\Sv^{(1)}_a\ra^2$,
   $\Sv^{(2)}_a\ra^4\dots$ \\[0.1cm]
\hline
   $\para{\EqBC}_{ab}\,\,\,$ & $\Sv^{(0)}_{ab}$& $\Sv^{(1)}_{ab}\ra^2$,  $\Sv^{(2)}_{ab}\ra^4\dots$ \\[0.1cm]
\hline
   ${\EqCtheta}_\myes\,\,\,$ &  $\Sw^{(1)}$& $\Sw^{(0)}$ and
   $\Sw^{(2)}\ra^2$,  $\Sw^{(3)}\ra^4\dots$  \\[0.1cm]
\hline
   ${\EqCtheta}_a\,\,\,$ &  $\stf{V}^{(1)}_a$&$\stf{V}^{(2)}_a \ra^2$, $\stf{V}^{(3)}_a
   \ra^4\dots$  \\[0.1cm]
\hline
${\EqCtheta}_{ab}\,\,\,$ &  $\stf{V}^{(0)}_{ab}$&
   $\stf{V}^{(1)}_{ab}\ra^2$ and $\stf{V}^{(2)}_{ab} \ra^4\dots$ \\[0.1cm]
\hline
   Gauge fixing \com{\eqref{Gauge}} &  $\stf{V}^{(0)}_a$& $\Sv^{(0)}_a$,
   $\stf{V}^{(1)}_a$, $\Sv^{(0)}\dots$\\[0.1cm]
\hline
\end{tabular}
\caption{Structure of boundary constraints. Each multipole of each
constraint is used to determine a variable as a function of other
variables. In each case, we specify which variable is determined and
report the essential variables on which it depends to emphasize the
structure of the recursive method. All these constraints need to be
truncated at a given power of $\ra$.}
\label{Table4}
\ifpre
\else
\end{center}
\fi
\end{table}

\begin{table}
\ifpre
\begin{tabular}{|c|c|c|}
\else
\begin{center}
\begin{tabular}{ccc}
\fi
\hline
Equation & Variable & Essential dependence\\[0.1cm]
\hline
   $\para{\EqD}^{(n+1)}_\myes\,\,\,$  &   $\Sv^{(n+2)}_\myes$& $P^{(n+1)}_\myes$, $\Sv^{(n+1)}_\myes$\\[0.1cm]
\hline
   $\para{\EqD}^{(n)}_a\,\,\,$  & $\Sv^{(n+1)}_a$& $\Sv^{(n)}_a$, $P^{(n)}_a$\\[0.1cm]
\hline
   $\para{\EqD}^{(n)}_{ab}\,\,\,$  & $\Sv^{(n+1)}_{ab}$&   $\Sv^{(n)}_{ab}$, $P^{(n)}_{ab}$\\[0.1cm]
\hline
   $\omeg{\EqD}^{(n+1)}\,\,\,$& $\Sw^{(n+1)}$& $P^{(n+1)}_\myes$, $\Sw^{(n)}$\\[0.1cm]
\hline
   $\stf{\EqD}^{(n+1)}_a\,\,\,$ & $\stf{V}^{(n+2)}_a$&   $\stf{V}^{(n+1)}_a$\\[0.1cm]
\hline
   $\stf{\EqD}^{(n)}_{ab}\,\,\,$ & $\stf{V}^{(n+1)}_{ab}$&  $\stf{V}^{(n)}_{ab}$\\[0.1cm]
\hline
   ${\EqCD}^{(n)}_\myes$ & $P^{(n+1)}_\myes$& $P^{(n)}_\myes$ \\[0.1cm]
\hline
   ${\EqCD}^{(n)}_a$ & $P^{(n+1)}_a$& $P^{(n)}_a$ \\[0.1cm]
\hline
   ${\EqCD}^{(n)}_{ab}$ & $P^{(n+1)}_{ab}$& $P^{(n)}_{ab}$\\[0.1cm]
\hline
\end{tabular}
\caption{Structure of dependence for the constraints obtained either
  from the higher moments of the Navier-Stokes equation or from the
  incompressibility constraint. We have  indicated only the essential
  dependence, so as to emphasize clearly the
  structure of the recursive method, but it depends in general on the
  full set of lower order variables.}
\label{Table5}
\ifpre
\else
\end{center}
\fi
\end{table}

\subsubsection{Quadrupoles of constraints}\label{ResultConstraints}

Let us first examine the quadrupoles of the constraints for which we
need only the lowest order expressions. From the quadrupole of the orthoradial constraint at lowest order,
that is, from ${\EqCtheta}^{(0)}_{ab}$ we get that
\prebool{\bea\label{EqV0ab}
\stf{V}^{(0)}_{ab} &=& \ra^2\left[-\mytfrac{3}{2}\stf{V}^{(1)}_{ab}-\mytfrac{3}{8}\kappa_{\langle
    a}\tkappa_{b \rangle}\Sw\right.\\
&&-\left.\mytfrac{3}{16}\kappa_{\langle a}\kappa_{b \rangle} \partial_s \Sv+\mytfrac{3}{8} \kappa_{\langle a}
  (\Sv \partial_s\kappa_{b \rangle}+\partial_s\omega_{b \rangle})\right]+\calO{\epsilon_\ra}\nonumber
\eea}
{\be\label{EqV0ab}
\stf{V}^{(0)}_{ab} = +\ra^2\left[-\mytfrac{3}{2}\stf{V}^{(1)}_{ab}-\mytfrac{3}{8}\kappa_{\langle
    a}\tkappa_{b \rangle}\Sw-\mytfrac{3}{16}\kappa_{\langle a}\kappa_{b \rangle} \partial_s \Sv+\mytfrac{3}{8} \kappa_{\langle a}
  (\Sv \partial_s\kappa_{b \rangle}+\partial_s\omega_{b \rangle})\right]+\calO{\epsilon_\ra}
\ee}
where we recall that the notation $\langle a_1\dots a_n\rangle $ means
the STF part and the notation $( a_1\dots a_n) $ means the symmetric part. 
Actually, the property that surface tension terms start at order
$\epsilon_\ra$ and viscous terms start at order $\epsilon_\ra^2$ arises for all $\stf{V}^{(0)}_L $ with $\ell
\geq 2$. 

This property is fortunate since conversely from the lowest
order of the quadrupole of the radial constraint
${\EqCr}^{(0)}_{ab}$ we find that $P_{ab}^{(0)}$ depends on
$\stf{V}^{(0)}_{ab} /\ra^2$. Once this dependence is replaced, we get
\prebool{\bea
&&P_{ab}^{(0)} =\frac{\ST}{\ra}\left(3\rb_{ab}+\kappa_{\langle a}
  \kappa_{b \rangle} \right) +\mytfrac{3}{2}\kappa_{\langle a}\tkappa_{b
  \rangle}\Sw\\
&&+\mytfrac{3}{4}\kappa_{\langle a} \kappa_{b \rangle} \partial_s \Sv-\mytfrac{3}{2} \kappa_{\langle a}
  (\Sv\partial_s\kappa_{b \rangle}+\partial_s\omega_{b \rangle})+\calO{\epsilon_\ra}\nonumber.
\eea}
{\bea
P_{ab}^{(0)} &=&\frac{\ST}{\ra}\left(3\rb_{ab}+\kappa_{\langle a}
  \kappa_{b \rangle} \right) +\mytfrac{3}{2}\kappa_{\langle a}\tkappa_{b
  \rangle}\Sw+\mytfrac{3}{4}\kappa_{\langle a} \kappa_{b
  \rangle} \partial_s \Sv\\
&&-\mytfrac{3}{2} \kappa_{\langle a}
  (\Sv\partial_s\kappa_{b \rangle}+\partial_s\omega_{b \rangle})+\calO{\epsilon_\ra}.
\eea}

Finally from the quadrupole of the longitudinal constraint $\para{\EqBC}^{(0)}_{ab}$ we get
\be
\Sv_{ab}^{(0)} =0+ \calO{\epsilon_\ra^2}\,.
\ee

\subsubsection{Monopoles and dipoles of constraints}\label{Secmonopdipole}

We now examine the monopoles and dipoles of the constraints. The
lowest orders have been found already in~\S~\eqref{SecCons1} when deriving the viscous string
model, and they need to be used to replace variables appearing in the
order $\epsilon_\ra^2$ terms. Once this is done, then from the
constraints $\para{\EqBC}^{(\leq 1)}_\myes$ and $\para{\EqBC}^{(\leq
  1)}_a$ we get that the constraints \eqref{RadialConstraintLowCurved}
should be supplemented by
\prebool{\beas
\Sv^{(1)}&\supset& \mytfrac{1}{32}\ra^2\left[ -64 {{\Sv}^{(2)}} - 16 {{\Sv}^{(1)}_a} \widetilde{\kappa}^{a} + 6 {\cal H} \kappa_{a} \kappa^{a} \partial_s{{\Sv}}\right. \nonumber \\ 
&& \qquad+ 84 {\cal H} \partial_s {\cal H} \partial_s{{\Sv}} - 12 {\cal H} {\Sv} \kappa^{a} \partial_s{\kappa_{a}} - 12 {\cal H} \kappa^{a} \partial_s{{\omega}_{a}} \nonumber \\ 
&& \qquad+ 6 \partial_s{{\Sv}} \partial_s^{2} {\cal H} + 84 {\cal H}^2 \partial^2_s{{\Sv}} + 12 \partial_s {\cal H} \partial^2_s{{\Sv}} \nonumber \\ 
&& \qquad \left. + 48 {\cal H}^3 \partial_s{{\Sv}} + 20 {\cal H} \partial^3_s{{\Sv}} + \partial^4_s{{\Sv}}\right]\\
\Sv^{(0)}_a&\supset&\mytfrac{1}{4} \ra^2\left[-12 {{\Sv}^{(1)}_{a}} + 10 {\cal H} \kappa_{a} {\Sw} - 11 {\cal H} \widetilde{\kappa}_{a} \partial_s{{\Sv}} \right.\nonumber \\ 
&& \qquad\left. + 10 {\cal H} {\Sv} \partial_s{\widetilde{\kappa}_{a}}+ 10 {\cal H} \partial_s{\widetilde{\omega}_{a}} -  \widetilde{\kappa}_{a} \partial^2_s{{\Sv}}\right]\,.
\eeas}
{\beas
\Sv^{(1)}&\supset& \mytfrac{1}{32}\ra^2\left[ -64 {{\Sv}^{(2)}_\myes} - 16 {{\Sv}^{(1)a}} \widetilde{\kappa}_{a} + 6 {\cal H} \kappa_{a} \kappa^{a} \partial_s{{\Sv}}\right.+ 84 {\cal H} \partial_s {\cal H} \partial_s{{\Sv}} - 12 {\cal H} {\Sv} \kappa^{a} \partial_s{\kappa_{a}}  \\ 
&& \quad - 12 {\cal H} \kappa^{a} \partial_s{{\omega}_{a}}+ 6 \partial_s{{\Sv}} \partial_s^{2} {\cal H} + 84 {\cal H}^2 \partial^2_s{{\Sv}} + 12 \partial_s {\cal H} \partial^2_s{{\Sv}} \left. + 48 {\cal H}^3 \partial_s{{\Sv}} + 20 {\cal H} \partial^3_s{{\Sv}} + \partial^4_s{{\Sv}}\right]\nonumber\\
\Sv^{(0)}_a&\supset&\mytfrac{1}{4} \ra^2\left[-12 {{\Sv}^{(1)}_{a}} + 10 {\cal H} \kappa_{a} {\Sw} - 11 {\cal H} \widetilde{\kappa}_{a} \partial_s{{\Sv}} \right.\left. + 10 {\cal H} {\Sv} \partial_s{\widetilde{\kappa}_{a}}+ 10 {\cal H} \partial_s{\widetilde{\omega}_{a}} -  \widetilde{\kappa}_{a} \partial^2_s{{\Sv}}\right]\,.
\eeas}
From the constraint ${\EqCtheta}^{(\leq 1)}_\myes$ we get an expression for $\Sw^{(1)}$ but we do not need any higher order
contribution since the dynamical equation for $\Sw$~\eqref{Eqkappalowdtw} is
already part of the first corrections. However, from
${\EqCtheta}^{(\leq 1)}_a$ we get the first corrections for the constraint on $\stf{V}^{(1)}_a$. Finally, from the constraints ${\EqCr}^{(\leq 1)}_\myes$ and
${\EqCr}^{(\leq 1)}_a$ we obtain the corrections on the constraints
for $P^{(0)}_\myes$ and $P^{(0)}_a$. These are rather
large expressions, and we gathered them in Appendix~\ref{SecExtraConstraint}.

As in the symmetric case, the constraint for $\Sv^{(1)}$ now depends
on $\Sv^{(2)}$ when including order $\epsilon_\ra^2$ corrections, and
we shall thus need another constraint to replace it. In fact we also
need constraints for $\Sv^{(1)}_a$, $\stf{V}^{(2)}_a$, and $\stf{V}^{(1)}_{ab}$.
As in the axisymmetric case, these additional constraints will come from higher moments of the Navier-Stokes
equations and the incompressibility constraint (see
Table~\ref{Table5}). Since the expressions of these constraints can be rather large, we report them in Appendix~\ref{SecExtraConstraint}.

\subsubsection{Corrections for dynamical equations}

The evolution of the radius with the first corrections included is
given by
\prebool{\bea\label{FirstRadCurvedString}
\partial_t \ln \ra&\supset& {\ra}^2 (-\mytfrac{3}{2}{\cal
  H}^2 \partial_s{{\Sv}} - \mytfrac{3}{8} \partial_s {\cal
  H} \partial_s{{\Sv}}  \nonumber\\
&&\qquad - \mytfrac{5}{8} {\cal H} \partial^2_s{{\Sv}}- \mytfrac{1}{16} \partial^3_s{{\Sv}})\,.
\eea}
{\be\label{FirstRadCurvedString}
\partial_t \ln \ra\supset {\ra}^2 (-\mytfrac{3}{2}{\cal H}^2 \partial_s{{\Sv}} - \mytfrac{3}{8} \partial_s {\cal H} \partial_s{{\Sv}}  - \mytfrac{5}{8} {\cal H} \partial^2_s{{\Sv}}- \mytfrac{1}{16} \partial^3_s{{\Sv}})\,.
\ee}
It is striking that once all the constraints are properly replaced, we
reach the same expression as in the
axisymmetric case \eqref{dtRCorrectionsAxi}, even though the original expression deduced
from the monopole of ~\eqref{EqPropaR} is formally more complex.
In principle, we could apply the same method as in \S~\ref{SecCosserat} and use the average longitudinal velocity
instead of $\Sv^{(0)}$, thus changing the fundamental variable.

Concerning the corrections of the dynamical equations for $\Sv$ and
$U_a$, we report them in Appendix~\ref{SecFinal}. They are obtained from the replacement of all constraints, and then
the replacement of the lowest order dynamical equation to remove the
time derivatives appearing in the corrective terms. Several comments
are in order here.
\begin{itemize}
\item As for the lowest order string model, these equations could be recast in a covariant
form, that is using the canonical Cartesian basis, following the
method described in \S~\ref{SecCovariantMethod}. This is especially
straightforward when using the property \eqref{Magicds}, so we do not
write it explicitly. 
\item Odd powers of $\ra$ correspond to surface tension effects ($\ra^{-1}$ for the
lowest order and $\ra$ for the corrections), whereas even powers of
$\ra$ correspond to inertial and viscous effects ($\ra^0$ for the
lowest order and $\ra^2$ for the first corrections). 
\item  We remark that the quadrupoles of shape $\rb_{ab}$ which appear
  from surface tension effects do not retroact on the dynamical
  equations for $\Sv$ and $U_a$.
\item In practice, it proves easier to multiply the equations considered [e.g., boundary
constraint \eqref{VectorBoundaryConstraint} or Navier-Stokes \eqref{SecNS}] by $h$ or $h^2$, so as to avoid
unnecessary factors $h^{-1}=1/(1+\tkappa_a y^a)$ or $h^{-2}$ which would need to
be expanded in an infinite series in $\tkappa_a y^a$. For instance,
${\cal K}$ given by \eqref{calK} involves $h^{-1}$, but it is not the case for $h {\cal  K}$. All products of tensors fully contracted with vectors $y^a$ are
then handled thanks to \eqref{AbyyMagic}. Hence, it is in principle possible to find very general relations between
multipoles, as we did for instance with the incompressibility condition~\eqref{EqCLn}.
\end{itemize}

Note that the dynamical evolution of $\Sw$ needs to be obtained from
Eq.~(\ref{Eqkappalowdtw}). Similarly, the dynamical evolution of the
quadrupole now needs to be determined independently. From the quadrupole of
\eqref{EqPropaR} and using the constraint~\eqref{EqV0ab}, we find the simple lowest order dynamical equation
\prebool{\bea\label{dtRab}
(\partial_t +\Sv \partial_s) \rb_{ab} &=&-\stf{V}^{(1)}_{ab}+\rb_{ab} \partial_s \Sv\nonumber\\
&&-2{\eps_{(a}}^c\rb_{b) c}(\Sw-\para{\omega})\,,
\eea}
{\bea\label{dtRab}
(\partial_t +\Sv \partial_s) \rb_{ab} &=& -\stf{V}^{(1)}_{ab}+\rb_{ab} \partial_s \Sv-2{\eps_{(a}}^c\rb_{b) c}(\Sw-\para{\omega})\,,
\eea}
in which one should replace the constraint \eqref{Vh1ab}. 

Note that since the evolution of $\rb_{ab}$ is sourced by
$\stf{V}^{(1)}_{ab}$, then from~\eqref{Vh1ab} we see that it
  contains typical terms of the type $\kappa_{\langle a}
  \kappa_{b\rangle}\partial_s \Sv$. If we were to consider the dynamics of
  higher multipoles such as $\rb_L$, it would be sourced by terms of
  the type $\kappa_{\langle a_1} \dots \kappa_{a_\ell
    \rangle}\partial_s \Sv$ among other terms. For comparison, from \eqref{LowRadCurvedString} we see
  that $\ln \ra$ is typically sourced by $\partial_s \Sv$. Hence, the
  higher the shape multipole is, the more spatial derivatives are involved in its dynamics. This justifies why when considering
  corrections of order $\epsilon_\ra^n$ we need only to consider the
  multipoles $\rb_L$ with $\ell \leq n$. It also justifies {\it a
  posteriori} why we are working with the shape multipoles $\rb_L$
  (more precisely their dimensionally reduced variables $\dimless{\rb}_L$) as
  defined in \eqref{Defrb} and not the $\hrb_L$ of \eqref{Defhrb}, which are
  better suited to describe relative shape perturbations.

\subsection{Straight fibers with elliptic sections}\label{SecElliptic}

It is now easy to consider the case of straight fibers but with
non-circular sections. We need only to consider the special case
$\kappa^a=\omega^a=\para{\omega}=U^a=\para{U}=0$.
We recover immediately the dynamical equations found in the
axisymmetric case with the first corrections included
[\eqref{EqF1DLow}, \eqref{dtRCorrectionsAxi} and corrections~\eqref{dtvCorrectionsAxi}].
However, we also obtain the dynamical evolution of the shape quadrupole,
which evolves as an independent equation. Indeed, to retroact on the
dynamics of $\Sv$ we would need terms of the type $\rb_{ab} \rb^{ab}$ which would appear only when corrections of order $\epsilon_\ra^4$ are
included.

Eq.~\eqref{dtRab} restricted to straight fibers takes
formally the same form, except that the constraint \eqref{Vh1ab} now
needs to be also considered in that restriction when replaced.
The last term of Eq.~\eqref{dtRab}  is expected, as it just states that axial rotation $\Sw$
will rotate the ellipticity, but the sectional rotation of the
orthonormal basis ($\para{\omega}$) must also be taken into account
and subtracted. However it is only really an effect of the choice of
basis to measure components. Indeed we can rewrite it in a manifestly
covariant form following the method of \S~\ref{SecCovariantMethod}. Defining $\rb_{\mu\nu}\equiv \rb_{ab} {d^a}_\mu
{d^b}_\nu$ the shape quadrupole evolution is given by
\prebool{\bea
{P_\perp}_\mu^\alpha {P_\perp}_\nu^\beta[ (\partial_t +\Sv \partial_s) \rb_{\alpha\beta}] &=&
-\stf{V}^{(1)}_{ab} {d^a}_\mu {d^b}_\nu\\
&+&\rb_{\mu\nu} \partial_s \Sv-2 \Sw  T^\beta {\eps_{\beta(\mu}}^{\alpha}\rb_{\nu) \alpha}, \nonumber
\eea}
{\bea
{P_\perp}_\mu^\alpha {P_\perp}_\nu^\beta[ (\partial_t +\Sv \partial_s) \rb_{\alpha\beta}] &=&
-\stf{V}^{(1)}_{ab} {d^a}_\mu {d^b}_\nu+\rb_{\mu\nu} \partial_s \Sv-2 \Sw  T^\beta {\eps_{\beta(\mu}}^{\alpha}\rb_{\nu) \alpha}
\eea}
and we can check that the contribution from the orthonormal basis
rotation $\para{\omega}$ has disappeared.


Let us now compute the evolution of the relative ellipticity, that is, the
evolution of the dimensionless moments $\hrb_{ab}= \ra^2
\rb_{ab}$. By using the lowest order of the fiber radius
evolution \eqref{FirstRadCurvedString}, we finally find that the dimensionless
quadrupole evolves according to
\prebool{\bea\label{dtRabStraight}
(\partial_t +\Sv \partial_s) \hrb_{ab} &=&-\frac{\ST}{\ra}
\hrb_{ab}-2 {\cal H}\Sv \hrb_{ab} \\
&&-2{\varepsilon_{(a}}^c\hrb_{b) c}(\Sw-\para{\omega})\,.\nonumber
\eea}
{\bea\label{dtRabStraight}
(\partial_t +\Sv \partial_s) \hrb_{ab} &=&-\frac{\ST}{\ra}
\hrb_{ab}-2 {\cal H}\Sv \hrb_{ab} -2{\varepsilon_{(a}}^c\hrb_{b) c}(\Sw-\para{\omega})\,.
\eea}
In a stationary regime ($\partial_t \hrb_{ab}=0$) we see that there is a
competition between surface tension effects which tend to decrease
ellipticity and stretching (${\cal H} \leq 0$) which tends to increase
the relative contribution of ellipticity. Indeed if we assess the
evolution of $Q^2 \equiv \hrb_{ab}\hrb^{ab}$ from the previous equation, its
second line which is a purely rotational effect does not contribute
since ${\eps_a}^c \hrb_{cb} \hrb^{ba}=0$, and we get simply
\be
(\partial_t +\Sv \partial_s) Q =-\frac{\ST}{\ra}Q-2 {\cal H}\Sv Q\,. 
\ee
If it is true that stretching increases the ellipticity, the surface
tension effects encompassed in the first term on the right hand side would eventually dominate and damp ellipticity, as $\ra$ is reduced
by stretching.

\subsection{Alternative method for rotating frames}\label{SecAltRot}

In our formalism we have allowed for the possibility to be working in
a rotating frame. This was taken into account in the Navier-Stokes
equation, by adding the fictitious forces term~\eqref{Imu}. There is,
however, a simpler method to recover all expressions in a rotating
frame. First we derive the results in a non-rotating frame,
allowing to cut by approximately half the number of terms in the
final results, and then we relate all variables in the non-rotating frame
with their counterparts in the rotating frame. We denote by
$\galifr{\omega}^\tetr{i}$ the components of the rotation rate of the orthonormal basis
[defined by~\eqref{DefRotation}] in the non-rotating frame, and by
$\rotfr{\omega}^\tetr{i}$ its counterpart in the rotating
frame. Similarly, we define $\galifr{\totV}^\tetr{i}$ and
$\rotfr{\totV}^\tetr{i}$ as the velocity in the non-rotating frame and
rotating frame respectively, and adopt a similar notation for the FCL
velocity components $U^\tetr{i}$. These quantities are related by
\beas\label{FundChangeFrame}
\galifr{\totV}^\tetr{i}&=&\rotfr{\totV}^\tetr{i}+[\gr{\Omega}\times
\gr{x}]^\tetr{i}\\
\galifr{U}^\tetr{i}&=&\rotfr{U}^\tetr{i}+[\gr{\Omega}\times
\gr{R}]^\tetr{i}\slabel{FundChangeFrame2}\\
\galifr{\omega}^\tetr{i}&=&\rotfr{\omega}^\tetr{i}+\Omega^\tetr{i}\,,
\eeas
where we recall that the wedge products are performed according to~\eqref{scalwedge}.
In particular, when considering the relative velocity with respect to the FCL [$V^\mu$
defined in~\eqref{totVUV}], we note that the previous relations imply
that the moments of its longitudinal and sectional part are unchanged except for
\be\label{TransfoSTF}
\galifr{\Sw^{(0)}}=\rotfr{\Sw^{(0)}}+\para{\Omega}\,,\qquad \galifr{\Sv}_a^{(0)}=\rotfr{\Sv}_a^{(0)}+\widetilde{\Omega}_a\,.
\ee

In order to replace all variables referring to the non-rotating frame
in terms of the variables referring to the rotating frame, we also
need to be able to relate the time derivatives of the FCL velocity
components. Specifically, we need
\prebool{\beas\label{Timederframes}
\partial_t \galifr{U}^\tetr{i}&=&\partial_t \rotfr{U}^\tetr{i}\slabel{Timederframes1}\\
&&-\left[\rotfr{\omega}\times\left(\gr{\Omega}\times
\gr{R}\right)\right]^\tetr{i}+[\gr{\Omega}\times
\rotfr{U}]^\tetr{i}\,\nonumber\\
\partial_t\Omega^\tetr{i} &=& [\gr{\Omega}\times \rotfr{\omega}]^\tetr{i}\,,\slabel{Timederframes2}
\eeas}
{\beas\label{Timederframes}
\partial_t \galifr{U}^\tetr{i}&=&\partial_t \rotfr{U}^\tetr{i}\slabel{Timederframes1}-\left[\rotfr{\omega}\times\left(\gr{\Omega}\times
\gr{R}\right)\right]^\tetr{i}+[\gr{\Omega}\times
\rotfr{U}]^\tetr{i}\,\\
\partial_t\Omega^\tetr{i} &=& [\gr{\Omega}\times \rotfr{\omega}]^\tetr{i}\,,\slabel{Timederframes2}
\eeas}
where we emphasize that in these expressions, we are considering time derivatives of
components. The relation \eqref{Timederframes1} is obtained from the
definition~\eqref{DefU} (which reads as here $\partial_t R^\mu =
\galifr{U}^\mu$) and ~\eqref{FundChangeFrame2}, by using the definition $[\gr{\Omega}\times \gr{R}]^\tetr{i}\equiv {{d}^\tetr{i}}_\mu (\gr{\Omega}\times \gr{R})^\mu $ and the property
  \eqref{DefRotation}. The relation \eqref{Timederframes2} is obtained from the definition $\Omega^\tetr{i} \equiv {d^\tetr{i}}_\mu
  \Omega^\mu$ and the property {\eqref{DefRotation}. Finally, we also need the derivatives
\be\label{Spatderframes}
\partial_s \left([\gr{\Omega}\times
\gr{R}]^\tetr{i}\right) =
-\widetilde{\Omega}^\tetr{i}-[\gr{\kappa}\times(\gr{\Omega}\times
\gr{R})]^\tetr{i}\,.
\ee

Using \eqref{FundChangeFrame}, \eqref{TransfoSTF}, and
\eqref{Timederframes}, we were able to check that starting from the expressions
found in a non-rotating frame ($\Omega_a=\para{\Omega}=0$), we recover
the expressions valid in a rotating frame, for all constraints and
all dynamical equations. It is thus a healthy consistency check for the
validity and correctness of the method.

\subsection{Physical insights on constraints}

In this section we give physical interpretations for the various velocity
components. For simplicity, we neglect the effect of curvature, so as
to emphasize the physical effects more clearly.

\subsubsection{Pressure constraint and the Trouton ratio}\label{SecMonopolePressure}

Let us first analyze how the constraint for $\Sv^{(1)}$ \eqref{RadialConstraintLowCurved} arises at
lowest order. If we have a longitudinal velocity gradient ($\partial_s
\Sv \neq 0$), then we have a radial infall which is constrained by
incompressibility as we get $\shV=-\partial_s \Sv$ from~\eqref{uprime}. As the radial infall is proportional to the
radial distance $V^a \supset \shV y^a/2$, there is a radial gradient
in this radial velocity. Thanks to viscous forces, this creates a
component $\tau^{(\visc)}_{ab} \propto  \visc \delta_{ab}
\shV$. From the boundary condition (where we ignore surface tension
for simplicity), there is necessarily a pressure appearing to
compensate for this, $P \simeq \visc \shV $. Hence on the fiber section,
the longitudinal force per unit area is $\para{\VF} = \tau_{\tetr{3}\tetr{3}}=2\visc \partial_s \Sv-\visc \shV= 3
\visc \partial_s \Sv$, and we recover the standard Trouton enhancement
ratio~\citep{Trouton} $3/2$. Indeed, in the end we get a factor $3$
instead of the factor $2$ that would have been found if we had forgotten the pressure contribution. To summarize, in order to satisfy the boundary constraint, a
pressure must appear when we have a gradient in the longitudinal
velocity, and it transforms a $2$ into a $3$ in the longitudinal
viscous forces because this pressure also acts on the sections.

\subsubsection{The Hagen-Poiseuille  profile}\label{SecHPdiscussion}

If we now consider a gradient in the gradient of the longitudinal
velocity ($\partial^2_s \Sv \neq 0$), it will induce a gradient of
radial infall ($\partial_s \shV = -\partial_s^2 \Sv$). This gradient
of radial infall induces viscous forces per unit area applied on fiber sections, of the form $\VF^a = \tau^{\tetr{3}a}
\simeq \visc (\partial_s \shV/2) y^a$. We must realize that the
component $\tau^{\tetr{3}a}$ gives the sectional components of the
force per unit area applied onto the fiber sections, but $\tau_{\tetr{3}a} \yu^a$ gives also the
longitudinal forces applied on the boundary. A way to have a
longitudinal component of viscous forces on the fiber side, is to have
a HP profile, that is a parabolic profile. Indeed, a velocity profile
$\para{V} = \Sv^{(1)} r^2$ creates $\tau_{\tetr{3}a} \simeq \visc 2
\Sv^{(1)} y_a$ and if $\Sv^{(1)}=-\partial_s \shV/4 = \partial_s^2
\Sv/4$, then the boundary constraint is satisfied. If the fiber radius
is not constant along the fiber (${\cal H}\neq 0$), this reasoning is
slightly altered, thus explaining the corresponding contribution in \eqref{RadialConstraintLowCurved}.
To summarize, when we have a gradient of stretching, then we have a gradient of infall velocity which creates a component
$\tau_{\tetr{3}a}$ on the fiber side, which in turn needs to be canceled by a
HP profile so as to satisfy the boundary conditions, and eventually
when everything is taken into account, there is no sectional component for the forces per unit area on sections ($\VF_a\simeq0$).

\subsubsection{Dipolar longitudinal velocity}\label{SecDipVel}

Let us consider a simple case in which there is no longitudinal
velocity $\Sv=\Sv_\myes^{(0)}$, but we allow for a gradient in the
sectional components of the FCL velocity ($\partial_s U_a \neq
0$). This induces a force per unit area applied on sections of the form $\VF^a = \tau^{\tetr{3}a}
\simeq \visc \partial_s U_a $. Following the same reasoning as in the
previous section, we realize that the flow must adapt in a way which
creates an additional contribution for $\tau^{\tetr{3}a}$ in order to
satisfy the boundary constraint. If we consider a dipolar modulation
of the longitudinal velocity $\para{V}= \Sv_a^{(0)} y^a$, then it
creates a stress $\tau^{\tetr{3}a} = \visc \Sv_a^{(0)} $. If $\Sv^{(0)}_a= -\partial_s
U_a$, then the boundary constraint is satisfied, and given that we have
considered $\Sv=\Sv_\myes^{(0)}=0$, this is equivalent to the
constraint \eqref{RadialConstraintLowCurved2} once we use
\eqref{omegacomponent}. Physically, if neighbor sections slide along
each other (that is if they have a relative velocity which is sectional), the viscous forces force them to rotate so that they do
not slide along each other. To conclude, we might think that a gradient in the
sectional velocity would create a sectional force per unit area on
sections, but in fact the boundary condition would ensure that this is
not the case, just like in the previous section. Only when considering
higher order constraints will there be a sectional viscous force per unit area on sections, and this is
why the lowest order model is a string model where viscous forces per
area are necessarily longitudinal.

\subsubsection{Dipolar pressure constraint}

If we have a dipole modulation of the radial infall ($\shV_a^{(0)}\neq 0 $), then following the reasoning
of~\S~\ref{SecMonopolePressure}, it will induce a dipole in the
pressure so as to satisfy boundary constraints. Indeed the sectional
velocity contains $V^a = y^a/2(\shV_b^{(0)}y^b)$ and, thus, $\tau_{ab}
\supset \visc \delta_{ab}(\shV_c^{(0)}y^c) $. In order to compensate
for this component, we need a pressure gradient $P_a^{(0)} =
\visc \shV_a^{(0)}$, which once all other constraints are used gives~\eqref{P0alow}.
To be fully correct, we should also mention that $\shV_a^{(0)}$ also
contributes as $\tau_{ab} \supset \mu y_{(a} \shV_{b)}^{(0)}$, but there is also a parabolic sectional
component $V_a = \stf{V}_a^{(1)}r^2$ which contributes as $\tau_{ab}
\supset 4 \mu y_{(a} \stf{V}_{b)}^{(1)}$, and from the orthoradial
constraint they must cancel. Again, the bottom line is that at lowest
order the system adapts so that $\tau_{\tetr{3}a}\simeq 0$, but this
property also implies that there is no sectional component for the viscous forces
per unit area on sections.

\subsection{Comparison with rod models}

The methods based on rod models follow a slightly different logical
route in order to obtain a one-dimensional reduction. Indeed, in our
method we solve for the constraints and then use them to obtain
the volumic forces. In rod models, we solve instead only {\it some} of the boundary constraints, those needed to
get the pressure moments, and then we use them to compute the forces
per unit area applied on fiber sections. These are then integrated to get
total forces and total torques applied on sections, allowing to establish a momentum balance equation and an angular momentum balance
equation on a slice of fluid contained between two infinitesimally close sections. The information
of the boundary constraints which was not used explicitly is then used
implicitly because we use that no force is applied on the sides
of the infinitesimal slice, so the balance equations involve only the
forces and torques on the fiber sections. If surface tension effects are
considered, then they are of course added to the side of the
infinitesimal slice. In this section we collect the expressions for
forces and torques that we find with our formalism, so as to
facilitate comparisons with existing literature using rod models.

\subsubsection{Viscous forces} 

The total force on sections is
\be\label{TotalForceDefinition}
\gr{F} \simeq \int_{r=0}^{r=\ra} \gr{\VF} r \dd r \dd \theta \,,
\ee
where we recall the definition~\eqref{DefVF} for the forces per unit area on sections.
This relation is approximate because we have neglected the effect of
non-circular sections. If we were to take into account correctly the
effect of a non-circular
shape, then we would need to perform a change of variables as in
Appendix~\ref{AppAltShape}. At lowest order the total longitudinal force is given by
\be\label{Forces}
\para{F} \simeq \pi
\ra^2 \para{\VF}_\myes^{(0)}=\pi\ra^2\left(-\frac{\ST}{\ra}+3\mu \partial_s \Sv\right)\,.
\ee
Note that in order to write a momentum balance equation on an
infinitesimal slice, we should also consider {\it i)} long distance
forces $\gr{g}$ in the bulk of the slice, and {\it ii)} the effect of surface
tension on the side of the slice [e.g., Eq. (49) of
~\citet{Ribe2006}]. This can also be computed by integrating all the
volumic forces on the infinitesimal slice, as the total lineic force is
obtained from
\be\label{balancevolume}
\frac{\dd \gr{F}}{\dd s}^{\rm tot} \equiv \int_{r=0}^{r=\ra} \gr{f} h r \dd r \dd \theta \,,
\ee
where we recall that the total volumic forces are given by
\eqref{rawfmu}. The factor $h$, which mathematically is
$\sqrt{g_{ij}}$ with the metric \eqref{gij}, takes into account the fact that if the
fiber is curved, then for an infinitesimal slice there is more fluid
in the exterior of curvature and less in the interior of curvature. 
At lowest order we get simply
\be\label{TotF1}
\frac{\dd \gr{F}}{\dd s}^{\rm tot} =  \pi \ra^2 \gr{g} +2 \partial_s (\pi \ST \ra
\gr{T})+\partial_s (\para{F} \gr{T})\,.
\ee
The first term is the effect of gravity, the second is the
effect of surface tension on the side of the infinitesimal slice, and
the last is the net effect of viscous forces on sections. Using \eqref{Forces}, the total lineic force is simply
\be\label{TotF2}
\frac{\dd \gr{F}}{\dd s}^{\rm tot} = \pi \ra^2 \gr{g} + \partial_s (\pi \ST \ra
\gr{T})+3 \visc \partial_s (\pi \ra^2 \partial_s \Sv \gr{T})\,,
\ee
in agreement with the r.h.s. of \eqref{Convective}.
If we had ignored the effect due to the induced pressure on the
sections [given by the first term of \eqref{Forces}], we would have overestimated the total lineic force by a
factor $2$ as seen when comparing \eqref{TotF2} with \eqref{TotF1}. So
we can state that if for viscous forces the effect of the constrained
pressured is to enhance by a factor $3/2$ the total forces, then for
surface tension the effect of the constrained pressure is a reduction by a factor $2$.

\subsubsection{Viscous torques} 

Still neglecting the effect
of non-circular sections, the total torque applied on a fiber section is given by
\be
\gr{\Gamma} \simeq \int_{r=0}^{r=\ra} (y^a \gr{d_a}) \times \gr{\VF} \dd \theta r \dd r 
\ee
or in components (that is, using $\gr{\Gamma}=\Gamma^a
\gr{d}_a+\para{\Gamma} \gr{T}$)
\prebool{\bea
\Gamma^a &\simeq& -\int_{r=0}^{r=\ra}\widetilde{y}^a \para{\VF} \dd \theta r
\dd r\\
\para{\Gamma} &\simeq& \int_{r=0}^{r=\ra} \widetilde{y}_b \VF^b \dd \theta r \dd r \,.
\eea}
{\be 
\Gamma^a = -\int_{r=0}^{r=\ra}\widetilde{y}^a \para{\VF} \dd \theta r
\dd r\,,\qquad\para{\Gamma} = \int_{r=0}^{r=\ra} \widetilde{y}_b \VF^b \dd \theta
r \dd r \,.
\ee}
In order to obtain the lowest order expressions for the torque, we
only need to keep contributions which
are linear in $y^a$ in the components of $\gr{\VF}$, and we find
\be\label{Torques}
\Gamma_a \simeq \frac{\pi \ra^4}{4}
{\eps_{a}}^b \para{\VF}_b^{(0)}=- \frac{\pi \ra^4}{4}\widetilde{\para{\VF}}_a^{(0)}\,,\myquad \para{\Gamma} \simeq \frac{\pi
\ra^4}{2} \omeg{\VF}^{(0)}\,.
\ee
Note that $\stf{V}_{ab}^{(0)}$ does not contribute to the torque as it
corresponds to a shear flow inside the section.

The expression of torques takes a simple form when expressed in terms
of fluid vorticity. At lowest order, the vorticity components obtained
from~\eqref{varpiComponents} with the lowest order constraints
replaced are given by
\beas
{\myvort}^\tetr{3}&=& \Sw+\calO{\epsilon_\ra^2}\\
\myvort^a &=&\omega^a+\kappa^a \Sv +\calO{\epsilon_\ra^2}\,.
\eeas
The longitudinal component of the torque, is then found from
\bea\label{LongiTorque}
\omeg{\VF}^{(0)} \simeq\visc\left[\partial_s
  \Sw-\mytfrac{1}{2}\kappa^a\left(\widetilde{\omega}_a+\Sv_a^{(0)}\right)\right]
&=&\visc\left(\partial_s \Sw+\tkappa_a
  {\omega}^a\right)\nonumber\\
&=&\visc (\partial_s \myvort)^\tetr{3}
\eea
where in the second equality we have used the lowest order
constraint~\eqref{RadialConstraintLowCurved2} and in the third we used
the property~\eqref{dtXidtXi2}.  This component of the torque is
induced by twisting (longitudinal difference of vorticity) and its physical origin is thus obvious.

The sectional torque is found from
\bea\label{SectionalForce}
-\widetilde{\para{\VF}}_a^{(0)} &\simeq& -\frac{\ST}{\ra}\kappa_a +3\visc
\left[\partial_s (\omega_a+\kappa_a
  \Sv)-\mytfrac{3}{2}\kappa_a \partial_s \Sv
  -\Sw\tkappa_a\right]\nonumber\\
&=& -\frac{\ST}{\ra}\kappa_a +3\visc (\partial_s
\myvort)_a-\frac{9}{2}\visc \kappa_a \partial_s \Sv\,.
\eea
The physical origin of the second term is simple. The sectional component of vorticity corresponds to a rotation around a sectional axis. 
If we consider two neighbor sections which have different sectional
vorticities, as would happen if the fiber is bent, then the fluid located inside will be squeezed on one
side and stretched on the other side, that is, there will appear a dipole of
stretching. Then from the viscous forces induced, this creates a
sectional torque. With this naive view we would get a factor
$2$ and not a factor $3$, but as in~\S~\ref{SecMonopolePressure},
there is also a dipolar pressure which is induced to satisfy the
boundary conditions, and it implies again the appearance of the Trouton
factor enhancement $3/2$. 

The last term of~\eqref{SectionalForce} is more subtle and it has been
ignored in~\citet{Ribe2004,Ribe2006}. Indeed, {\it stretching} implies the viscous
force~\eqref{Forces}, {\it twisting} implies the longitudinal
torque~\eqref{LongiTorque}, and {\it bending} induces the second term of the
sectional torque \eqref{SectionalForce}, and these geometries have been
considered separately in these references even though they can have
mixed effects. However, the coupling of stretching with curvature
has been ignored, and it happens to have an effect on the torque which is contained in the last term in
\eqref{SectionalForce}. We stress that this effect arises at the same
order, and is not an order $\epsilon_\ra^2$ correction.

In a simple case, the physical origin of this last term can also be
understood. Let us ignore surface tension and consider a stationary
regime with no axial rotation ($\omega_a=\Sw=0$) and constant curvature
($\partial_s \kappa^a=0$). The expression of the sectional force
\eqref{SectionalForce} is simply
\be
-\widetilde{\para{\VF}}_a^{(0)} \simeq -\frac{3}{2}\visc
\kappa_a \partial_s \Sv\,.
\ee
As the fiber is stretched ($\partial_s\Sv>0$), there is a radial
infall since $\shV=-\partial_s \Sv$. Since the fiber is curved, the
particles in the exterior of curvature ($\tkappa_a y^a>0$) are
compressed while they move closer to the FCL, and conversely the
particles in the interior of curvature ($\tkappa_a y^a<0$) are
stretched while they move toward the FCL. As a result of viscous
forces, this creates a contribution to the sectional torque. And, as
usual, we get a factor $3/2$ enhancement, the Trouton ratio, exactly like
for the other contribution to the bending torque. 

The first term of~\eqref{SectionalForce} comes from surface tension
effects. If the FCL is curved, then $\tkappa^a$ points in the exterior of curvature (or $-\tkappa_a$
points toward the center of curvature of the FCL). This means that
extrinsic curvature is increased in the exterior (for points such that
$\tkappa_a y^a >0$) and decreased in the
interior (for points such that $\tkappa_a y^a <0$). From Young-Laplace law, this induces a dipole in pressure,
with more pressure in the outside than in the inside as can be seen on
the constraint~\eqref{P0alow}. This dipolar pressure distribution
creates in turn a sectional torque. 

However, if we perform an angular momentum balance equation on an
infinitesimal slice, one should also add {\it i)} the torque of long
distance forces $\gr{g}$ in the bulk of the slice and {\it ii)} the
contribution of surface tension on the side of the slice which also creates a torque [e.g., Eq. (50) of ~\citet{Ribe2006}]. Just as
for the momentum balance equation, the angular momentum balance
equation is best computed by integrating the torques of volumic forces
on the infinitesimal slice, as the total lineic torque is
obtained from
\be\label{balancevolumeGamma}
\frac{\dd \gr{\Gamma}}{\dd s}^{\rm tot} \equiv \int_{r=0}^{r=\ra} (y^a
\gr{d_a}) \times \gr{f} \,h \dd \theta r \dd r\,.
\ee
We find that it can be expressed as
\be \label{balancevolumeGamma2}
\frac{\dd \gr{\Gamma}}{\dd s}^{\rm tot} = \gr{T}\times \gr{F} +\frac{\dd \gr{\gamma}}{\dd s}\,,
\ee
where the first term involves the sectional part of the total viscous
forces defined in \eqref{TotalForceDefinition}, and the second term is given at lowest order by
\be\label{dgama}
\frac{\dd \gr{\gamma}}{\dd s} \simeq \partial_s \gr{\Gamma}+\frac{\pi
  \ra^4}{4}\left(\gr{\tkappa}\times \gr{g}+4\frac{\ST}{\ra} {\cal H} \gr{\kappa}\right)\,.
\ee
It appears clearly that the last term on the right-hand side comes from
surface tension effects on the side of the infinitesimal slice, whereas
the first term is the net effect of torques applied on the sections of the
infinitesimal slice. The middle term is the torque induced by gravity,
which comes from the fact that when the fiber is curved, then the center of
mass of an infinitesimal slice is not exactly on the FCL, but is
instead offset by $\ra^2 \gr{\tkappa}/4$. In components \eqref{dgama}
reads as simply
\beas
&&\left[\frac{\dd \gr{\gamma}}{\dd s} \right]^a \simeq \partial_s \Gamma^a -\tkappa^a \para{\Gamma}+\ST
\pi \ra^3 {\cal H} \kappa^a\\
&&{\left[\frac{\dd \gr{\gamma}}{\dd s} \right]}^{\tetr{3}} \simeq\partial_s \para{\Gamma}+\tkappa^a \Gamma_a\,,
\eeas
where the expressions  of the torques applied on fiber sections are
given by \eqref{Torques} with \eqref{LongiTorque} and
\eqref{SectionalForce}. For completeness, we report the explicit result which is
\prebool{\bea
&&\left[\frac{\dd \gr{\gamma}}{\dd s} \right]^a\simeq \frac{\pi
  \ra^4}{4} \Bigl(\para{g} \kappa_{a} + \frac{\ST {\cal H}
  \kappa_{a}}{{\ra}} - 12 \visc {\cal H} \widetilde{\kappa}_{a} {\Sw}- 2 \visc \kappa_{b} \kappa^{b} {\omega}_{a} \nonumber\\
&&\qquad+ 2 \visc \kappa_{a}
\kappa^{b} {\omega}_{b} - 6 \visc {\cal H}
\kappa_{a} \partial_s{{\Sv}} -
\frac{\ST \partial_s{\kappa_{a}}}{{\ra}} + 12 \visc {\cal H} {\Sv} \partial_s{\kappa_{a}} \nonumber\\
&&\qquad+ \tfrac{3}{2} \visc \partial_s{{\Sv}} \partial_s{\kappa_{a}}- 3 \visc {\Sw} \partial_s{\widetilde{\kappa}_{a}} - 5 \visc \widetilde{\kappa}_{a} \partial_s{{\Sw}}  + 12 \visc {\cal H} \partial_s{{\omega}_{a}} \nonumber \\ 
&& \qquad-  \tfrac{3}{2}
\visc \kappa_{a} \partial^2_s{{\Sv}} + 3 \visc
{\Sv} \partial^2_s{\kappa_{a}} + 3
\visc \partial^2_s{{\omega}_{a}}\Bigr)
\eea
\bea
{\left[\frac{\dd \gr{\gamma}}{\dd s} \right]}^{\tetr{3}}&\simeq&\frac{\pi
  \ra^4}{4}\Bigl(- g^{a} \kappa_{a} - 3 \visc \kappa_{a} \kappa^{a}
{\Sw} + 8 \visc {\cal H} \widetilde{\kappa}^{a} {\omega}_{a}\nonumber \\
&&+ 3 \visc {\Sv} \widetilde{\kappa}^{a} \partial_s{\kappa_{a}} + 2 \visc {\omega}^{a} \partial_s{\widetilde{\kappa}_{a}} + 8 \visc {\cal H} \partial_s{{\Sw}} \nonumber \\ 
&&+ 5 \visc \widetilde{\kappa}^{a} \partial_s{{\omega}_{a}} + 2 \visc \partial^2_s{{\Sw}}\Bigr)\,.
\eea}
{\bea
\left[\frac{\dd \gr{\gamma}}{\dd s} \right]^a&\simeq& \frac{\pi
  \ra^4}{4} \Bigl(\para{g} \kappa_{a} + \frac{\ST {\cal H}
  \kappa_{a}}{{\ra}}  -
\frac{\ST \partial_s{\kappa_{a}}}{{\ra}} - 12 \visc {\cal H} \widetilde{\kappa}_{a} {\Sw}- 2 \visc \kappa_{b} \kappa^{b} {\omega}_{a} + 2 \visc \kappa_{a}
\kappa^{b} {\omega}_{b} \nonumber\\
&&\qquad - 6 \visc {\cal H}
\kappa_{a} \partial_s{{\Sv}}+ 12 \visc {\cal H} {\Sv} \partial_s{\kappa_{a}} + \tfrac{3}{2} \visc \partial_s{{\Sv}} \partial_s{\kappa_{a}}- 3 \visc {\Sw} \partial_s{\widetilde{\kappa}_{a}} - 5 \visc \widetilde{\kappa}_{a} \partial_s{{\Sw}}  \nonumber \\ 
&&\qquad + 12 \visc {\cal H} \partial_s{{\omega}_{a}} -  \tfrac{3}{2}
\visc \kappa_{a} \partial^2_s{{\Sv}} + 3 \visc
{\Sv} \partial^2_s{\kappa_{a}} + 3
\visc \partial^2_s{{\omega}_{a}}\Bigr)\\
{\left[\frac{\dd \gr{\gamma}}{\dd s} \right]}^{\tetr{3}}&\simeq&\frac{\pi
  \ra^4}{4}\Bigl(- g^{a} \kappa_{a} - 3 \visc \kappa_{a} \kappa^{a}
{\Sw} + 8 \visc {\cal H} \widetilde{\kappa}^{a} {\omega}_{a}+ 3 \visc {\Sv} \widetilde{\kappa}^{a} \partial_s{\kappa_{a}} + 2 \visc {\omega}^{a} \partial_s{\widetilde{\kappa}_{a}} \nonumber \\ 
&&\qquad + 8 \visc {\cal H} \partial_s{{\Sw}}+ 5 \visc \widetilde{\kappa}^{a} \partial_s{{\omega}_{a}} + 2 \visc \partial^2_s{{\Sw}}\Bigr)\,.
\eea}

\subsubsection{Sectional forces and dipolar HP profile}

The component of the longitudinal velocity 
$\Sv_b^{(1)}r^2 y^b$ can be considered as a dipolar HP profile, as it
is a parabolic profile with a dipolar modulation. This velocity component induces a force per area on sections  $\VF^a = \tau^{\tetr{3}a}
\simeq \visc 2 y^a(\Sv_b^{(1)} y^b) +\visc \Sv_a^{(1)}r^2$ and it
remains undetermined by the boundary constraints. When averaged
over directions it contributes to the forces per unit area as $\VF^a
\simeq 2 \visc\Sv_a^{(1)}r^2$, and after integration over the whole
section, it contributes to the sectional part of the total viscous force. Indeed, the sectional
components of the total viscous force applied on sections [defined in \eqref{TotalForceDefinition}] are
\prebool{\bea\label{TxN}
&&F_a\simeq\pi\ra^4\Bigl(-2 \visc {{\Sv}^{(1)}_a} + 3 \visc {\cal H}  \kappa_{a} {\Sw} + \tfrac{1}{4} \visc  \kappa^{b} \widetilde{\kappa}_{a} {\omega}_{b}  \\ 
&& \quad-  \tfrac{15}{4} \visc {\cal H}  \widetilde{\kappa}_{a} \partial_s{{\Sv}} + \tfrac{1}{4} \visc  {\Sw} \partial_s{\kappa_{a}} + 3 \visc {\cal H}  {\Sv} \partial_s{\widetilde{\kappa}_{a}} \nonumber \\ 
&& \quad+ \tfrac{1}{8} \visc  \partial_s{{\Sv}} \partial_s{\widetilde{\kappa}_{a}} + \tfrac{1}{2} \visc  \kappa_{a} \partial_s{{\Sw}} + 3 \visc {\cal H}  \partial_s{\widetilde{\omega}_{a}} \nonumber \\ 
&& \quad -  \tfrac{3}{8} \visc  \widetilde{\kappa}_{a} \partial^2_s{{\Sv}} + \tfrac{1}{4} \visc  {\Sv} \partial^2_s{\widetilde{\kappa}_{a}} -  \tfrac{1}{4} \visc  \kappa_{b} \kappa^{b} \widetilde{\omega}_{a}+ \tfrac{1}{4} \visc \partial^2_s{\widetilde{\omega}_{a}}\Bigr)\,.\nonumber
\eea}
{\bea\label{TxN}
F_a&\simeq&\pi\ra^4\Bigl(-2 \visc  {{\Sv}^{(1)}_a} + 3 \visc {\cal H} \kappa_{a} {\Sw} + \tfrac{1}{4} \visc \kappa^{b} \widetilde{\kappa}_{a} {\omega}_{b} -  \tfrac{15}{4} \visc {\cal H} \widetilde{\kappa}_{a} \partial_s{{\Sv}} + \tfrac{1}{4} \visc {\Sw} \partial_s{\kappa_{a}} \nonumber \\ 
&& \qquad+ 3 \visc {\cal H} {\Sv} \partial_s{\widetilde{\kappa}_{a}} +
\tfrac{1}{8}
\visc \partial_s{{\Sv}} \partial_s{\widetilde{\kappa}_{a}} +
\tfrac{1}{2} \visc \kappa_{a} \partial_s{{\Sw}} + 3 \visc {\cal H}  \partial_s{\widetilde{\omega}_{a}} -  \tfrac{3}{8} \visc\widetilde{\kappa}_{a} \partial^2_s{{\Sv}} \nonumber \\ 
&&\qquad + \tfrac{1}{4} \visc {\Sv} \partial^2_s{\widetilde{\kappa}_{a}} -  \tfrac{1}{4} \visc \kappa_{b} \kappa^{b} \widetilde{\omega}_{a}+ \tfrac{1}{4} \visc \partial^2_s{\widetilde{\omega}_{a}}\Bigr)\,.
\eea}

As the rotation of sections is constrained by \eqref{RadialConstraintLowCurved2}, then an
angular momentum balance equation would in fact determine the value of
$\Sv_a^{(1)}$ because part of the torque balance equation \eqref{balancevolumeGamma2} comes from $\gr{T}\times
\gr{F}$.  In our method it is determined from $\para{\EqD}^{(0)}_a$
(see Table.~\ref{Table5}), that is, from the dipole of the longitudinal
part of the Navier-Stokes equation. Indeed it
determines the rate of change of the local dipolar longitudinal
velocity $\Sv_a^{(0)}$ which is related to the vorticity and the local
rotation rate of the fluid on the FCL [see the discussion which
follows \eqref{RadialConstraintLowCurved}], and it thus contains the
same information as the angular momentum balance equation used in rod
models. The expression obtained for
$\Sv_a^{(1)}$ is reported in Appendix \ref{SecExtraConstraint}, and
once replaced in Eq.~(\ref{TxN}), the sectional components of the
total viscous force applied on sections read as
\bea\label{Fa}
&&F_a\simeq\pi\ra^4\Bigl( -\tfrac{\ST}{2 \ra} \partial_s \widetilde{\kappa}_a+ 3 \visc {\cal H}  \kappa_{a} {\Sw} + \tfrac{3}{4} \visc  \kappa^{b} \widetilde{\kappa}_{a} {\omega}_{b}  \\ 
&& \quad-  \tfrac{9}{2} \visc {\cal H}  \widetilde{\kappa}_{a} \partial_s{{\Sv}} + \tfrac{3}{4} \visc  {\Sw} \partial_s{\kappa_{a}} + 3 \visc {\cal H}  {\Sv} \partial_s{\widetilde{\kappa}_{a}} \nonumber \\ 
&& \quad- \tfrac{3}{8} \visc  \partial_s{{\Sv}} \partial_s{\widetilde{\kappa}_{a}} + \tfrac{5}{4} \visc  \kappa_{a} \partial_s{{\Sw}} + 3 \visc {\cal H}  \partial_s{\widetilde{\omega}_{a}} -\tfrac{1}{2}\Sv \Sw \kappa_a \nonumber \\ 
&& \quad -  \tfrac{21}{8} \visc
\widetilde{\kappa}_{a} \partial^2_s{{\Sv}} + \tfrac{3}{4} \visc
{\Sv} \partial^2_s{\widetilde{\kappa}_{a}} -  \tfrac{3}{4} \visc
\kappa_{b} \kappa^{b} \widetilde{\omega}_{a}+ \tfrac{3}{4}
\visc \partial^2_s{\widetilde{\omega}_{a}}\nonumber\\
&& \quad-\tfrac{1}{2}\Sw \omega_a
-\tfrac{1}{4}\visc \widetilde{\kappa}_b \omega^b
\kappa_a+\tfrac{3}{4}\widetilde{\kappa}_a \Sv \partial_s
\Sv+\tfrac{3}{4}\widetilde{\omega}_a \partial_s \Sv\Bigr)\,.\nonumber
\eea

\subsubsection{Rod models and their validity}

As explained in the previous sections, the constitutive relations of
rod models are based on the determination of forces per unit area on
fiber sections, and boundary constraints are only used explicitly to determine
the pressure profile. The boundary constraints are then used
implicitly in balance equations. 

At lowest order, only the momentum balance equation is used, and we
find \eqref{MOmentumBalance} which is also exactly what is found in
the viscous string model. This equation determines the motion of the
FCL, and thus the rotation rate $\omega^a$ can be inferred from
it. Given that the fiber vorticity is constrained to match the
rotation rate of the fluid located on the FCL (see \S~\ref{SecCons1}), the angular momentum balance equation is in
fact used to determine the sectional components of the viscous forces
per unit area $\VF^a$, as explained in the previous section, and it corresponds to the addition of an order
$\epsilon_\ra^2$ correction. 

To summarize, the rod model amounts to using
Eq.~(\ref{MOmentumBalance}), formally exactly like the viscous
string model, but it differs from it in the expression of the
force used on the right hand side which has sectional
components. Hence the rod model corresponds to
\be\label{MOmentumBalance2}
{\cal D}_t \gr{\totV}_{\rm    Cen}\simeq\gr{g}+\frac{1}{\pi
  \ra^2}\partial_s \gr{F}\,,
\ee
with the force given by
\bea\label{TotalForceRod}
\gr{F}&\equiv&(\pi \ra^2 3 \partial_s \Sv +\ST \pi \ra )\gr{T} + F^a \gr{d}_a\,,
\eea
and where we recall the definition of the convective derivative ${\cal
  D}_t\equiv \partial_t+\Sv \partial_s$. The components of the left-hand side of Eq.~(\ref{MOmentumBalance2}) are obtained from Eqs.~(\ref{ConvectiveDerivativeComponents}) exactly like for the
string model. The components of the first term on the right-hand side
are obtained from Eqs.~(\ref{Convective}) as in the string model. However, the rod model differs from the string model thanks to
the sectional components of the total viscous force which are reported
in Eq.~(\ref{Fa}). Note that from the relations of
\S~\ref{SecEssential} we must use
\be
\partial_s(F^a \gr{d}_a) = \partial_s F^a \gr{d}_a + \gr{T} \tilde
\kappa_a F^a
\ee
so as to evaluate the last term in
Eq.~(\ref{MOmentumBalance2}). Furthermore, the dynamics of $\Sw$ is
given by the lowest order dynamical equation (\ref{Eqkappalowdtw}) as
in our model, and it is required since $\Sw$ appears in
Eq.~(\ref{Fa}). In the rod model of Ref.~\citet{Ribe2006}, this is
equivalently found from the longitudinal part of the momentum balance
equation, even though it does not appear so explicitly as the physical case studied is stationary. 

Finally, we already mentioned that for straight fibers the model is improved by
considering the full expression of the boundary curvature~\cite{EggersDupont} given by
Eq.~(\ref{PSTexact}) instead of the lowest order $1/\ra$. We can use a
similar ansatz for curved fibers by noting that in Eq.~(\ref{TotalForceRod}), the term $\ST \pi \ra$ is in fact $\ST \pi
{\cal K}_\emptyset \ra^2 $ at lowest order in $\epsilon_\ra$. Hence
from Eq. (\ref{KR}) an improved rod model is obtained by the replacement
\be\label{ImprovedRodModel}
\ST \pi \ra \to \ST \pi \ra [1-\ra^2(\tfrac{3}{2}{\cal
  H}^2+\tfrac{1}{2}\kappa_a \kappa^a +\partial_s {\cal H})]
\ee
in Eq.~(\ref{TotalForceRod}). This improved rod model is
necessary to compute in Ref.~\cite{PitrouPRE2} the Rayleigh-Plateau instability of a viscous
fibers.


The validity of rod models is limited by the fact that we are considering one correction while
discarding other sources of corrections, whereas in our approach we
systematically consider all corrections of order
$\epsilon_\ra^2$. Among the effects ignored there are the following:
\begin{itemize}
\item the difference between the velocity of the central line and the
  velocity of the fluid on the central line (see ~\S~\ref{GaugeFix});
\item the HP profile induced by the constraint~\eqref{RadialConstraintLowCurved1} which mixes the fluid particles belonging to
  neighboring sections;
\item the shape moments which are sourced and invalidate the
  assumption that the sections remain circular.
\end{itemize}

Furthermore, even though it is computationally involved, our approach
allows to find the corrections up to any order when rod models would
fail because of the impossibility to deal with the mixing sections. In
principle, we have a clear recursive algorithm made of constraints
replacements in fundamental dynamical equations.

However, there are cases in which rod models capture the essential
corrections. First, shape moments appear in all intermediary
expressions but do not appear in the final dynamical equations
(\ref{Finalv}) and (\ref{FinalUa}), hence, if we are not interested in
the fiber sections shape but only on the central line, they can be ignored. Furthermore, if we are considering the steady motion in
a rotating frame, and if the Rosby number is very small, that is, for
high rotation rates, then it does not matter if we have ignored most
of the corrective effects. Indeed, the boundary constraints do not involve the
frame rotation, and frame rotation enters essentially only in the
determination of
$\Sv_a^{(1)}$ from $\para{\EqD}^{(0)}_a$, or equivalently in the
determination of the total sectional forces since it is related through
\eqref{TxN}. In this regime, the fast rotation induces a sectional
force and its expression should be captured correctly by an angular
momentum balance equation thanks to \eqref{balancevolumeGamma2}, provided the last term
in~\eqref{SectionalForce} is correctly included. In the end, rod models take only some corrective terms, but if the system is considered in a
fast-rotating frame, these retained corrective terms should also be the
most important ones, and rod models should lead to a reliable extension of the viscous string model.

\section{Conclusion}

We have developed all the theoretical tools which are required to
obtain a general one-dimensional description of curved fibers. A
concrete application for toroidal viscous fibers is presented
separately in \citet{PitrouPRE2}. From a theoretical point of view, our $2+1$ splitting, our use of fiber adapted
coordinates, and more importantly our parametrization of the velocity field in terms of
STF tensors allow for a clear discussion about constraints and dynamical
equations. It avoids the cluttered component by component expressions
which are usual in such context~\citep{Wilmott1992}, and it
bears a more transparent geometrical meaning since all quantities are
analyzed in terms of their monopole, dipole, quadrupole, and higher
order multipoles. We find that it is the natural language which allows
to overcome the complexity of equations for curved viscous
fibers. Indeed, the use of an adapted formalism is the key to
understand in depth apparently complex problems. From a practical or computational point of view, this STF
based approach is very powerful as it is possible to handle tensors
with appropriate abstract tensor packages, and to this end we used
{\it xAct}~\citep{xAct}. The corresponding notebooks are available upon
request from the author.

The main results of this article are the following:
\begin{itemize}
\item We have recovered the standard results of axisymmetric fibers at
lowest order in \eqref{EvolRCircLow} and \eqref{EqF1DLow}, including
also axial rotation.
\item The first corrections for this model are collected in \eqref{dtAllCorrectionsAxi}.
\item We extended these results to include a second set of corrections
  and these can be found in Appendix~\ref{AppOrder2}.
\item The main purpose of this article was to develop a formalism for curved fibers and we first rederived the viscous string
model whose central equations are \eqref{AllEqskappalow}, \eqref{InertialTerms} and
\eqref{LowRadCurvedString}. Its covariant formulation is summarized in
\eqref{ConsCov} and \eqref{DynCov}.
\item We found the first corrections for curved fibers in full
  generality in \eqref{Finalv}, \eqref{FinalUa}, and
  \eqref{FirstRadCurvedString}, and these are relevant when
  $\epsilon_\ra$ is not so small since they are of order $\epsilon_\ra^2$.
\item Elliptic shape perturbations are sourced at that order and their dynamics is governed at that order by \eqref{dtRab}
with \eqref{Vh1ab} replaced. 
\item In particular, when restricting to straight fibers, the dynamical
  equation for the evolution of elliptic shape perturbations takes the
  simple form \eqref{dtRabStraight}.
\item Finally, when comparing with rod models methods, we have exhibited
a missing term in the expression \eqref{SectionalForce} for the torque
applied on fiber sections.
\end{itemize}

\begin{acknowledgments}

I would like to thank G. Faye for his help on the
irreducible representations of ${\rm SO}(2)$, and R. Gy and F. Vianey for
their encouragement to write this article. I also thank J. Eggers and
N. Ribe for comments on earlier versions of this article. This research was initiated when the author was working for Saint-Gobain Recherche.
\end{acknowledgments}

\ifpre
\else
\bibliographystyle{jfm}
\fi
\bibliography{BiblioJets}

\appendix

\section{STF formalism}\label{STF}

\subsection{Extraction of STF tensors}

Integrals on directions are simply
\bea\label{nnnIntegrals}
\int \frac{\dd \theta}{2\pi}\yu^{a_1}\dots \yu^{a_{2n}} &=&
\frac{(2n-1)!!}{(2n)!!} \delta^{(a_1 a_2}\dots
\delta^{a_{2n-1}a_{2n})}\prebool{\nonumber}{}\\
\int \frac{\dd \theta}{2\pi}\yu^{a_1}\dots \yu^{a_{2n+1}} &=&0\,.
\eea
Here, $\delta^{(a_1 a_2}\dots \delta^{a_{2n-1}a_{2n})}$ means that the indices need to be fully
symmetrized among the $(2n-1)!!$ possible permutations. The lowest non-vanishing integrals are

\begin{align}
\int\frac{\dd \theta}{2\pi}\yu^a \yu^b=&\frac{1}{2}\delta^{ab}\\
\int\frac{\dd \theta}{2\pi}\yu^a \yu^b \yu^c \yu^d=&\frac{1}{8}\left(\delta^{ab}\delta^{cd}+\delta^{ac}\delta^{bd}+\delta^{ad}\delta^{bc}\right)\,.
\end{align}
We recall that in general for STF tensors, we can use the multi-index notation $K \equiv a_1 \dots
a_k$ or $L \equiv b_1 \dots b_\ell$. However, in order to avoid confusion on multi-indices in this
section and the next one, we use the weaker multi-index notation $a_K\equiv a_1 \dots a_{k}$ or $b_L\equiv b_1 \dots b_{\ell}$.
Let us define 
\be\label{IJtensor}
I^{a_L}_{b_L} \equiv \delta^{a_1}_{\langle b_1} \dots
\delta^{a_\ell}_{b_\ell \rangle}\,,\myquad J^{a_L}_{b_L} \equiv
{\epsilon^{a_1}}_{\langle b_1}\delta^{a_2}_{b_2} \dots
\delta^{a_\ell}_{b_\ell \rangle}\,,
\ee
where we recall that when indices are enclosed in $\langle \dots
\rangle$ we must take the symmetric trace-free part. Clearly, these quantities are
just related by
\be
{\epsilon^{a_1}}_c\, I^{c a_{L-1}}_{b_L} = J^{a_L}_{b_L} \,,\qquad {\epsilon^{a_1}}_c\, J^{c a_{L-1}}_{b_L} =- I^{a_L}_{b_L} \,.
\ee
They have the interesting properties
\beas
I^{c a_{L-1}}_{c b_{L-1}} &=& I^{a_{L-1}}_{b_{L-1}} \qquad\,\, I^a_a=2 \qquad
I^a_b {\epsilon_a}^b=0\\
J^{c a_{L-1}}_{c b_{L-1}} &=& J^{a_{L-1}}_{b_{L-1}}\qquad J^a_a=0
\qquad J^a_b
{\epsilon_a}^b=2\,,
\eeas
\be
I^{a_L}_{b_L} I^{b_L}_{a_L} = 2\,,\myquad
J^{a_L}_{b_L} J^{b_L}_{a_L} =2\,,\myquad
I^{a_L}_{b_L} J^{b_L}_{a_L} = 0\,.
\ee
The tensors~\eqref{IJtensor} are used when computing the following
integrals on direction vectors
\beas\label{nnIntegrals}
2^\ell \int\frac{\dd \theta}{2\pi} n^{a_L} n_{\langle b_K
  \rangle} &=&\delta^{\ell}_k\,  I^{a_L}_{b_L}\\
2^\ell \int\frac{\dd \theta}{2\pi} {\epsilon^a}_{c}n^c n^{a_{L-1}} n_{\langle b_K
  \rangle} &=& \delta^{\ell}_k\, J^{a_L}_{b_L}\,.
\eeas

It is then immediate to show that for a scalar function expanded in STF tensors as
\be
S=\sum_\ell S_L n^{\langle L \rangle}
\ee
then the STF moments can be extracted through
\be
S_L =2^{-\ell} \int \frac{\dd \theta}{2 \pi} n_{\langle L \rangle} S\,.
\ee
This type of integral is very well suited for a tensor computer algebra system such as
{\it xAct}~\citep{xAct} since we need only to implement the rules~\eqref{nnnIntegrals}.

\subsection{Products of STF tensor}\label{AppOrdSTF}

As explained in~\S~\ref{SecIrreps}, STF tensors in two-dimensions are
irreducible representations of ${\rm SO}(2)$. If the tensor has $\ell$
indices, then it is in the representation $D_\ell$. When we have a product of two STF tensors of rank $\ell$ and $\ell'$, it means we have the tensor
product of the representations $D_\ell \otimes
D_{\ell'}$. This is not irreducible, but it can be  decomposed in irreducible
representations. Given that the dimension of $D_\ell$ is $2$ (except for
$D_0$ for which the dimension is 1), then this tensor product is of dimension $2\times 2
= 4$. When decomposed in irreducible representations it is either of
the form $D_{|\ell-\ell'|}\oplus D_{\ell+\ell'}$ if
$\ell\neq \ell'$, or $D_{\ell+\ell'}\oplus D_0 \oplus D_0$ if
$\ell=\ell'$. When counting the dimensions, it is the statement that
$2\times2=2+2$ in the former case, and $2\times 2=2+1+1$ in the latter case.

In order to see in practice how this decomposition is performed, let
us consider two STF tensors $A_K$ and $B_L$. 
If we assume first that $k<\ell$, then\be\label{ABgeneral}
A_{a_K} B_{b_L} = A_{\langle a_K} B_{b_L\rangle } + A^{c_K} B_{c_K  \langle b_{L-K}} I^{a_K}_{b_K \rangle }\,.
\ee
Under this form, we have indeed decomposed the product into two
irreducible parts $D_{|k-\ell|}$ and $D_{|k+\ell|}$ (that is
$2\times2=2+2$), which are respectively the STF tensors $A^{c_K} B_{c_K
  \langle b_{L-K} \rangle}$ and $A_{\langle a_K} B_{b_L\rangle }$.

However, if the tensors are of equal rank ($k=\ell$), this gets
slightly different since
\prebool{\bea\label{ABequal}
A_{a_L} B_{b_L} &=& A_{\langle a_L} B_{b_L \rangle} +\frac{1}{2} A^{c_L}
B_{c_L} I^{a_L}_{b_L} \nonumber\\
&&+\frac{1}{2} {\epsilon^{c }}_{d} A_{c
  c_{L-1}} B^{d c_{L-1}} J^{a_L}_{b_L}\,.
\eea}
{\bea\label{ABequal}
A_{a_L} B_{b_L} &=& A_{\langle a_L} B_{b_L \rangle} +\frac{1}{2} A^{c_L}
B_{c_L} I^{a_L}_{b_L} +\frac{1}{2} {\epsilon^{c }}_{d} A_{c
  c_{L-1}} B^{d c_{L-1}} J^{a_L}_{b_L}\,.
\eea}
In that case we have decomposed the product as a sum of two scalar functions
(both corresponding to the representation $D_0$) and an element of $D_{2\ell}$  (that is $2\times2 =
2+1+1$) which are respectively  $A^{c_L}
B_{c_L} $,  ${\epsilon^{c }}_{d} A_{c c_{L-1}} B^{d c_{L-1}}$ and  the STF tensor $A_{\langle a_L} B_{b_L \rangle} $.

In both cases we get (with $k\le\ell$)
\be\label{AbyyMagic}
A_{K}B_Ly^K y^L = A_{\langle K} B_{L \rangle} y^K y^L + \frac{r^{2k}}{2^k}
A^K B_{K c_{L-K}}y^{c_{L-K}}.
\ee
If we consider the expansion~\eqref{DefVaExpand}, we can first remove
the traces in the $L$ indices to recast the expansion as
\be
V_a(y^i,t) = \sum_{\ell=0}^\infty \sum_{n=0}^\infty  V^{(n)}_{a \langle L
  \rangle}(s,t) y^L r^{2n}\,.
\ee
Each $V^{(n)}_{a\langle L\rangle}$ can be handled exactly as a product
of tensors $A_a$ and $B_L$. If $\ell > 1$ we
can use the decomposition~\eqref{ABgeneral}, but if $\ell=1$ we must
use \eqref{ABequal}. Combining all the terms in the
sum~\eqref{DefVaExpand} we finally conclude that the decomposition of
a $2$-vector field in terms of irreps is necessarily of the form~\eqref{IrrepsVa}.

\section{Alternate shape representation}\label{AppAltShape}

We can consider the following STF moments
\be\label{DefML}
{\cal M}_{L} \equiv 2^\ell \int y_{\langle L \rangle} \bar \rho(y^b)  \dd^2 y^b\,,
\ee
which are integrals on the fiber section which should capture its
shape. $\bar \rho$ is a step function which is unity if there is a fluid
particle and vanishes otherwise. These moments are built just like the
material moments of constant density extended objects, or like the
electric moments of uniformly charged extended objects~\citep{Jackson}. These moments can be related to the
moments of radial dimensions $\rb_L$ defined in \eqref{Defrb}, but
the relation is non linear. To see this, let us change variables and define rescaled coordinates
\be
z^a \equiv \frac{y^a}{1+\hrb_L \yu^L}\qquad y^a = z^a (1+\hrb_L \yu^L)\,.
\ee
The Jacobian of the transformation is
\be
\dd^2 y^b =J \dd^2 z^b\,,\qquad J=\left(1 + \hrb_L \yu^L\right)^2 
\ee
and the integrals~\eqref{DefML} are recast as
\be\label{RelMultipoles}
{\cal M}_{L} = 2^\ell \frac{2\pi \ra^{\ell+2}}{\ell+2}\int
\left(1+\hrb_M \yu^M\right)^{\ell+2} \yu_{\langle L \rangle} \frac{\dd \theta}{2\pi}\,.
\ee
For the monopole we obtain
\be
{\cal M}_\myes = \pi \ra^2 (1+ \sum_{\ell=1}^\infty 2^{-\ell }\hrb_L
\hrb^L)\simeq \pi \ra^2\,,
\ee
which is just the area of the section. If $\ell > 0$ then the other geometric multipoles are simply
approximated by (keeping only linear terms)
\be
{\cal M}_{L} \simeq 2\pi \ra^{\ell+2} \hrb_{L}= 2\pi \ra^{2(\ell+1)} \rb_{L}\,,
\ee
where \eqref{nnIntegrals} was used to compute the integral \eqref{RelMultipoles}.

The kinematic equation, giving the evolution in time of these moments
is found from the conservation equation of the density function $\bar \rho$
\be
\partial_t \bar \rho+\totV_R^i \partial_i
\bar \rho =\partial_t \bar \rho +\totV_R^3 \partial_s \bar \rho + \totV_R^a \partial_a
\bar \rho = 0\,.
\ee
Indeed integrating over directions as in~\eqref{DefML}, and after
integrations by parts, we get simply
\prebool{\bea
\partial_t {\cal M}^{L} &=&- \int y^{\langle L
  \rangle}( \totV_R^3 \partial_s \bar \rho  -\bar\rho \partial_a \totV_R^a)
\,\dd^2 y\nonumber\\
&&+\ell \int y^{\langle a_{L-1}}\totV_R^{a_\ell \rangle} \bar \rho\dd^2 y\,.
\eea}
{\bea
\partial_t {\cal M}^{L} &=&- \int y^{\langle L
  \rangle}( \totV_R^3 \partial_s \bar \rho  -\bar\rho \partial_a \totV_R^a)
\,\dd^2 y+\ell \int y^{\langle a_{L-1}}\totV_R^{a_\ell \rangle} \bar \rho\dd^2 y\,.
\eea} 
$\totV_R^a$ needs to be expressed in terms of its
multipoles. To this end, we first express it in terms of $V^a$ from the relation
\eqref{VRVC2}, and then use the expansion~\eqref{IrrepsVa} for $V^a$. As for $\totV_R^3$, we should first  use
that $\totV_R^3 =\totV_R^\tetr{3}/h$, expand $1/h=1/(1+\tkappa_a y^a)$
which brings increasing powers of $\tkappa_a y^a$, and then we should
relate it to $\para{V}$ from \eqref{VRVC2} so as to use the expansion
\eqref{ScalarSTF} for $\para{V}$. The angular integrals can
then be performed with \eqref{nnIntegrals}. The resulting
dynamical equations for these shape multipoles ${\cal M}^{L}$ are
rather complicated because of the high powers in the curvature
vector $\kappa^a$, but they are linear in both the velocity multipoles
and the shape multipoles ${\cal M}^{L}$. Instead, the dynamical
equation~\eqref{EqPropaR} was still linear in velocity multipoles, but
extremely non-linear in the shape multipoles $\rb_L$. Hence, it is
not surprising that the relation between the two types of
multipoles~\eqref{RelMultipoles} is very non-linear. Since the
normal vector and thus the extrinsic curvature are more easily
expressed with the multipoles $\rb_L$ as seen on~\eqref{NormalVector},
we chose to work with these multipoles so as to be able to include
surface tension effects.

\section{Velocity of the coincident point}\label{AppGalilee}

Let us define the space-time Cartesian coordinates $X^{\hat{\mu}}=(t,x^\mu)$ with
$\hat{\mu}=0,1,2,3$ and the space-time fiber adapted coordinates $Y^{\hat{\imath}}=(t,y^{\hat{\imath}})$ with $\hat{\imath} = 0,1,2,3$. Then each coordinate system is
  a function of the other one, that is we have the functions $X^{\hat{\mu}}(Y^{\hat{\imath}})$ and $Y^{\hat{\imath}}(X^{\hat{\mu}})$ which are related by
\be
\frac{\partial X^{\hat{\mu}}}{\partial Y^{\hat{\imath}}}
\frac{\partial Y^{\hat{\imath}}}{\partial X^{\hat{\nu}}} =
\delta^{\hat\mu}_{\hat\nu}\,,\myquad \frac{\partial
  Y^{\hat{\imath}}}{\partial X^{\hat{\mu}}} \frac{\partial
  X^{\hat{\mu}}}{\partial Y^{\hat{\jmath}}} = \delta^{\hat{\imath}}_{\hat{\jmath}}\,.
\ee
In particular, using the first relation for $\hat{\mu}=\mu$ and
$\hat{\nu} = 0$, and the second relation with $\hat{\imath}=i$ and
$\hat{\jmath}=0$ we get
\beas
0&=&\left.\frac{\partial x^\mu}{\partial t}\right|_{y}+\left.\frac{\partial
  x^\mu}{\partial y} \frac{\partial y^i}{\partial
  t}\right|_{x}=\left.\frac{\partial x^\mu}{\partial t}\right|_{y}+\left.{d_i}^\mu \frac{\partial y^i}{\partial
  t}\right|_{x}\\
0&=&\left.\frac{\partial y^i}{\partial t}\right|_{x}+\left.\frac{\partial
  y^i}{\partial x^\mu} \frac{\partial x^\mu}{\partial t}\right|_{y}
= \left.\frac{\partial y^i}{\partial t}\right|_{x}+\left.{d^i}_\mu\frac{\partial x^\mu}{\partial t}\right|_{y}
\eeas
and we recover~\eqref{VCDef}.

\section{Cartan structure relation}\label{AppCartan}

In this appendix, we build on the four dimensional perspective of the
previous section. Let us define the space-time tetrad 
\be
d_\tetr{\hat{\imath}} = (d_0,d_\tetr{i}). 
\ee
It is made from the spatial orthonormal basis on
which we have added a time directed vector 
\be
d_0^{\hat{\mu}} = \delta_0^{\hat{\mu}}, \myquad\Rightarrow\myquad  \partial_s d_0=\partial_t d_0=0\,.
\ee 
Let us define the infinitesimal rotation matrices
\be
[J_i]_{jk} \equiv \eta_{ijk}\myquad \Rightarrow \myquad [J_i , J_j] = -\eta_{ijk} J_k
\ee
where $\eta_{ijk}$ is the permutation symbol with $\eta_{123} = 1$ and
where the sum on the index $k$ is implied. We can define an operator valued (rotation valued) one-form in the
four dimensional classical space-time by
\be
{\bm \Omega}\equiv \kappa^\tetr{i} J_i \,\dd s+ \omega^\tetr{i} J_i \,\dd t\,.
\ee
The components of this form in the Cartesian canonical basis are
\be
{\bm \Omega}= {\Omega}_{\hat{\mu}} \dd x^{\hat{\mu}}\qquad {\Omega}_{\hat{\mu}}\equiv \kappa^\tetr{i} J_i \delta^3_{\hat{\mu}}+ \omega^\tetr{i} J_i \delta^0_{\hat{\mu}}\,.
\ee
This form is clearly the connection form of the Cartan formalism since
(still with the sum on $k$ implied)
\prebool{\bea
&&\partial_s d_{\tetr{{j}}}  =[{\Omega}_{3}]_{jk} d_{\tetr{{k}}} \Leftrightarrow\partial_s
d_{\tetr{{j}}} = \kappa^\tetr{i} [J_i]_{jk} d_{\tetr{{k}}}\Leftrightarrow \partial_s
d_{\tetr{{j}}}  = \kappa \times d_{\tetr{{j}}} \nonumber\\
&&\partial_t d_{\tetr{{j}}}  =[{\Omega}_{0}]_{jk} d_{\tetr{{k}}} \Leftrightarrow\,\, \partial_t
d_{\tetr{{j}}} = \omega^\tetr{i} [J_i]_{jk} d_{\tetr{{k}}}\Leftrightarrow\partial_t
d_{\tetr{{j}}}  = \omega \times d_{\tetr{{j}}} \nonumber
\eea}
{\beas
&&\partial_s d_{\tetr{{j}}}  =[{\Omega}_{3}]_{jk} d_{\tetr{{k}}} \Leftrightarrow\partial_s
d_{\tetr{{j}}} = \kappa^\tetr{i} [J_i]_{jk} d_{\tetr{{k}}}\Leftrightarrow \partial_s
d_{\tetr{{j}}}  = \kappa \times d_{\tetr{{j}}}\\
&&\partial_t d_{\tetr{{j}}}  =[{\Omega}_{0}]_{jk} d_{\tetr{{k}}} \Leftrightarrow\,\, \partial_t
d_{\tetr{{j}}} = \omega^\tetr{i} [J_i]_{jk} d_{\tetr{{k}}}\Leftrightarrow\partial_t
d_{\tetr{{j}}}  = \omega \times d_{\tetr{{j}}} 
\eeas}
which in a four dimensional perspective reads exactly as the first Cartan
structure equation
\be
\partial_{\hat{\mu}} d_{\tetr{\hat{\jmath}}} =
[{\Omega}_{\hat{\mu}}]_{{\hat{\jmath}} {\hat{k}}} d_{\tetr{\hat{k}}} \,.
\ee

Then, since the classical space-time is flat, the second Cartan
structure equation reads as~\citep{Nakahara}
\be
\dd {\bm \Omega} + {\bm \Omega} \wedge {\bm \Omega}  = 0\,.
\ee
Expressed explicitly in a basis of two-forms $\dd y^{\hat \imath}
\wedge \dd y^{\hat \jmath}=\dd y^{\hat \imath} \otimes \dd y^{\hat
  \jmath}-\dd y^{\hat \jmath} \otimes \dd y^{\hat \imath}$, the terms
of this equation read as
\beas\label{dOOO}
\dd {\bm \Omega } &=& \mytfrac{1}{2}\left[\partial_t \kappa^\tetr{i}
  -\partial_s \omega^\tetr{i} \right]J_i\,\dd t \wedge \dd s\prebool{\nonumber}{}\\
{\bm \Omega } \wedge {\bm \Omega }  &=&\mytfrac{1}{2}\omega^\tetr{i}
\kappa^\tetr{j}[J_i,J_j]\dd t \wedge \dd s=\mytfrac{1}{2}[\kappa \times
\omega]^\tetr{i} J_i\,\dd t \wedge \dd s\prebool{\nonumber}{}
\eeas
and therefore we recover the structure relation~\eqref{RiemannCurvature}.

\allowdisplaybreaks

\ifpre
\begin{widetext}
\else
\fi

\section{Second set of corrections for axisymmetric viscous fibers}\label{AppOrder2}

As discussed in~\S~\ref{SecStringStraight}, the dynamical equation
\eqref{dtwlowAxi} is part of the first set of corrections, implying
that \eqref{dtwCorrectionsAxi} is in fact part of the second set of corrections,
so we only report the second corrections for $\partial_t \Sv$
and $\partial_t \ln \ra$. The second corrections to the dynamical equation for $\Sv$ are
\begin{align}\label{Eqdv2}
\partial_t \Sv \supset&\ra^4\Bigl[
- \mytfrac{1}{2} \mathcal{H}^3 \Sw^2
 -  \mytfrac{1}{4} \mathcal{H} \Sw^2 \partial_s\mathcal{H}
 + 48 \mathcal{H}^5 \partial_s\Sv
 + \mytfrac{543}{4} \mathcal{H}^3 \partial_s\mathcal{H} \partial_s\Sv
 + \mytfrac{453}{8} \mathcal{H} (\partial_s\mathcal{H})^2 \partial_s\Sv\nonumber\\
& \qquad
 + \mytfrac{117}{8} \mathcal{H}^3 (\partial_s\Sv)^2
+ \mytfrac{147}{16} \mathcal{H} \partial_s\mathcal{H} (\partial_s\Sv)^2
 + \mytfrac{1}{2} \mathcal{H} (\partial_s\Sv)^3
 -  \mytfrac{11}{4} \mathcal{H}^2 \Sw \partial_s\Sw
 -  \mytfrac{1}{6} \Sw \partial_s\mathcal{H} \partial_s\Sw\nonumber\\
& \qquad
 -  \mytfrac{7}{6} \mathcal{H} (\partial_s\Sw)^2
 + \mytfrac{315}{8} \mathcal{H}^2 \partial_s\Sv \partial_s^2\mathcal{H}
+ 12 \partial_s\mathcal{H} \partial_s\Sv \partial_s^2\mathcal{H}
 + \mytfrac{5}{16} (\partial_s\Sv)^2 \partial_s^2\mathcal{H}
 + \mytfrac{321}{4} \mathcal{H}^4 \partial_s^2\Sv\nonumber\\
& \qquad
 + \mytfrac{1}{24} \Sw^2 \partial_s^2\Sv
 + \mytfrac{951}{8} \mathcal{H}^2 \partial_s\mathcal{H} \partial_s^2\Sv
 + \mytfrac{33}{2} (\partial_s\mathcal{H})^2 \partial_s^2\Sv 
+ \mytfrac{327}{16} \mathcal{H}^2 \partial_s\Sv \partial_s^2\Sv
 + \mytfrac{45}{16} \partial_s\mathcal{H} \partial_s\Sv \partial_s^2\Sv\nonumber\\
& \qquad
 + \mytfrac{17}{32} (\partial_s\Sv)^2 \partial_s^2\Sv
 + \mytfrac{339}{16} \mathcal{H} \partial_s^2\mathcal{H} \partial_s^2\Sv
 + \mytfrac{67}{16} \mathcal{H} (\partial_s^2\Sv)^2
 -  \mytfrac{7}{6} \mathcal{H} \Sw \partial_s^2\Sw 
 -  \mytfrac{19}{48} \partial_s\Sw \partial_s^2\Sw\nonumber\\
& \qquad
 + \mytfrac{93}{16} \mathcal{H} \partial_s\Sv \partial_s^3\mathcal{H}
 + \mytfrac{15}{8} \partial_s^2\Sv \partial_s^3\mathcal{H}
 + \mytfrac{135}{4} \mathcal{H}^3 \partial_s^3\Sv
 + \mytfrac{207}{8} \mathcal{H} \partial_s\mathcal{H} \partial_s^3\Sv
 + \mytfrac{7}{2} \mathcal{H} \partial_s\Sv \partial_s^3\Sv \nonumber\\
& \qquad
 + \mytfrac{47}{16} \partial_s^2\mathcal{H} \partial_s^3\Sv
 + \mytfrac{33}{64} \partial_s^2\Sv \partial_s^3\Sv
 -  \mytfrac{5}{48} \Sw \partial_s^3\Sw
 + \mytfrac{15}{32} \partial_s\Sv \partial_s^4\mathcal{H}
 + \mytfrac{63}{16} \mathcal{H}^2 \partial_s^4\Sv\nonumber\\
& \qquad
 + \mytfrac{15}{8} \partial_s\mathcal{H} \partial_s^4\Sv
 -  \mytfrac{3}{64} \partial_s\Sv \partial_s^4\Sv+ \mytfrac{1}{48} \partial_s^6\Sv\nonumber\\
&\qquad+\ST \Bigl(\mytfrac{41}{8} \mathcal{H}^3 \partial_s^2{\cal K}
 + \mytfrac{53}{16} \mathcal{H} \partial_s\mathcal{H} \partial_s^2{\cal K}
 + \mytfrac{3}{8} \mathcal{H} \partial_s\Sv \partial_s^2{\cal K}
 + \mytfrac{1}{4} \partial_s^2\mathcal{H} \partial_s^2{\cal K}
 + \mytfrac{1}{8} \partial_s^2\Sv \partial_s^2{\cal K}\nonumber\\
& \qquad\quad
 + \mytfrac{65}{16} \mathcal{H}^2 \partial_s^3{\cal K}
 -  \mytfrac{1}{16} \mathcal{H} \Sv \partial_s^3{\cal K}
+ \mytfrac{3}{4} \partial_s\mathcal{H} \partial_s^3{\cal K}
 + \mytfrac{11}{96} \partial_s\Sv \partial_s^3{\cal K}
 -  \mytfrac{1}{16} \mathcal{H} \partial_t{\partial_s^2{\cal K}}\nonumber\\
& \qquad\quad
 + \mytfrac{29}{32} \mathcal{H} \partial_s^4{\cal K}
 -  \mytfrac{1}{48} \Sv \partial_s^4{\cal K}
 -  \mytfrac{1}{48} \partial_t{\partial_s^3{\cal K}}
 + \mytfrac{3}{64} \partial_s^5{\cal K}\Bigr)\Bigr]\,.
\end{align}
As for the radius evolution, it should be corrected at that order by
\begin{align}\label{EqdR2}
\partial_t \ln \ra \supset&\ra^4\Bigl[ 
 - \mytfrac{9}{4} \mathcal{H}^4 \partial_s\Sv
 -  \mytfrac{63}{16} \mathcal{H}^2 \partial_s\mathcal{H} \partial_s\Sv
 -  \mytfrac{15}{32} (\partial_s\mathcal{H})^2 \partial_s\Sv
 -  \mytfrac{3}{8} \mathcal{H}^2 (\partial_s\Sv)^2
 -  \mytfrac{1}{16} \partial_s\mathcal{H} (\partial_s\Sv)^2\nonumber\\
&\qquad
 + \mytfrac{1}{8} \mathcal{H} \Sw \partial_s\Sw
+ \mytfrac{1}{48} (\partial_s\Sw)^2
 -  \mytfrac{3}{4} \mathcal{H} \partial_s\Sv \partial_s^2\mathcal{H}
 -  \mytfrac{51}{16} \mathcal{H}^3 \partial_s^2\Sv
 -  \mytfrac{63}{32} \mathcal{H} \partial_s\mathcal{H} \partial_s^2\Sv\nonumber\\
&\qquad
 -  \mytfrac{13}{32} \mathcal{H} \partial_s\Sv \partial_s^2\Sv
 -  \mytfrac{9}{64} \partial_s^2\mathcal{H} \partial_s^2\Sv
 -  \mytfrac{3}{64} (\partial_s^2\Sv)^2
+ \mytfrac{1}{48} \Sw \partial_s^2\Sw
 -  \mytfrac{3}{64} \partial_s\Sv \partial_s^3\mathcal{H}\nonumber\\
&\qquad
 -  \mytfrac{27}{32} \mathcal{H}^2 \partial_s^3\Sv
 -  \mytfrac{5}{32} \partial_s\mathcal{H} \partial_s^3\Sv
 -  \mytfrac{3}{64} \partial_s\Sv \partial_s^3\Sv
 -  \mytfrac{1}{64} \mathcal{H} \partial_s^4\Sv
 + \mytfrac{1}{128} \partial_s^5\Sv\nonumber\\
&\qquad-\ST \Bigl(\mytfrac{3}{16} \mathcal{H}^2 \partial_s^2{\cal K}
 +  \mytfrac{1}{32} \partial_s\mathcal{H} \partial_s^2{\cal K}
 +  \mytfrac{3}{32} \mathcal{H} \partial_s^3{\cal K}
 +  \mytfrac{1}{96} \partial_s^4{\cal K}\Bigr)\Bigr]\,.
\end{align}

\section{Higher order constraints for curved fibers}\label{SecExtraConstraint}

From the boundary constraint~\eqref{VectorBoundaryConstraint} we get
the additional contributions
\begin{align}
\stf{V}^{(1)}_a\supset&\ra^2\Bigl(
- \mytfrac{12}{5} {{\stf{V}}^{(2)}_{a}} 
- 3 {\cal H} {{\Sv}^{(1)}_{a}} 
-  \mytfrac{21}{80} \varepsilon_{a}{}^{c} {{\stf{V}}^{(1)}_{bc}}\kappa^{b} 
+ \mytfrac{7}{2} {\cal H}^2 \kappa_{a} {\Sw} 
+ \mytfrac{3}{128} \kappa_{a} \kappa_{b} \kappa^{b} {\Sw} \nonumber\\
&+ \mytfrac{63}{40} {\cal H} \kappa^{b} \widetilde{\kappa}_{a}{\omega}_{b} 
-  \mytfrac{63}{40} {\cal H} \kappa_{b} \kappa^{b}\widetilde{\omega}_{a} 
+ \mytfrac{5}{8} \kappa_{a} {\Sw} \partial_s {\cal H} 
-  \mytfrac{19}{4} {\cal H}^2 \widetilde{\kappa}_{a} \partial_s{{\Sv}} 
-  \mytfrac{3}{256} \kappa_{b} \kappa^{b}\widetilde{\kappa}_{a} \partial_s{{\Sv}} \nonumber\\
&-  \mytfrac{31}{32} \widetilde{\kappa}_{a} \partial_s {\cal  H} \partial_s{{\Sv}} 
-  \mytfrac{3}{5} \partial_s{{{\Sv}^{(1)}_{a}}} 
+ {\cal H} {\Sw} \partial_s{\kappa_{a}} 
+ \mytfrac{7}{2} {\cal H}^2 {\Sv} \partial_s{\widetilde{\kappa}_{a}} 
+ \mytfrac{3}{128} {\Sv} \kappa_{b} \kappa^{b} \partial_s{\widetilde{\kappa}_{a}} \nonumber\\
&+ \mytfrac{5}{8} {\Sv} \partial_s {\cal H} \partial_s{\widetilde{\kappa}_{a}} 
+ \mytfrac{1}{8} {\cal  H} \partial_s{{\Sv}} \partial_s{\widetilde{\kappa}_{a}} 
+ \mytfrac{103}{40} {\cal H} \kappa_{a} \partial_s{{\Sw}}
+ \mytfrac{7}{2} {\cal H}^2 \partial_s{\widetilde{\omega}_{a}} 
+ \mytfrac{3}{128} \kappa_{b} \kappa^{b} \partial_s{\widetilde{\omega}_{a}} \nonumber\\
&+ \mytfrac{5}{8} \partial_s {\cal H} \partial_s{\widetilde{\omega}_{a}}
 - \mytfrac{1}{16} \partial_s{\widetilde{\kappa}_{a}} \partial^2_s{{\Sv}} 
+ {\cal H} {\Sv} \partial^2_s{\widetilde{\kappa}_{a}} 
+ {\cal H} \partial^2_s{\widetilde{\omega}_{a}}  
-  \mytfrac{53}{32} {\cal H}\widetilde{\kappa}_{a} \partial^2_s{{\Sv}} 
-  \mytfrac{7}{64} \widetilde{\kappa}_{a} \partial^3_s{{\Sv}}\Bigr)\\
P^{(0)}_\myes\supset&-\ST \ra\left(\mytfrac{3}{2}{\cal
    H}^2+\mytfrac{1}{2}\kappa_a \kappa^a+\partial_s{\cal H}\right)\nonumber\\
&+\ra^2 \Bigl(- {P^{(1)}_\myes}
 - 3 {\cal H}^2 \partial_s{{\Sv}}
 -  \mytfrac{1}{4} \kappa_{a} \kappa^{a} \partial_s{{\Sv}}
 -  \mytfrac{9}{4} \partial_s {\cal H} \partial_s{{\Sv}}
 + \mytfrac{1}{2} {\Sv} \kappa^{a} \partial_s{\kappa_{a}}
 + \mytfrac{1}{2} \kappa^{a} \partial_s{{\omega}_{a}}\nonumber\\
& \qquad -  \mytfrac{9}{4} {\cal H} \partial^2_s{{\Sv}}
 -  \mytfrac{3}{8} \partial^3_s{{\Sv}}\Bigr)\\
P^{(0)}_a\supset&\ST\ra\Bigl(- \mytfrac{3}{2} \varepsilon_{a}{}^{c} \mathcal{R}_{bc} \kappa^{b}
 + \mytfrac{7}{2} {\cal H}^2 \widetilde{\kappa}_{a}
 + \mytfrac{3}{4} \kappa_{b} \kappa^{b} \widetilde{\kappa}_{a}
 + 2 \widetilde{\kappa}_{a} \partial_s {\cal H}
 + {\cal H} \partial_s{\widetilde{\kappa}_{a}}\Bigr)\\
+&\ra^2\Bigl(- {P^{(1)}_{a}}
 -  \mytfrac{8}{5} {{\stf{V}}^{(2)}_{a}}
 + 4 {\cal H} {{\Sv}^{(1)}_{a}}
 + \mytfrac{9}{20} \varepsilon_{a}{}^{c} {{\stf{V}}^{(1)}_{bc}} \kappa^{b}
 - 5 {\cal H}^2 \kappa_{a} {\Sw}
 -  \mytfrac{15}{32} \kappa_{a} \kappa_{b} \kappa^{b} {\Sw}\nonumber\\
&\qquad 
 + \mytfrac{13}{10} {\cal H} \kappa^{b} \widetilde{\kappa}_{a} {\omega}_{b}
 -  \mytfrac{13}{10} {\cal H} \kappa_{b} \kappa^{b}
 \widetilde{\omega}_{a}
-  \mytfrac{5}{2} \kappa_{a} {\Sw} \partial_s {\cal H}
 + \mytfrac{17}{2} {\cal H}^2 \widetilde{\kappa}_{a} \partial_s{{\Sv}}
 + \mytfrac{31}{64} \kappa_{b} \kappa^{b} \widetilde{\kappa}_{a} \partial_s{{\Sv}}\nonumber\\
&\qquad 
 + \mytfrac{43}{8} \widetilde{\kappa}_{a} \partial_s {\cal H} \partial_s{{\Sv}}
 + \mytfrac{8}{5} \partial_s{{{\Sv}^{(1)}_{a}}}
 - 2 {\cal H} {\Sw} \partial_s{\kappa_{a}}
 -  \mytfrac{1}{2} {\Sv} \kappa^{b}
 \widetilde{\kappa}_{a} \partial_s{\kappa_{b}}
- 5 {\cal H}^2 {\Sv} \partial_s{\widetilde{\kappa}_{a}}\nonumber\\
&\qquad 
 -  \mytfrac{15}{32} {\Sv} \kappa_{b} \kappa^{b} \partial_s{\widetilde{\kappa}_{a}}
 -  \mytfrac{5}{2} {\Sv} \partial_s {\cal H} \partial_s{\widetilde{\kappa}_{a}}
 + \mytfrac{1}{2} {\cal H} \partial_s{{\Sv}} \partial_s{\widetilde{\kappa}_{a}}
 -  \mytfrac{7}{10} {\cal H} \kappa_{a} \partial_s{{\Sw}}
 -  \mytfrac{1}{2} \kappa^{b} \widetilde{\kappa}_{a} \partial_s{{\omega}_{b}}\nonumber\\
&\qquad 
 - 5 {\cal H}^2 \partial_s{\widetilde{\omega}_{a}}
-  \mytfrac{15}{32} \kappa_{b} \kappa^{b} \partial_s{\widetilde{\omega}_{a}}
 -  \mytfrac{5}{2} \partial_s {\cal H} \partial_s{\widetilde{\omega}_{a}}
 + \mytfrac{41}{8} {\cal H} \widetilde{\kappa}_{a} \partial^2_s{{\Sv}}
 + \mytfrac{1}{4} \partial_s{\widetilde{\kappa}_{a}} \partial^2_s{{\Sv}}\nonumber\\
&\qquad 
 - 2 {\cal H} {\Sv} \partial^2_s{\widetilde{\kappa}_{a}}
 - 2 {\cal H} \partial^2_s{\widetilde{\omega}_{a}}
 + \mytfrac{11}{16} \widetilde{\kappa}_{a} \partial^3_s{{\Sv}}\Bigr)\nonumber\,.
\end{align}

From the higher order of the Navier-Stokes equation that we consider
as constraints, we get
\begin{align}
\Sv^{(2)}=&\frac{\ST}{\ra}\left(
- \mytfrac{1}{16} {\cal H}^3
 -  \mytfrac{5}{64} {\cal H} \kappa_{a} \kappa^{a}
 -  \mytfrac{7}{64} \kappa^{a} \partial_s{\kappa_{a}}
 + \mytfrac{1}{32} \partial_s^{2} {\cal H}\right) \nonumber \\
&
- \mytfrac{3}{16} {\cal H} (\partial_s{{\Sv}})^2
 -  \mytfrac{1}{64} \para{g} \kappa_{a} \kappa^{a}
 + \mytfrac{3}{32} {\cal H} ({\Sv})^2 \kappa_{a} \kappa^{a}
 + \mytfrac{3}{16} {\cal H} {\Sv} \kappa^{a} {\Omega}_{a}
 + \mytfrac{3}{32} {\cal H} {\Omega}_{a} {\Omega}^{a}\nonumber\\
& + \mytfrac{1}{32} U^{a} \kappa^{b} \widetilde{\kappa}_{a} {\Omega}_{b}
 + \mytfrac{1}{64} \kappa_{a} \kappa^{a} \overline{I}
 + \mytfrac{3}{32} \kappa^{a} {\Sw} \widetilde{\Omega}_{a}
 -  \mytfrac{1}{64} \kappa^{a} \para{\Omega} \widetilde{\Omega}_{a}
 + \mytfrac{1}{32} U^{a} \kappa_{a} \kappa^{b} \widetilde{\Omega}_{b}\nonumber\\
& + \mytfrac{3}{16} {\cal H} {\Sv} \kappa^{a} {\omega}_{a}
 -  \mytfrac{1}{32} \widetilde{\kappa}^{a} {\Sw} {\omega}_{a}
 + \mytfrac{3}{32} \widetilde{\kappa}^{a} \para{\Omega} {\omega}_{a}
 + \mytfrac{3}{16} {\cal H} {\Omega}^{a} {\omega}_{a}
 + \mytfrac{3}{32} {\cal H} {\omega}_{a} {\omega}^{a}
 + \mytfrac{1}{64} \widetilde{\kappa}^{a} {\omega}_{a} \bar{\omega}\nonumber\\
& + \mytfrac{1}{64} U^{a} \kappa_{b} \kappa^{b} \widetilde{\omega}_{a}
 -  \mytfrac{9}{64} {\cal H} \kappa_{a} \kappa^{a} \partial_s{{\Sv}}
 + \mytfrac{7}{64} {\Sv} \kappa_{a} \kappa^{a} \partial_s{{\Sv}}
 + \mytfrac{7}{64} \kappa^{a} {\Omega}_{a} \partial_s{{\Sv}}
 + \mytfrac{7}{64} \kappa^{a} {\omega}_{a} \partial_s{{\Sv}}\nonumber\\
& + \mytfrac{9}{16} {\cal H} \partial_s {\cal H} \partial_s{{\Sv}}
 + \mytfrac{5}{64} ({\Sv})^2 \kappa^{a} \partial_s{\kappa_{a}}
 + \mytfrac{3}{32} \widetilde{\kappa}^{a} {\Sw} \partial_s{\kappa_{a}}
 + \mytfrac{3}{32} {\Sv} {\Omega}^{a} \partial_s{\kappa_{a}}
 + \mytfrac{3}{32} {\Sv} {\omega}^{a} \partial_s{\kappa_{a}}\nonumber\\
& -  \mytfrac{3}{64} \kappa^{a} \partial_s{{\Sv}} \partial_s{\kappa_{a}}
 + \mytfrac{3}{64} {\Sv} \partial_s{\kappa_{a}} \partial_s{\kappa^{a}}
 + \mytfrac{1}{16} {\Sw} \partial_s{{\Sw}}
 + \mytfrac{1}{16} \para{\Omega} \partial_s{{\Sw}}
 + \mytfrac{1}{16} {\Sv} \kappa^{a} \partial_s{{\omega}_{a}}\nonumber\\
& + \mytfrac{3}{32} {\Omega}^{a} \partial_s{{\omega}_{a}}
 + \mytfrac{3}{32} {\omega}^{a} \partial_s{{\omega}_{a}}
 + \mytfrac{3}{64} \partial_s{\kappa^{a}} \partial_s{{\omega}_{a}}
 + \mytfrac{1}{64} \kappa_{a} \kappa^{a} \partial_t \bar{U}
 -  \mytfrac{1}{64} \kappa^{a} \partial_t {\omega}_{a}\nonumber\\
& + \mytfrac{9}{16} {\cal H}^2 \partial^2_s{{\Sv}}
 -  \mytfrac{33}{128} \kappa_{a} \kappa^{a} \partial^2_s{{\Sv}}
 -  \mytfrac{9}{64} \partial_s{{\Sv}} \partial^2_s{{\Sv}}
 + \mytfrac{9}{64} {\Sv} \kappa^{a} \partial^2_s{\kappa_{a}}
 + \mytfrac{9}{64} \kappa^{a} \partial^2_s{{\omega}_{a}}
 + \mytfrac{9}{32} {\cal H} \partial^3_s{{\Sv}}\nonumber\\
& + \mytfrac{3}{64} \partial^4_s{{\Sv}}\\
\Sv^{(1)}_a=&\frac{\ST}{4\ra} \partial_s{\widetilde{\kappa}_{a}}+\mytfrac{1}{4} {\Sv} \kappa_{a} {\Sw}
 + \mytfrac{1}{4} {\Sv} \kappa_{a} \para{\Omega}
 + \mytfrac{1}{4} {\Sw} {\Omega}_{a}
 + \mytfrac{1}{4} \para{\Omega} {\Omega}_{a}
 + \mytfrac{1}{4} {\Sw} {\omega}_{a}
 + \mytfrac{1}{4} \para{\Omega} {\omega}_{a}
 -  \mytfrac{1}{4} \kappa^{b} \widetilde{\kappa}_{a} {\omega}_{b}
 \nonumber\\
& + \mytfrac{1}{8} \kappa_{a} \widetilde{\kappa}^{b} {\omega}_{b}+ \mytfrac{1}{4} \kappa_{b} \kappa^{b} \widetilde{\omega}_{a}
 + \mytfrac{3}{8} {\cal H} \widetilde{\kappa}_{a} \partial_s{{\Sv}}
 -  \mytfrac{3}{8} {\Sv} \widetilde{\kappa}_{a} \partial_s{{\Sv}}
 -  \mytfrac{3}{8} \widetilde{\Omega}_{a} \partial_s{{\Sv}}
 -  \mytfrac{3}{8} \widetilde{\omega}_{a} \partial_s{{\Sv}}
 -  \mytfrac{1}{4} {\Sw} \partial_s{\kappa_{a}}\nonumber\\
& + \mytfrac{1}{4} \partial_s{{\Sv}} \partial_s{\widetilde{\kappa}_{a}}
 -  \mytfrac{3}{8} \kappa_{a} \partial_s{{\Sw}}
 + \mytfrac{9}{8} \widetilde{\kappa}_{a} \partial^2_s{{\Sv}}
 -  \mytfrac{1}{4} {\Sv} \partial^2_s{\widetilde{\kappa}_{a}}
 -  \mytfrac{1}{4} \partial^2_s{\widetilde{\omega}_{a}}\\
\stf{V}^{(2)}_a=&\frac{\ST}{\ra}\left(- \mytfrac{5}{32} \varepsilon_{a}{}^{c} \mathcal{R}_ {bc} \kappa^{b}
 + \mytfrac{5}{128} {\cal H}^2 \widetilde{\kappa}_{a}
 + \mytfrac{5}{128} \kappa_{b} \kappa^{b} \widetilde{\kappa}_{a}
 -  \mytfrac{5}{128} \widetilde{\kappa}_ {a} \partial_s {\cal H}
 -  \mytfrac{1}{64} {\cal H} \partial_s{\widetilde{\kappa}_{a}}
 + \mytfrac{1}{64} \partial^2_s{\widetilde{\kappa}_{a}}\right)\nonumber\\
&- \mytfrac{1}{32} \varepsilon_{a}{}^{c} {{\stf{V}}^{(1)}_{bc}} \kappa^{b}
 -  \mytfrac{5}{192} \Omega^2 \widetilde{\kappa}_{a}
 -  \mytfrac{5}{64} ({\Sw})^2 \widetilde{\kappa}_{a}
 -  \mytfrac{1}{256} (\partial_s{{\Sv}})^2 \widetilde{\kappa}_{a}
 -  \mytfrac{5}{192} ({\Sv})^2 \kappa_{b} \kappa^{b} \widetilde{\kappa}_{a}\nonumber\\
& + \mytfrac{25}{256} \kappa_{a} \kappa_{b} \kappa^{b} {\Sw}
 -  \mytfrac{1}{16} \widetilde{\kappa}_{a} {\Sw} \para{\Omega}
 + \mytfrac{1}{24} \widetilde{\kappa}_{a} \para{\Omega}^2
 -  \mytfrac{3}{32} {\Sv} \kappa^{b} \widetilde{\kappa}_{a} {\Omega}_{b}
 -  \mytfrac{1}{24} \widetilde{\kappa}_{a} {\Omega}_{b} {\Omega}^{b}
 + \mytfrac{1}{24} {\Sv} \kappa_{b} \kappa^{b} \widetilde{\Omega}_{a}\nonumber\\
& + \mytfrac{1}{24} \kappa^{b} {\Omega}_{b} \widetilde{\Omega}_{a}
 + \mytfrac{3}{16} {\cal H} \kappa^{b} \widetilde{\kappa}_{a} {\omega}_{b}
 -  \mytfrac{5}{48} {\Sv} \kappa^{b} \widetilde{\kappa}_{a} {\omega}_{b}
 -  \mytfrac{7}{48} \widetilde{\kappa}_{a} {\Omega}^{b} {\omega}_{b}
 + \mytfrac{1}{24} \kappa^{b} \widetilde{\Omega}_{a} {\omega}_{b}\nonumber\\
& -  \mytfrac{5}{64} \widetilde{\kappa}_{a} {\omega}_{b} {\omega}^{b}
 -  \mytfrac{3}{16} {\cal H} \kappa_{b} \kappa^{b} \widetilde{\omega}_{a}
 + \mytfrac{5}{96} {\Sv} \kappa_{b} \kappa^{b} \widetilde{\omega}_{a}
 + \mytfrac{5}{96} \kappa^{b} {\Omega}_{b} \widetilde{\omega}_{a}
 + \mytfrac{5}{96} \kappa^{b} {\omega}_{b} \widetilde{\omega}_{a}\nonumber\\
& -  \mytfrac{5}{32} {\cal H} {\Sv} \widetilde{\kappa}_{a} \partial_s{{\Sv}}
 + \mytfrac{235}{1536} \kappa_{b} \kappa^{b} \widetilde{\kappa}_{a} \partial_s{{\Sv}}
 -  \mytfrac{1}{16} \kappa_{a} {\Sw} \partial_s{{\Sv}}
 + \mytfrac{5}{64} \kappa_{a} \para{\Omega} \partial_s{{\Sv}}
 -  \mytfrac{5}{32} {\cal H} \widetilde{\Omega}_{a} \partial_s{{\Sv}}\nonumber\\
& -  \mytfrac{5}{32} {\cal H} \widetilde{\omega}_{a} \partial_s{{\Sv}}
 + \mytfrac{9}{64} \widetilde{\kappa}_{a} \partial_s {\cal H} \partial_s{{\Sv}}
 + \mytfrac{3}{32} {\Sv} {\Sw} \partial_s{\kappa_{a}}
 + \mytfrac{3}{32} {\Sv} \para{\Omega} \partial_s{\kappa_{a}}
 -  \mytfrac{35}{384} {\Sv} \kappa^{b} \widetilde{\kappa}_{a} \partial_s{\kappa_{b}}\nonumber\\
& -  \mytfrac{7}{96} \widetilde{\kappa}_{a} {\omega}^{b} \partial_s{\kappa_{b}}
 + \mytfrac{19}{96} \kappa^{b} \widetilde{\omega}_{a} \partial_s{\kappa_{b}}
 + \mytfrac{25}{256} {\Sv} \kappa_{b} \kappa^{b} \partial_s{\widetilde{\kappa}_{a}}
 -  \mytfrac{1}{8} \kappa^{b} {\omega}_{b} \partial_s{\widetilde{\kappa}_{a}}
 + \mytfrac{7}{32} {\cal H} \partial_s{{\Sv}} \partial_s{\widetilde{\kappa}_{a}}\nonumber\\
& -  \mytfrac{9}{64} {\Sv} \partial_s{{\Sv}} \partial_s{\widetilde{\kappa}_{a}}
 + \mytfrac{3}{16} {\cal H} \kappa_{a} \partial_s{{\Sw}}
 -  \mytfrac{1}{96} {\Sv} \kappa_{a} \partial_s{{\Sw}}
 -  \mytfrac{1}{96} {\Omega}_{a} \partial_s{{\Sw}}
 -  \mytfrac{1}{96} {\omega}_{a} \partial_s{{\Sw}}
 -  \mytfrac{5}{48} \partial_s{\kappa_{a}} \partial_s{{\Sw}}\nonumber\\
& + \mytfrac{3}{32} {\Sw} \partial_s{{\omega}_{a}}
 + \mytfrac{3}{32} \para{\Omega} \partial_s{{\omega}_{a}}
 -  \mytfrac{83}{384} \kappa^{b} \widetilde{\kappa}_{a} \partial_s{{\omega}_{b}}
 + \mytfrac{57}{256} \kappa_{b} \kappa^{b} \partial_s{\widetilde{\omega}_{a}}
 -  \mytfrac{9}{64} \partial_s{{\Sv}} \partial_s{\widetilde{\omega}_{a}}\nonumber\\
& + \mytfrac{9}{64} {\cal H} \widetilde{\kappa}_{a} \partial^2_s{{\Sv}}
 -  \mytfrac{1}{16} {\Sv} \widetilde{\kappa}_{a} \partial^2_s{{\Sv}}
 -  \mytfrac{1}{16} \widetilde{\Omega}_{a} \partial^2_s{{\Sv}}
 -  \mytfrac{1}{16} \widetilde{\omega}_{a} \partial^2_s{{\Sv}}
 + \mytfrac{13}{64} \partial_s{\widetilde{\kappa}_{a}} \partial^2_s{{\Sv}}
 -  \mytfrac{1}{64} {\Sw} \partial^2_s{\kappa_{a}}\nonumber\\
& + \mytfrac{5}{128} \partial_s{{\Sv}} \partial^2_s{\widetilde{\kappa}_{a}}
 -  \mytfrac{9}{64} \kappa_{a} \partial^2_s{{\Sw}}
 + \mytfrac{9}{128} \widetilde{\kappa}_{a} \partial^3_s{{\Sv}}
 -  \mytfrac{1}{64} {\Sv} \partial^3_s{\widetilde{\kappa}_{a}}
 -  \mytfrac{1}{64} \partial^3_s{\widetilde{\omega}_{a}}\\
\stf{V}^{(1)}_{ab}=&\frac{\ST}{\ra}\left(\mathcal{R}_{ab} +\mytfrac{1}{3} \kappa_{\langle a} \kappa_{b \rangle
  } \right)+\kappa_{\langle a} \kappa_{b \rangle} \left[\mytfrac{1}{6}
  ({\Sv})^2 + \mytfrac{5}{8} \partial_s{{\Sv}} \right]
+\mytfrac{1}{4}\kappa_{\langle a} \tkappa_{b \rangle}  {\Sw}  -  \mytfrac{1}{4} \Sv\kappa_{\langle a}\partial_s{\kappa_{b\rangle}}\nonumber\\
&-  \mytfrac{1}{4} \kappa_{\langle a} \partial_s{\omega}_{b  \rangle}+\mytfrac{1}{6}\left({\omega}_{\langle a}+{\Omega}_{\langle a}\right) \left({\omega}_{b\rangle}+{\Omega}_{b\rangle}\right)+ \frac{1}{3}{\Sv} \kappa_{\langle a} \left({\omega}_{b \rangle}+{\Omega}_{b \rangle}\right) \label{Vh1ab}\,.
\end{align}
From the incompressibility constraint \eqref{EqCD} we also obtain
\begin{align}
P_\myes^{(1)}=&\frac{\ST}{4\ra}\left(- {\cal H}^2 -
  \kappa_{a} \kappa^{a} + \partial_s {\cal
    H}\right) \\
&+\mytfrac{1}{2} \Omega^2
 + \mytfrac{1}{2} ({\Sw})^2
 -  \mytfrac{3}{8} (\partial_s{{\Sv}})^2
 + \mytfrac{1}{2} ({\Sv})^2 \kappa_{a} \kappa^{a}
 + {\Sw} \para{\Omega}
 + {\Sv} \kappa^{a} {\Omega}_{a}
 + {\Sv} \kappa^{a} {\omega}_{a}
 + {\Omega}^{a} {\omega}_{a}
 + \mytfrac{1}{2} {\omega}_{a} {\omega}^{a}\nonumber\\
& -  \mytfrac{1}{8} \kappa_{a} \kappa^{a} \partial_s{{\Sv}}
 + \mytfrac{1}{4} {\Sv} \kappa^{a} \partial_s{\kappa_{a}}
 + \mytfrac{1}{4} \kappa^{a} \partial_s{{\omega}_{a}}
 + \mytfrac{1}{4} \partial^3_s{{\Sv}}\nonumber\\
P_a^{(1)}=&\frac{\ST}{\ra}\left( - \mytfrac{3}{4} \varepsilon_{a}{}^{c} \mathcal{R}_ {bc} \kappa^{b}
 + \mytfrac{3}{16} {\cal H}^2 \widetilde{\kappa}_{a}
 + \mytfrac{3}{16} \kappa_{b} \kappa^{b} \widetilde{\kappa}_{a}
 -  \mytfrac{3}{16} \widetilde{\kappa}_ {a} \partial_s {\cal H}
 + \mytfrac{1}{8} {\cal H} \partial_s{\widetilde{\kappa}_{a}}
 -  \mytfrac{1}{8} \partial^2_s{\widetilde{\kappa}_{a}}\right)\\
&- \mytfrac{1}{8} \Omega^2 \widetilde{\kappa}_{a}
 -  \mytfrac{3}{8} ({\Sw})^2 \widetilde{\kappa}_{a}
 + \mytfrac{9}{32} (\partial_s{{\Sv}})^2 \widetilde{\kappa}_{a}
 -  \mytfrac{1}{8} ({\Sv})^2 \kappa_{b} \kappa^{b} \widetilde{\kappa}_{a}
 -  \mytfrac{3}{16} \kappa_{a} \kappa_{b} \kappa^{b} {\Sw}
 -  \mytfrac{1}{2} \widetilde{\kappa}_{a} {\Sw} \para{\Omega}\nonumber\\
& -  \mytfrac{1}{4} {\Sv} \kappa^{b} \widetilde{\kappa}_{a} {\Omega}_{b}
 -  \mytfrac{1}{2} {\Sv} \kappa^{b} \widetilde{\kappa}_{a} {\omega}_{b}
 -  \mytfrac{1}{2} \widetilde{\kappa}_{a} {\Omega}^{c} {\omega}_{c}
 -  \mytfrac{3}{8} \widetilde{\kappa}_{a} {\omega}_{c} {\omega}^{c}
 + \mytfrac{1}{4} {\Sv} \kappa_{b} \kappa^{b} \widetilde{\omega}_{a}
 + \mytfrac{1}{4} \kappa^{b} {\Omega}_{b} \widetilde{\omega}_{a}\nonumber\\
& + \mytfrac{1}{4} \kappa^{b} {\omega}_{b} \widetilde{\omega}_{a}
 + \mytfrac{3}{4} {\cal H} {\Sv} \widetilde{\kappa}_{a} \partial_s{{\Sv}}
 + \mytfrac{1}{8} \kappa_{b} \kappa^{b} \widetilde{\kappa}_{a} \partial_s{{\Sv}}
 -  \mytfrac{1}{2} \kappa_{a} {\Sw} \partial_s{{\Sv}}
 -  \mytfrac{1}{8} \kappa_{a} \para{\Omega} \partial_s{{\Sv}}
 + \mytfrac{3}{4} {\cal H} \widetilde{\Omega}_{a} \partial_s{{\Sv}}\nonumber\\
& + \mytfrac{3}{4} {\cal H} \widetilde{\omega}_{a} \partial_s{{\Sv}}
 + \mytfrac{1}{4} {\Sv} {\Sw} \partial_s{\kappa_{a}}
 + \mytfrac{1}{4} {\Sv} \para{\Omega} \partial_s{\kappa_{a}}
 -  \mytfrac{1}{16} {\Sv} \kappa^{b} \widetilde{\kappa}_{a} \partial_s{\kappa_{b}}
 -  \mytfrac{3}{16} {\Sv} \kappa_{b} \kappa^{b} \partial_s{\widetilde{\kappa}_{a}}
 -  \mytfrac{3}{8} {\Sv} \partial_s{{\Sv}} \partial_s{\widetilde{\kappa}_{a}}\nonumber\\
& -  \mytfrac{1}{4} {\Sv} \kappa_{a} \partial_s{{\Sw}}
 -  \mytfrac{1}{4} {\Omega}_{a} \partial_s{{\Sw}}
 -  \mytfrac{1}{4} {\omega}_{a} \partial_s{{\Sw}}
 + \mytfrac{1}{4} \partial_s{\kappa_{a}} \partial_s{{\Sw}}
 + \mytfrac{1}{4} {\Sw} \partial_s{{\omega}_{a}}
 + \mytfrac{1}{4} \para{\Omega} \partial_s{{\omega}_{a}}\nonumber\\
& -  \mytfrac{1}{16} \kappa^{b} \widetilde{\kappa}_{a} \partial_s{{\omega}_{b}}
 -  \mytfrac{3}{16} \kappa_{b} \kappa^{b} \partial_s{\widetilde{\omega}_{a}}
 -  \mytfrac{3}{8} \partial_s{{\Sv}} \partial_s{\widetilde{\omega}_{a}}
 + \mytfrac{1}{4} {\Sv} \widetilde{\kappa}_{a} \partial^2_s{{\Sv}}
 + \mytfrac{1}{4} \widetilde{\Omega}_{a} \partial^2_s{{\Sv}}
 + \mytfrac{1}{4} \widetilde{\omega}_{a} \partial^2_s{{\Sv}}\nonumber\\
& -  \mytfrac{1}{8} \partial_s{\widetilde{\kappa}_{a}} \partial^2_s{{\Sv}}
 + \mytfrac{1}{8} {\Sw} \partial^2_s{\kappa_{a}}
 + \mytfrac{3}{16} \partial_s{{\Sv}} \partial^2_s{\widetilde{\kappa}_{a}}
 + \mytfrac{1}{8} \kappa_{a} \partial^2_s{{\Sw}}
 -  \mytfrac{3}{8} \widetilde{\kappa}_{a} \partial^3_s{{\Sv}}
 + \mytfrac{1}{8} {\Sv} \partial^3_s{\widetilde{\kappa}_{a}}
 + \mytfrac{1}{8} \partial^3_s{\widetilde{\omega}_{a}}\,.\nonumber
\end{align}

\section{Higher order corrections for curved fibers}\label{SecFinal}

Once the boundary constraints of~\S~\ref{Secmonopdipole} and
Appendix~\ref{SecExtraConstraint} are replaced, the dynamical evolution for the longitudinal velocity and the FCL velocity are given by

\begin{align}
\partial_t \Sv =&\para{g}
 + \frac{\ST{\cal H}}{{\ra}}
 -  \overline{I}
 - 2 U^{a} \widetilde{\Omega}_{a}
 -  U^{a} \widetilde{\omega}_{a}
 + 6 {\cal H} \partial_s{{\Sv}}
 -  {\Sv} \partial_s{{\Sv}}
 -  \partial_t \bar{U}
+ 3 \partial^2_s{{\Sv}}
 + {\ST\ra} (\mytfrac{5}{4} {\cal H}^3
 \nonumber\\
& + \mytfrac{1}{4} {\cal H} \kappa_{a} \kappa^{a}+ \mytfrac{15}{4} {\cal H} \partial_s {\cal H}
 + \mytfrac{1}{2} \kappa^{a} \partial_s{\kappa_{a}}
 + \mytfrac{5}{4} \partial_s^{2} {\cal H})
 + {\ra}^2 \Bigl(\Omega^2 {\cal H}
 - 8 {{\Sv}^{(2)}_\myes}
 + {\cal H} {\Sv}^2 \kappa_{a} \kappa^{a}\nonumber\\
& - 5 {{\Sv}^{(1)a}} \widetilde{\kappa}_{a}
 + {\cal H} {\Sw}^2
 + 2 {\cal H} {\Sw} \para{\Omega}
 + 2 {\cal H} {\Sv} \kappa^{a} {\Omega}_{a}
 + \mytfrac{1}{2} \kappa^{a} {\Sw} \widetilde{\Omega}_{a}
 + 2 {\cal H} {\Sv} \kappa^{a} {\omega}_{a}
 + \mytfrac{1}{2} \widetilde{\kappa}^{a} {\Sw} {\omega}_{a}\nonumber\\
& + \widetilde{\kappa}^{a} \para{\Omega} {\omega}_{a}
 + 2 {\cal H} {\Omega}^{a} {\omega}_{a}
 + {\cal H} {\omega}_{a} {\omega}^{a}
 + 12 {\cal H}^3 \partial_s{{\Sv}}
 -  \mytfrac{9}{4} {\cal H} \kappa_{a} \kappa^{a} \partial_s{{\Sv}}
 + \mytfrac{5}{4} {\Sv} \kappa_{a} \kappa^{a} \partial_s{{\Sv}}\nonumber\\
& + \mytfrac{5}{4} \kappa^{a} {\Omega}_{a} \partial_s{{\Sv}}
 + \mytfrac{5}{4} \kappa^{a} {\omega}_{a} \partial_s{{\Sv}}
 + 21 {\cal H} \partial_s {\cal H} \partial_s{{\Sv}}
 -  \mytfrac{3}{4} {\cal H} (\partial_s{{\Sv}})^2
 + \mytfrac{3}{2} {\cal H} {\Sv} \kappa^{a} \partial_s{\kappa_{a}}\nonumber\\
& + \mytfrac{1}{2} {\Sv}^2 \kappa^{a} \partial_s{\kappa_{a}}
 + \mytfrac{1}{4} \widetilde{\kappa}^{a} {\Sw} \partial_s{\kappa_{a}}
 + \mytfrac{1}{2} {\Sv} {\Omega}^{a} \partial_s{\kappa_{a}}
 + \mytfrac{1}{2} {\Sv} {\omega}^{a} \partial_s{\kappa_{a}}
 + \mytfrac{1}{8} \kappa^{a} \partial_s{{\Sv}} \partial_s{\kappa_{a}}
 + {\Sw} \partial_s{{\Sw}}\nonumber\\
& + \para{\Omega} \partial_s{{\Sw}}
 + \mytfrac{3}{2} {\cal H} \kappa^{a} \partial_s{{\omega}_{a}}
 + \mytfrac{1}{2} {\Sv} \kappa^{a} \partial_s{{\omega}_{a}}
 + \mytfrac{1}{2} {\Omega}^{a} \partial_s{{\omega}_{a}}
 + \mytfrac{1}{2} {\omega}^{a} \partial_s{{\omega}_{a}}
 + 3 \partial_s{{\Sv}} \partial_s^{2} {\cal H}\nonumber\\
& + 18 {\cal H}^2 \partial^2_s{{\Sv}}
 -  \mytfrac{3}{8} \kappa_{a} \kappa^{a} \partial^2_s{{\Sv}}
 + 6 \partial_s {\cal H} \partial^2_s{{\Sv}}
 -  \mytfrac{3}{4} \partial_s{{\Sv}} \partial^2_s{{\Sv}}
 + \mytfrac{1}{4} {\Sv} \kappa^{a} \partial^2_s{\kappa_{a}}
 + \mytfrac{1}{4} \kappa^{a} \partial^2_s{{\omega}_{a}}\nonumber\\
& + 6 {\cal H} \partial^3_s{{\Sv}}
 + \mytfrac{3}{4} \partial^4_s{{\Sv}}\Bigr)+\calO{\epsilon_\ra^3}\\
(\partial_t U)_a =& g_{a}
 -  \frac{\ST\widetilde{\kappa}_{a}}{{\ra}}
 + ({\Sv})^2 \widetilde{\kappa}_{a}
 - 2 \tilde{U}_{a} \para{\Omega}
 -  I_{a}
 + 2 \bar{U} \widetilde{\Omega}_{a}
 + 2 {\Sv} \widetilde{\Omega}_{a}
 + 2 {\Sv} \widetilde{\omega}_{a}
 - 3 \widetilde{\kappa}_{a} \partial_s{{\Sv}}\nonumber\\
& + {\ST\ra} (\mytfrac{3}{4} \varepsilon_{a}{}^{c} \mathcal{R}_{bc} \kappa^{b}
 -  \mytfrac{53}{16} {\cal H}^2 \widetilde{\kappa}_{a}
 -  \mytfrac{9}{16} \kappa_{b} \kappa^{b} \widetilde{\kappa}_{a}
 -  \mytfrac{35}{16} \widetilde{\kappa}_{a} \partial_s {\cal H}
 -  \mytfrac{7}{8} {\cal H} \partial_s{\widetilde{\kappa}_{a}}
 -  \mytfrac{1}{8} \partial^2_s{\widetilde{\kappa}_{a}})\nonumber\\
& + {\ra}^2 \Bigl(- \mytfrac{24}{5} {{\stf{V}}^{(2)}_{a}}
 - 8 {\cal H} {{\Sv}^{(1)}_{a}}
 -  \mytfrac{33}{20} \varepsilon_{a}{}^{c} {{\stf{V}}^{(1)}_{bc}} \kappa^{b}
 -  \mytfrac{1}{8} \Omega^2 \widetilde{\kappa}_{a}
 -  \mytfrac{1}{8} {\Sv}^2 \kappa_{b} \kappa^{b} \widetilde{\kappa}_{a}
 + 12 {\cal H}^2 \kappa_{a} {\Sw}\nonumber\\
& + \mytfrac{9}{32} \kappa_{a} \kappa_{b} \kappa^{b} {\Sw}
 -  \mytfrac{5}{8} \widetilde{\kappa}_{a} {\Sw}^2
 -  \widetilde{\kappa}_{a} {\Sw} \para{\Omega}
 -  \mytfrac{1}{4} {\Sv} \kappa^{b} \widetilde{\kappa}_{a} {\Omega}_{b}
 -  \mytfrac{1}{4} {\Sv} \kappa^{b} \widetilde{\kappa}_{a} {\omega}_{b}
 + \mytfrac{29}{10} {\cal H} \kappa_{a} \widetilde{\kappa}^{b} {\omega}_{b}\nonumber\\
& -  \mytfrac{1}{2} \widetilde{\kappa}_{a} {\Omega}^{b} {\omega}_{b}
 -  \mytfrac{3}{8} \widetilde{\kappa}_{a} {\omega}_{b} {\omega}^{b}
 + \mytfrac{1}{4} \kappa^{b} {\Omega}_{b} \widetilde{\omega}_{a}
 + \mytfrac{1}{4} \kappa^{b} {\omega}_{b} \widetilde{\omega}_{a}
 + 3 \kappa_{a} {\Sw} \partial_s {\cal H}
 - 18 {\cal H}^2 \widetilde{\kappa}_{a} \partial_s{{\Sv}}\nonumber\\
& + \mytfrac{3}{4} {\cal H} {\Sv} \widetilde{\kappa}_{a} \partial_s{{\Sv}}
 -  \mytfrac{19}{64} \kappa_{b} \kappa^{b} \widetilde{\kappa}_{a} \partial_s{{\Sv}}
 -  \mytfrac{1}{4} \kappa_{a} {\Sw} \partial_s{{\Sv}}
 -  \mytfrac{3}{8} \kappa_{a} \para{\Omega} \partial_s{{\Sv}}
 + \mytfrac{3}{4} {\cal H} \widetilde{\Omega}_{a} \partial_s{{\Sv}}\nonumber\\
& + \mytfrac{3}{4} {\cal H} \widetilde{\omega}_{a} \partial_s{{\Sv}}
 -  \mytfrac{51}{8} \widetilde{\kappa}_{a} \partial_s {\cal H} \partial_s{{\Sv}}
 + \mytfrac{3}{32} \widetilde{\kappa}_{a} (\partial_s{{\Sv}})^2
 -  \mytfrac{6}{5} \partial_s{{{\Sv}^{(1)}_{a}}}
 + 4 {\cal H} {\Sw} \partial_s{\kappa_{a}}
 + \mytfrac{1}{4} {\Sv} {\Sw} \partial_s{\kappa_{a}}\nonumber\\
& + \mytfrac{3}{4} {\Sv} \para{\Omega} \partial_s{\kappa_{a}}
 + \mytfrac{19}{32} {\Sv} \kappa^{b} \widetilde{\kappa}_{a} \partial_s{\kappa_{b}}
 -  \mytfrac{9}{32} {\Sv} \kappa_{a} \widetilde{\kappa}^{b} \partial_s{\kappa_{b}}
 + 12 {\cal H}^2 {\Sv} \partial_s{\widetilde{\kappa}_{a}}
 + 3 {\Sv} \partial_s {\cal H} \partial_s{\widetilde{\kappa}_{a}}\nonumber\\
& + \mytfrac{1}{2} {\cal H} \partial_s{{\Sv}} \partial_s{\widetilde{\kappa}_{a}}
 -  \mytfrac{1}{8} {\Sv} \partial_s{{\Sv}} \partial_s{\widetilde{\kappa}_{a}}
 + \mytfrac{69}{10} {\cal H} \kappa_{a} \partial_s{{\Sw}}
 -  \mytfrac{1}{2} {\Sv} \kappa_{a} \partial_s{{\Sw}}
 -  \mytfrac{1}{4} {\Omega}_{a} \partial_s{{\Sw}}
 -  \mytfrac{1}{4} {\omega}_{a} \partial_s{{\Sw}}\nonumber\\
& + \mytfrac{3}{4} \partial_s{\kappa_{a}} \partial_s{{\Sw}}
 + \mytfrac{1}{4} {\Sw} \partial_s{{\omega}_{a}}
 + \mytfrac{3}{4} \para{\Omega} \partial_s{{\omega}_{a}}
 + \mytfrac{1}{4} \bar{\omega} \partial_s{{\omega}_{a}}
 + \mytfrac{19}{32} \kappa^{b} \widetilde{\kappa}_{a} \partial_s{{\omega}_{b}}
 -  \mytfrac{9}{32} \kappa_{a} \widetilde{\kappa}^{b} \partial_s{{\omega}_{b}}\nonumber\\
& + 12 {\cal H}^2 \partial_s{\widetilde{\omega}_{a}}
 + 3 \partial_s {\cal H} \partial_s{\widetilde{\omega}_{a}}
 + \mytfrac{1}{8} \partial_s{{\Sv}} \partial_s{\widetilde{\omega}_{a}}
 -  \mytfrac{1}{4} \partial_s{\widetilde{\kappa}_{a}} \partial_t {\Sv}
 -  \mytfrac{1}{4} \kappa_{a} \partial_t {\Sw}
 -  \mytfrac{63}{8} {\cal H} \widetilde{\kappa}_{a} \partial^2_s{{\Sv}}\nonumber\\
& + \mytfrac{3}{8} {\Sv} \widetilde{\kappa}_{a} \partial^2_s{{\Sv}}
 + \mytfrac{1}{4} \widetilde{\Omega}_{a} \partial^2_s{{\Sv}}
 + \mytfrac{1}{4} \widetilde{\omega}_{a} \partial^2_s{{\Sv}}
 -  \mytfrac{3}{8} \partial_s{\widetilde{\kappa}_{a}} \partial^2_s{{\Sv}}
 + \mytfrac{3}{8} {\Sw} \partial^2_s{\kappa_{a}}
 + 4 {\cal H} {\Sv} \partial^2_s{\widetilde{\kappa}_{a}}\nonumber\\
& -  \mytfrac{1}{4} {\Sv}^2 \partial^2_s{\widetilde{\kappa}_{a}}
 + \mytfrac{9}{16} \partial_s{{\Sv}} \partial^2_s{\widetilde{\kappa}_{a}}
 + \mytfrac{3}{8} \kappa_{a} \partial^2_s{{\Sw}}
 + 4 {\cal H} \partial^2_s{\widetilde{\omega}_{a}}
 -  \mytfrac{1}{2} {\Sv} \partial^2_s{\widetilde{\omega}_{a}}
 + \mytfrac{1}{8} \widetilde{\kappa}_{a} \partial_s \partial_t {\Sv}\nonumber\\
& -  \mytfrac{1}{4} \partial_s \partial_t \widetilde{\omega}_{a}
 -  \mytfrac{21}{16} \widetilde{\kappa}_{a} \partial^3_s{{\Sv}}
 + \mytfrac{3}{8} {\Sv} \partial^3_s{\widetilde{\kappa}_{a}}
 + \mytfrac{3}{8} \partial^3_s{\widetilde{\omega}_{a}}\Bigr)+\calO{\epsilon_\ra^3}\,.
\end{align}

If the higher moments of Navier-Stokes equation and the
incompressibility constraint are then used to replace the unknown
variables as summarized in Table~\ref{Table5}, and also if time derivatives in
corrective terms are replaced using the lower order dynamical
equations, we obtain the closed and final results with order
$\epsilon_\ra^2$ corrections included
\begin{align}
\partial_t \Sv =&\para{g}
 + \frac{\ST{\cal H}}{{\ra}}
 -  \overline{I}
 - 2 U^{a} \widetilde{\Omega}_{a}
 -  U^{a} \widetilde{\omega}_{a}
 + 6 {\cal H} \partial_s{{\Sv}}
 -  {\Sv} \partial_s{{\Sv}}
 -  \partial_t \bar{U}
+ 3 \partial^2_s{{\Sv}}
 + \ST{\ra} (\mytfrac{7}{4} {\cal H}^3
 \nonumber\\
& 
+ \mytfrac{3}{4} {\cal H} \kappa_{a} \kappa^{a}
+ \mytfrac{15}{4} {\cal H} \partial_s {\cal H}
 + \mytfrac{1}{4} \kappa^{a} \partial_s{\kappa_{a}}
 + \partial_s^{2} {\cal H})
 + {\ra}^2 \Bigl(\Omega^2 {\cal H}
 + \mytfrac{1}{4} {\cal H} {\Sv}^2 \kappa_{a} \kappa^{a}
 + {\cal H} {\Sw}^2\nonumber\\
& + 2 {\cal H} {\Sw} \para{\Omega}
 + \mytfrac{1}{2} {\cal H} {\Sv} \kappa^{a} {\Omega}_{a}
 -  \mytfrac{3}{4} {\cal H} {\Omega}_{a} {\Omega}^{a}
 + \kappa^{a} {\Sw} \widetilde{\Omega}_{a}
 + \mytfrac{5}{4} \kappa^{a} \para{\Omega} \widetilde{\Omega}_{a}
 + \mytfrac{1}{2} {\cal H} {\Sv} \kappa^{a} {\omega}_{a}\nonumber\\
& -  \mytfrac{1}{2} \widetilde{\kappa}^{a} {\Sw} {\omega}_{a}
 -  \mytfrac{3}{4} \widetilde{\kappa}^{a} \para{\Omega} {\omega}_{a}
 + \mytfrac{1}{2} {\cal H} {\Omega}^{a} {\omega}_{a}
 + \mytfrac{1}{4} {\cal H} {\omega}_{a} {\omega}^{a}
 + 12 {\cal H}^3 \partial_s{{\Sv}}
 - 3 {\cal H} \kappa_{a} \kappa^{a} \partial_s{{\Sv}}\nonumber\\
& + 2 {\Sv} \kappa_{a} \kappa^{a} \partial_s{{\Sv}}
 + 2 \kappa^{a} {\Omega}_{a} \partial_s{{\Sv}}
 + 2 \kappa^{a} {\omega}_{a} \partial_s{{\Sv}}
 + \mytfrac{33}{2} {\cal H} \partial_s {\cal H} \partial_s{{\Sv}}
 + \mytfrac{3}{4} {\cal H} (\partial_s{{\Sv}})^2\nonumber\\
& + \mytfrac{3}{2} {\cal H} {\Sv} \kappa^{a} \partial_s{\kappa_{a}}
 -  \mytfrac{1}{4} {\Sv}^2 \kappa^{a} \partial_s{\kappa_{a}}
 + \mytfrac{3}{4} \widetilde{\kappa}^{a} {\Sw} \partial_s{\kappa_{a}}
 -  \mytfrac{1}{4} {\Sv} {\Omega}^{a} \partial_s{\kappa_{a}}
 -  \mytfrac{1}{4} {\Sv} {\omega}^{a} \partial_s{\kappa_{a}}\nonumber\\
& -  \mytfrac{3}{8} \kappa^{a} \partial_s{{\Sv}} \partial_s{\kappa_{a}}
 -  \mytfrac{3}{8} {\Sv} \partial_s{\kappa_{a}} \partial_s{\kappa^{a}}
 + \mytfrac{1}{2} {\Sw} \partial_s{{\Sw}}
 + \mytfrac{1}{2} \para{\Omega} \partial_s{{\Sw}}
 + \mytfrac{3}{2} {\cal H} \kappa^{a} \partial_s{{\omega}_{a}}
 -  \mytfrac{1}{4} {\Sv} \kappa^{a} \partial_s{{\omega}_{a}}\nonumber\\
& -  \mytfrac{1}{4} {\Omega}^{a} \partial_s{{\omega}_{a}}
 -  \mytfrac{1}{4} {\omega}^{a} \partial_s{{\omega}_{a}}
 -  \mytfrac{3}{8} \partial_s{\kappa^{a}} \partial_s{{\omega}_{a}}
 + 3 \partial_s{{\Sv}} \partial_s^{2} {\cal H}
 + \mytfrac{27}{2} {\cal H}^2 \partial^2_s{{\Sv}}
 -  \mytfrac{57}{16} \kappa_{a} \kappa^{a} \partial^2_s{{\Sv}}\nonumber\\
& + 6 \partial_s {\cal H} \partial^2_s{{\Sv}}
 + \mytfrac{3}{8} \partial_s{{\Sv}} \partial^2_s{{\Sv}}
 + \mytfrac{3}{8} {\Sv} \kappa^{a} \partial^2_s{\kappa_{a}}
 + \mytfrac{3}{8} \kappa^{a} \partial^2_s{{\omega}_{a}}
 + \mytfrac{15}{4} {\cal H} \partial^3_s{{\Sv}}
 + \mytfrac{3}{8} \partial^4_s{{\Sv}}\Bigr)+\calO{\epsilon_\ra^3}\label{Finalv}\\
(\partial_t U)_a =&g_{a}
 -  \frac{\ST \widetilde{\kappa}_{a}}{{\ra}}
 + ({\Sv})^2 \widetilde{\kappa}_{a}
 - 2 \tilde{U}_{a} \para{\Omega}
 -  I_{a}
 + 2 \bar{U} \widetilde{\Omega}_{a}
 + 2 {\Sv} \widetilde{\Omega}_{a}
 + 2 {\Sv} \widetilde{\omega}_{a}
 - 3 \widetilde{\kappa}_{a} \partial_s{{\Sv}}\nonumber\\
& + \ST{\ra} ( -  \mytfrac{31}{8} {\cal H}^2 \widetilde{\kappa}_{a}
 -  \mytfrac{7}{8} \kappa_{b} \kappa^{b} \widetilde{\kappa}_{a}
 -  \mytfrac{13}{8} \widetilde{\kappa}_{a} \partial_s {\cal H}
 -  \mytfrac{9}{4} {\cal H} \partial_s{\widetilde{\kappa}_{a}}
 -  \mytfrac{3}{4} \partial^2_s{\widetilde{\kappa}_{a}})\nonumber\\
& + {\ra}^2 \Bigl(\mytfrac{1}{2} {\Sv}^2 \kappa_{b} \kappa^{b} \widetilde{\kappa}_{a}
 + 12 {\cal H}^2 \kappa_{a} {\Sw}
 - 2 {\cal H} {\Sv} \kappa_{a} {\Sw}
 -  \mytfrac{1}{4} \widetilde{\kappa}_{a} {\Sw}^2
 - 2 {\cal H} {\Sv} \kappa_{a} \para{\Omega}
 -  \widetilde{\kappa}_{a} {\Sw} \para{\Omega}\nonumber\\
& -  \mytfrac{3}{4} \widetilde{\kappa}_{a} \para{\Omega}^2
 - 2 {\cal H} {\Sw} {\Omega}_{a}
 - 2 {\cal H} \para{\Omega} {\Omega}_{a}
 + {\Sv} \kappa^{b} \widetilde{\kappa}_{a} {\Omega}_{b}
 + \widetilde{\kappa}_{a} {\Omega}_{b} {\Omega}^{b}
 -  \mytfrac{1}{2} \kappa^{b} {\Omega}_{b} \widetilde{\Omega}_{a}\nonumber\\
& -  \mytfrac{1}{2} {\Sv} \kappa_{a} \kappa^{b} \widetilde{\Omega}_{b}
 - 2 {\cal H} {\Sw} {\omega}_{a}
 - 2 {\cal H} \para{\Omega} {\omega}_{a}
 + {\Sv} \kappa^{b} \widetilde{\kappa}_{a} {\omega}_{b}
 + 2 {\cal H} \kappa_{a} \widetilde{\kappa}^{b} {\omega}_{b}
 + \mytfrac{3}{2} \widetilde{\kappa}_{a} {\Omega}^{b} {\omega}_{b}\nonumber\\
& -  \mytfrac{1}{2} \kappa^{b} \widetilde{\Omega}_{a} {\omega}_{b}
 + \mytfrac{1}{2} \widetilde{\kappa}_{a} {\omega}_{b} {\omega}^{b}
 + 3 \kappa_{a} {\Sw} \partial_s {\cal H}
 - 21 {\cal H}^2 \widetilde{\kappa}_{a} \partial_s{{\Sv}}
 + \mytfrac{9}{2} {\cal H} {\Sv} \widetilde{\kappa}_{a} \partial_s{{\Sv}}\nonumber\\
& -  \mytfrac{27}{16} \kappa_{b} \kappa^{b} \widetilde{\kappa}_{a} \partial_s{{\Sv}}
 -  \mytfrac{1}{2} \kappa_{a} {\Sw} \partial_s{{\Sv}}
 -  \mytfrac{11}{4} \kappa_{a} \para{\Omega} \partial_s{{\Sv}}
 + \mytfrac{9}{2} {\cal H} \widetilde{\Omega}_{a} \partial_s{{\Sv}}
 + \mytfrac{9}{2} {\cal H} \widetilde{\omega}_{a} \partial_s{{\Sv}}\nonumber\\
& -  \mytfrac{27}{4} \widetilde{\kappa}_{a} \partial_s {\cal H} \partial_s{{\Sv}}
 + \mytfrac{15}{16} \widetilde{\kappa}_{a} (\partial_s{{\Sv}})^2
 + 6 {\cal H} {\Sw} \partial_s{\kappa_{a}}
 -  \mytfrac{1}{2} {\Sv} {\Sw} \partial_s{\kappa_{a}}
 -  \mytfrac{1}{2} {\Sv} \para{\Omega} \partial_s{\kappa_{a}}\nonumber\\
& + \mytfrac{3}{8} {\Sv} \kappa^{b} \widetilde{\kappa}_{a} \partial_s{\kappa_{b}}
 + \mytfrac{1}{2} \widetilde{\kappa}_{a} {\omega}^{b} \partial_s{\kappa_{b}}
 -  \kappa^{b} \widetilde{\omega}_{a} \partial_s{\kappa_{b}}
 + 12 {\cal H}^2 {\Sv} \partial_s{\widetilde{\kappa}_{a}}
 + \mytfrac{1}{2} \kappa^{b} {\omega}_{b} \partial_s{\widetilde{\kappa}_{a}}\nonumber\\
& + 3 {\Sv} \partial_s {\cal H} \partial_s{\widetilde{\kappa}_{a}}
 -  \mytfrac{9}{2} {\cal H} \partial_s{{\Sv}} \partial_s{\widetilde{\kappa}_{a}}
 + \mytfrac{9}{4} {\Sv} \partial_s{{\Sv}} \partial_s{\widetilde{\kappa}_{a}}
 + 8 {\cal H} \kappa_{a} \partial_s{{\Sw}}
 -  \mytfrac{1}{2} {\Sv} \kappa_{a} \partial_s{{\Sw}}\nonumber\\
& -  \mytfrac{1}{2} {\Omega}_{a} \partial_s{{\Sw}}
 -  \mytfrac{1}{2} {\omega}_{a} \partial_s{{\Sw}}
 + 2 \partial_s{\kappa_{a}} \partial_s{{\Sw}}
 -  \mytfrac{1}{2} {\Sw} \partial_s{{\omega}_{a}}
 -  \mytfrac{1}{2} \para{\Omega} \partial_s{{\omega}_{a}}
 + \mytfrac{3}{8} \kappa^{b} \widetilde{\kappa}_{a} \partial_s{{\omega}_{b}}\nonumber\\
& + \mytfrac{1}{2} \kappa_{a} \widetilde{\kappa}^{b} \partial_s{{\omega}_{b}}
 + 12 {\cal H}^2 \partial_s{\widetilde{\omega}_{a}}
 + 3 \partial_s {\cal H} \partial_s{\widetilde{\omega}_{a}}
 + \mytfrac{9}{4} \partial_s{{\Sv}} \partial_s{\widetilde{\omega}_{a}}
 -  \mytfrac{69}{4} {\cal H} \widetilde{\kappa}_{a} \partial^2_s{{\Sv}}\nonumber\\
& + \mytfrac{3}{2} {\Sv} \widetilde{\kappa}_{a} \partial^2_s{{\Sv}}
 + \mytfrac{3}{2} \widetilde{\Omega}_{a} \partial^2_s{{\Sv}}
 + \mytfrac{3}{2} \widetilde{\omega}_{a} \partial^2_s{{\Sv}}
 -  \mytfrac{21}{4} \partial_s{\widetilde{\kappa}_{a}} \partial^2_s{{\Sv}}
 + \mytfrac{3}{4} {\Sw} \partial^2_s{\kappa_{a}}
 + 6 {\cal H} {\Sv} \partial^2_s{\widetilde{\kappa}_{a}}\nonumber\\
& -  \mytfrac{3}{8} \partial_s{{\Sv}} \partial^2_s{\widetilde{\kappa}_{a}}
 + \mytfrac{5}{4} \kappa_{a} \partial^2_s{{\Sw}}
 + 6 {\cal H} \partial^2_s{\widetilde{\omega}_{a}}
 -  \mytfrac{27}{8} \widetilde{\kappa}_{a} \partial^3_s{{\Sv}}
 + \mytfrac{3}{4} {\Sv} \partial^3_s{\widetilde{\kappa}_{a}}
 + \mytfrac{3}{4} \partial^3_s{\widetilde{\omega}_{a}}\Bigr)+\calO{\epsilon_\ra^3}. \label{FinalUa}
\end{align}

\ifpre
\end{widetext}
\else
\fi

\end{document}